\documentclass[a4paper,fleqn,amssymb,usenatbib,useAMS,usedcolumns]{mnras}
\usepackage{times} 
\usepackage{latexsym,amssymb,amsmath,amsfonts}
\usepackage{rotating} 
\usepackage{float}
\usepackage{graphicx}
\usepackage{natbib}
\usepackage{longtable}
\usepackage{url}
\usepackage{dcolumn}
\usepackage{textcomp}
\usepackage{multicol}
\usepackage{bm}
\usepackage{pdflscape}
\usepackage[T1]{fontenc}
\usepackage{ae,aecompl}
\usepackage[caption=false]{subfig}
\begin{document}
\newcommand{\hi}{\mbox{H\,{\sc i}}}
\newcommand{\mgii}{\mbox{Mg\,{\sc ii}}}
\newcommand{\mgi}{\mbox{Mg\,{\sc i}}}
\newcommand{\feii}{\mbox{Fe\,{\sc ii}}}
\newcommand{\oi}{\mbox{O\,{\sc i}}}
\newcommand{\cii}{\mbox{C\,{\sc ii}}}
\newcommand{\ci}{\mbox{C\,{\sc i}}}
\newcommand{\sii}{\mbox{Si\,{\sc ii}}}
\newcommand{\znii}{\mbox{Zn~{\sc ii}}}
\newcommand{\caii}{\mbox{Ca\,{\sc ii}}}
\newcommand{\nai}{\mbox{Na\,{\sc i}}}
\def\h2{$\rm H_2$}
\def\Nh2{$N$(H${_2}$)}
\def\chin{$\chi^2_{\nu}$}
\def\chiu{$\chi_{\rm UV}$}
\def\sys{J0441$-$4313~}
\def\lya{\ensuremath{{\rm Ly}\alpha}}
\def\lymana{\ensuremath{{\rm Lyman}-\alpha}}
\def\kms{km\,s$^{-1}$}
\def\cms{cm$^{-2}$}
\def\cc{cm$^{-3}$}
\def\zabs{$z_{\rm abs}$}
\def\zem{$z_{\rm em}$}
\def\nhi{$N$($\hi$)}
\def\ln{log~$N$}
\def\nh{$n_{\rm H}$}
\def\ne{$n_{e}$}
\def\21{21-cm}
\def\ts{T$_{s}$}
\def\th{T$_{01}$}
\def\t0{T$_{0}$}
\def\ll{$\lambda\lambda$}
\def\l{$\lambda$}
\def\fc{$C_{f}$}
\def\c21{$C_{21}$}
\def\mjb{mJy~beam$^{-1}$}
\def\taudv{$\int\tau dv$}
\def\taup{$\tau_{\rm p}$}
\def\ha{H\,$\alpha$}
\def\hb{H\,$\beta$}
\def\oi{[O\,{\sc i}]}
\def\oii{[O\,{\sc ii}]}
\def\oiii{[O\,{\sc iii}]}
\def\nii{[N\,{\sc ii}]}
\def\sii{[S\,{\sc ii}]}
\def\taudvl{$\int\tau dv^{3\sigma}_{10}$}
\def\taudv{$\int\tau dv$}
\def\vshift{$v_{\rm shift}$}
\def\wmg{$W_{\mgii}$}
\def\wfe{$W_{\feii}$}
\def\dgi{$\Delta (g-i)$}
\def\ebv{$E(B-V)$}
%
%
\title[Neutral gas in merging galaxies]{
{Prevalence of neutral gas in centres of merging galaxies
}
\author[R. Dutta et al.]{R. Dutta$^1$ \thanks{E-mail: rdutta@eso.org}, R. Srianand$^2$\thanks{E-mail: anand@iucaa.in} and N. Gupta$^2$\thanks{E-mail: ngupta@iucaa.in} \\ 
$^1$ European Southern Observatory, Karl-Schwarzschild-Str. 2, D-85748 Garching Near Munich, Germany \\
$^2$ Inter-University Centre for Astronomy and Astrophysics, Post Bag 4, Ganeshkhind, Pune 411007, India \\
} 
}
\date{Accepted. Received; in original form }
\pubyear{}
\maketitle
\label{firstpage}
\pagerange{\pageref{firstpage}--\pageref{lastpage}}
%
%
\begin {abstract}  
\par\noindent
We present Giant Metrewave Radio Telescope and Very Large Array observations of \hi\ \21\ absorption in ten $z\le$0.2 
galaxy mergers that host strong radio sources. Seven of these mergers show strong absorption [with \nhi\ $\sim10^{21-22}$\,\cms,
for spin temperature of 100 K and unit covering factor], leading to a total detection rate of $83 \pm 17$\% in low-$z$ 
radio-loud galaxy mergers. This is $\sim3-4$ times higher than that found for intrinsic \hi\ \21\ absorption in low-$z$ 
radio sources in general. The fraction of intrinsic absorbers that are associated with mergers increases with increasing
\nhi\ threshold, i.e. $\sim$40\% and 100\% of the absorbers with \nhi\ $>10^{21}$\,\cms\ and $>10^{22}$\,\cms\ arise from 
mergers, respectively. The distribution of \nhi\ among the mergers is significantly different from that found in isolated 
systems, with mergers giving rise to six times stronger absorption on average. The fraction of redshifted absorption components 
(with respect to the systemic velocity of the radio source obtained from optical emission lines) among mergers is found to 
be higher by two to three times compared to that found for non-interacting systems. Follow-up spatially-resolved multi-wavelength
spectroscopy is essential to understand the exact connection between the presence of circumnuclear neutral gas and the 
AGN activity in these mergers.
\end {abstract}  
%
%
%
%
\section{Introduction} 
\label{sec_introduction}  
According to the hierarchical model of structure formation, galaxies evolve through interactions and mergers \citep{white1978,hopkins2006}. 
Galaxy interactions and mergers have significant effects on the star formation, morphology, kinematics and physical conditions of gas, both 
in the core and outskirts of galaxies \citep[e.g.][]{toomre1972,barnes1996,mihos1996,cox2004,cox2008,soto2012,moreno2015,hani2018}. Major 
galaxy mergers can funnel large quantities of gas to the central regions of galaxies, triggering intense bursts of star formation and fueling 
Active Galactic Nuclei (AGNs) \citep{springel2005,hopkins2008}. Feedback from the AGNs, in the form of ejection or heating of the gas, can 
in turn regulate or quench star formation in the galaxies. This negative AGN feedback is usually a necessary ingredient in semi-analytical 
and numerical simulations to reproduce the observed galaxy properties \citep[e.g.][]{croton2006,schawinski2006,samui2007}. AGN outflows or 
jets can also provide positive feedback by compressing gas clouds to high densities as they propagate through the interstellar medium and 
triggering star formation \citep[e.g.][]{chambers1987,gaibler2012,maiolino2017}. Further, AGN feedback can connect the properties of the 
central supermassive black hole (SMBH) to that of its host galaxy, and therefore, explain the observed correlation between the SMBH mass 
and galaxy mass or velocity dispersion \citep{ferrarese2000,gebhardt2000}. 

While the above studies have shown that the AGN feedback, both positive and negative, has a strong impact on the evolution of its host galaxy and 
the environment, what is yet to be well understood is the mechanism that triggers the AGN activity. Major mergers, secular evolution, and hot halo 
accretion are among the processes believed to be responsible for triggering the AGN activity \citep[see][and references therein]{alexander2012}.
While mergers may not be the sole or dominant trigger of AGN activity, with major mergers thought to trigger only the most luminous AGNs 
\citep{treister2012}, their role in this process is still important to investigate. The connection between galaxy mergers and AGN activity 
remains debatable, with some studies having found evidence for such a connection for both radio-loud and radio-quiet AGNs 
\citep[e.g.][]{combes2009,ellison2011,ramos2012,tadhunter2012,villar2012,khabiboulline2014,satyapal2014,weston2017}, 
and some studies having found no such evidence at different redshifts \citep[e.g.][]{schmitt2001,cisternas2011,scott2014,villforth2014}. 
It has been proposed that one of the reasons for studies at shorter wavelengths (optical) not finding a connection between mergers 
and AGNs, contrary to studies at higher wavelengths (infrared), could be the strong dust obscuration in the central regions of some mergers \citep{satyapal2017,weston2017}. 
There is evidence that the AGN can be fueled by infall of cold gas clouds \citep{morganti2009,fathi2013,tremblay2016,maccagni2018}. 
Therefore, it is of great importance to study the properties of gas in the circumnuclear discs and tori in AGN hosts associated with 
galaxy mergers, in order to test the connection between AGN activity and galaxy mergers. 

In the case of radio-loud AGNs, the circumnuclear atomic gas can be probed using \hi\ \21\ absorption. There have been several 
searches of associated \hi\ \21\ absorption in samples of radio-loud AGNs 
\citep[e.g.][]{vangorkom1989,carilli1998a,peck2000,morganti2001,vermeulen2003,gupta2006,chandola2011,allison2012,chandola2013,gereb2015,aditya2016,maccagni2017}.
Interestingly, at low redshifts, where it is possible to identify mergers by visual inspection, radio galaxies with very strong 
intrinsic absorption are usually observed to be undergoing major mergers. This suggests that mergers can trigger inflow of large
volumes of neutral gas to the central regions of galaxies. For example, based on the high concentration of \hi\ gas in the circumnuclear
region of a radio source undergoing merger and signature of infalling cold gas found in sub-arcsecond-scale spectroscopy, \citet{srianand2015}
have conjectured that the radio source may have been triggered by the gas infall due to the ongoing merger. However, the \hi\ \21\ 
absorption from mergers are reported in the literature either as individual detections or as part of samples where good quality optical
imaging and spectroscopic data are not available. Therefore, it is not possible to draw any general conclusions about the relationship
between mergers and AGN activity based on these observations. 

As a first step to address the above issue, we have undertaken a pilot project to search for \hi\ \21\ absorption in radio sources at 
$z\lesssim0.2$ that are associated with mergers. The broad objective is to shed light on the merger-AGN connection through neutral gas 
absorption. Our immediate objectives are to ascertain what fraction of merging galaxies associated with strong radio sources are detected
in \hi\ \21\ absorption, check for the association of mergers and very strong absorption with high \nhi\ values of $\sim10^{21-22}$\,\cms, 
and estimate what fraction of merging galaxies have inflowing/outflowing \hi\ gas. Consequently, using detailed analysis of multi-wavelength 
observations, we propose to understand the origin of the absorbing gas detected in the circumnuclear regions of mergers, and hence 
investigate the AGN feeding and feedback processes ongoing in these mergers.

This paper is arranged as follows. We describe our sample, observations and data reduction in Section~\ref{sec_observations}.  
We present the results from our search for neutral gas absorption in mergers and discuss individual systems from our sample in 
Section~\ref{sec_results}. We discuss the results by combining our sample with studies in the literature in Section~\ref{sec_discussion}. 
Finally, we summarize our results in Section~\ref{sec_summary}. Throughout this paper, we adopt a flat $\Lambda$-cold dark 
matter cosmology with $H_{\rm 0}$ = 70\,\kms~Mpc$^{-1}$ and $\Omega_{\rm M}$ = 0.30. Note that we use `absorption/absorber' 
to refer to \hi\ \21\ absorption/absorber from hereon unless otherwise mentioned.
%
%
\section{Sample \& Observations}  
\label{sec_observations}  
\subsection{Sample}
\label{sec_sample}
We compiled the merger sample using the Sloan Digital Sky Survey \citep[SDSS;][]{york2000}, the Faint Images of the Radio Sky 
at Twenty-Centimeters \citep[FIRST;][]{white1997}, the NRAO VLA Sky Survey \citep[NVSS;][]{condon1998}, and the NASA/IPAC 
Extragalactic Database (NED). We searched for radio sources that are brighter than 50 mJy at 1.4 GHz, having most of the radio 
flux in compact components (to facilitate a sensitive absorption search), and that are associated with galaxy mergers. 
We restricted our search to galaxies with spectroscopic redshift $\lesssim0.2$, so that we can robustly identify galaxy mergers 
through visual inspection of the optical images. We selected the following types of systems as mergers $-$ (i) an individual 
galaxy with a very disturbed morphology and sometimes with double nuclei, which is likely to be in the final stages of merging; 
(ii) a close pair of interacting galaxies which are at a projected separation $\lesssim$10~kpc or which show signs of strong tidal 
disturbances. In addition, we cross-matched samples of local galaxy mergers,\footnote{as compiled in \citet{comerford2015,barcos2017,satyapal2017,weston2017}} 
mainly identified from SDSS and the Wide-field Infrared Survey Explorer \citep[WISE;][]{wright2010} catalogs, with radio sources 
from FIRST and NVSS. Our final sample consists of ten merging systems, whose details are provided in Table~\ref{tab:sample}.
\begin{table*}
\caption{Details of the galaxy mergers in our sample.}
\centering
\begin{tabular}{cccccccc}
\hline
Source & Coordinates & $z$ & Projected  & Velocity   & $S$(1.4~GHz)  & Radio      & Spectral \\
       & (J2000)     &     & separation & separation & (mJy)         & morphology & index    \\
       & RA~~~~~~Dec &     & (kpc)      & (\kms)     &               &            &          \\
(1)    & (2)         & (3) & (4)        & (5)        & (6)           & (7)        & (8)      \\
\hline
J0054$+$7305 & 00:54:03.99 $+$73:05:05.40 & 0.01570 & 5   & ---    & 124 & R & 0.82 \\
             & 00:54:04.53 $+$73:05:19.80 & ---     &     &        &     &   &      \\
J1036$+$0221 & 10:36:31.96 $+$02:21:45.89 & 0.04990 & 2   & 170    & 202 & C & 0.46 \\
             & 10:36:31.88 $+$02:21:44.10 & 0.05049 &     &        &     &   &      \\
J1100$+$1002 & 11:00:17.98 $+$10:02:56.84 & 0.03624 & 13  & $-$90  & 140 & C & 0.33 \\
             & 11:00:19.10 $+$10:02:50.76 & 0.03594 &     &        &  31 & R & ---  \\
J1108$-$1015 & 11:08:26.51 $-$10:15:21.70 & 0.0273  & 6   & $-$820 & 324 & R & 0.65 \\
             & 11:08:26.93 $-$10:15:31.80 & 0.0245  &     &        &     &   &      \\
J1214$+$2931 & 12:14:17.80 $+$29:31:43.42 & 0.06350 & 8   & $-$70  &  77 & R & ---  \\
             & 12:14:18.25 $+$29:31:46.70 & 0.06326 &     &        &     &   &      \\
J1315$+$6207 & 13:15:35.10 $+$62:07:28.43 & 0.03083 & 21  & 140    &  46 & C & 0.67 \\
             & 13:15:30.72 $+$62:07:44.84 & 0.03130 &     &        &   8 & R & ---  \\
J1320$+$3408 & 13:20:35.40 $+$34:08:21.75 & 0.02306 & --- & ---    & 104 & R & 0.56 \\
J1356$+$1026 & 13:56:46.12 $+$10:26:09.09 & 0.12313 & 2   & ---    &  60 & C & ---  \\
             & 13:56:46.12 $+$10:26:08.00 & ---     &     &        &     &   &      \\
J1356$+$1822 & 13:56:02.89 $+$18:22:18.29 & 0.05036 & 4   & ---    & 368 & C & 0.98 \\
             & 13:56:02.63 $+$18:22:17.68 & ---     &     &        &     &   &      \\
J2054$+$0041 & 20:54:49.61 $+$00:41:53.07 & 0.20276 & 11  & $-$350 & 377 & C & 0.15 \\
             & 20:54:49.64 $+$00:41:49.85 & 0.20136 &     &        &     &   &      \\
\hline
\end{tabular}
\label{tab:sample}
\begin{flushleft}
{\it Notes.}
Column 1: name of the galaxy merger used throughout this work. 
Column 2: J2000 coordinates of the galaxies undergoing merger. 
Column 3: redshift measured using data from NED (for J0054$+$7305 and J1356$+$1822), SALT (for J1036$+$0221, J1100$+$1002, J1108$-$1015 and J2054$+$0041), 
and SDSS (for J1214$+$2931, J1315$+$6207, J1320$+$3408 and J1356$+$1026).
Column 4: projected separation between the two galaxy nuclei in kpc, obtained from the measured angular separation. 
Note that separate nuclei cannot be distinguished in J1320$+$3408 based on available optical images. 
Column 5: line-of-sight velocity separation in \kms\ between the two galaxy nuclei, taking the radio source or stronger radio source as the reference
(whenever SDSS spectra or our SALT spectra are available for both nuclei). 
Column 6: total flux density at 1.4~GHz in mJy of the radio source. In the case of J1100$+$1002 and J1315$+$6207, both the galaxies show radio emission.
Column 7: Morphology of the radio source at arcsecond-scales $-$ `C' is for compact and `R' is for resolved. 
The radio parameters given in columns 6 and 7 are from FIRST, except in the case of J0054$+$7305 and J1108$-$1015, where they are from NVSS.
Column 8: Spectral index of the radio source estimated from 1.4 GHz and 5 GHz fluxes whenever available in NED.
\end{flushleft}
\end{table*}
\subsection{Radio Observations}
\label{sec_radioobs}
The radio observations of nine sources were carried out using the Giant Metrewave Radio Telescope (GMRT) and that of one source 
(J2054$+$0041) were carried out using the Karl G. Jansky Very Large Array (VLA). The details of the observations are given in 
Table~\ref{tab:obslog}. The GMRT observations (Proposal IDs: 27\_015, 29\_079, 32\_038, 33\_027) were carried out using the L-band 
receivers with either the 4 MHz or 16 MHz baseband bandwidth split into 512 channels. We reobserved two sources, J0054$+$7305 and 
J1036$+$0221, with a larger bandwidth to confirm the broad and shallow absorption detected towards them. The spectra of these two 
sources presented here are from the observations obtained with a larger bandwidth. 

We searched for both \hi\ \21\ and OH 18-cm absorption towards J2054$+$0041 using L-band receivers on the VLA (Proposal IDs: 15A-176, 
17B-130). The \hi\ observations were carried out in A-configuration and the OH observations were carried out in B-configuration. 
The observations used a single 8 MHz sub-band, which was split into 4096 channels for the \hi\ data, and 512 channels for the OH data. 

All of the above data were acquired in two polarization products. In each of these observations, the pointing centre was at the coordinates
of the radio source, and the band was centred at the redshifted \hi\ \21\ (or OH 18-cm) line frequency. Standard calibrators were regularly
observed during each of the observations for calibration of flux density, bandpass, and phase. The data were reduced using the National 
Radio Astronomy Observatory (NRAO) Astronomical Image Processing System ({\sc aips}) following standard procedures, as described in \citet{dutta2016}.
The absorption spectra were extracted from spectral cubes that were obtained by imaging the continuum-subtracted data using {\tt ROBUST=0} 
weighting. In case of the system J0054$+$7305, where \hi\ \21\ emission is detected, the spectral cubes were deconvolved using a mask true 
for pixels with absolute flux greater than three times the single channel noise.
\begin{table*}
\caption{Radio observation log of the galaxy mergers.}
\centering
\begin{tabular}{cccccc}
\hline
Source & Date & Time & Central   & Spectral & Channel \\
       &      &      & Frequency & coverage & width   \\
       &      & (h)  & (MHz)     & (\kms)   & (\kms)  \\
(1)    & (2)  & (3)  & (4)       & (5)      & (6)     \\
\hline
J0054$+$7305 & 17 Jan 2016 & 3.0 & 1398.5 &  900 &   2 \\
             & 08 May 2017 & 2.5 & 1398.5 & 3500 &   7 \\
J1036$+$0221 & 25 Aug 2013 & 2.4 & 1352.3 &  900 &   2 \\
             & 31 Aug 2014 & 2.5 & 1352.3 & 3500 &   7 \\
J1100$+$1002 & 24 Aug 2013 & 3.2 & 1371.0 &  900 &   2 \\
J1108$-$1015 & 17 Jan 2016 & 3.0 & 1382.7 &  900 &   2 \\
J1214$+$2931 & 24 Dec 2014 & 5.9 & 1335.9 &  900 &   2 \\
J1315$+$6207 & 26 Feb 2018 & 4.2 & 1378.0 & 3500 &   7 \\
J1320$+$3408 & 31 Aug 2014 & 2.6 & 1389.3 & 3500 &   7 \\
J1356$+$1026 & 27 Feb 2018 & 4.0 & 1264.7 & 4000 &   8 \\
J1356$+$1822 & 30 Dec 2017 & 2.9 & 1352.2 & 3500 &   7 \\
J2054$+$0041 & 20 Jun 2015 & 1.2 & 1182.2 & 2000 & 0.5 \\
(OH 18-cm) & 14$-$22 Jan 2018 & 4.5 & 1385.0 & 1700 & 3 \\
\hline
\end{tabular}
\label{tab:obslog}
\begin{flushleft}
{\it Notes.} Column 1: galaxy merger name. Column 2: date of observation. Column 3: on-source observing time in h. 
Column 4: central frequency of the observing band in MHz. Column 5: spectral coverage in \kms. Column 6: channel width in \kms. \\
\end{flushleft}
\end{table*}
\subsection{Optical Observations}
\label{sec_opticalobs}
We obtained the optical spectra of four sources in our sample (J1036$+$0221, J1100$+$1002, J1108$-$1015 and J2054$+$0041) 
using the Robert Stobie Spectrograph (RSS) on the Southern African Large Telescope (SALT) in long-slit mode, under the programs
2014-2-SCI-017 (PI: Dutta) and 2013-1-IUCAA-002 (PI: Srianand). The observations of J1036$+$0221, J1100$+$1002 and J1108$-$1015 
were carried out with a 1.5$''$ slit and the PG0900 grating (coverage $\sim$4200-7200\,\AA), while that of J2054$+$0041 were 
carried out with a 2$''$ slit and the PG1300 grating (coverage $\sim$6600-8600\,\AA). The spectral settings were chosen to cover 
the expected wavelength ranges of the nebular emission lines. The spectral resolution ranges over $\sim$200$-$300\,\kms. The seeing 
was between 1.4$''$ to 2$''$. Each source was observed in one or two slit position angles, and each observation was split into two 
20 mins exposures. In the case of J2054$+$0041, observations were carried out with dithering to remove fringing effects in the red 
part. The data were reduced using the SALT science pipeline \citep{crawford2010}, i.e. {\tt PYSALT}\footnote{{\tt PYSALT} user package 
is the primary reduction and analysis software tools for the SALT telescope (http://pysalt.salt.ac.za/).}, and standard {\tt IRAF} 
tasks\footnote{IRAF is distributed by the National Optical Astronomy Observatories, which are operated by the Association of 
Universities for Research in Astronomy, Inc., under cooperative agreement with the National Science Foundation.}, as described 
in \citet{dutta2017a}. The spectrophotometric standard stars, LTT~7379 and LTT~9239, were used for flux calibration.
%
%
\section{Results}
\label{sec_results}
\begin{figure*}
\subfloat[J0054$+$7305]{\includegraphics[width=0.26\textwidth, bb=90 30 760 710, clip=true]{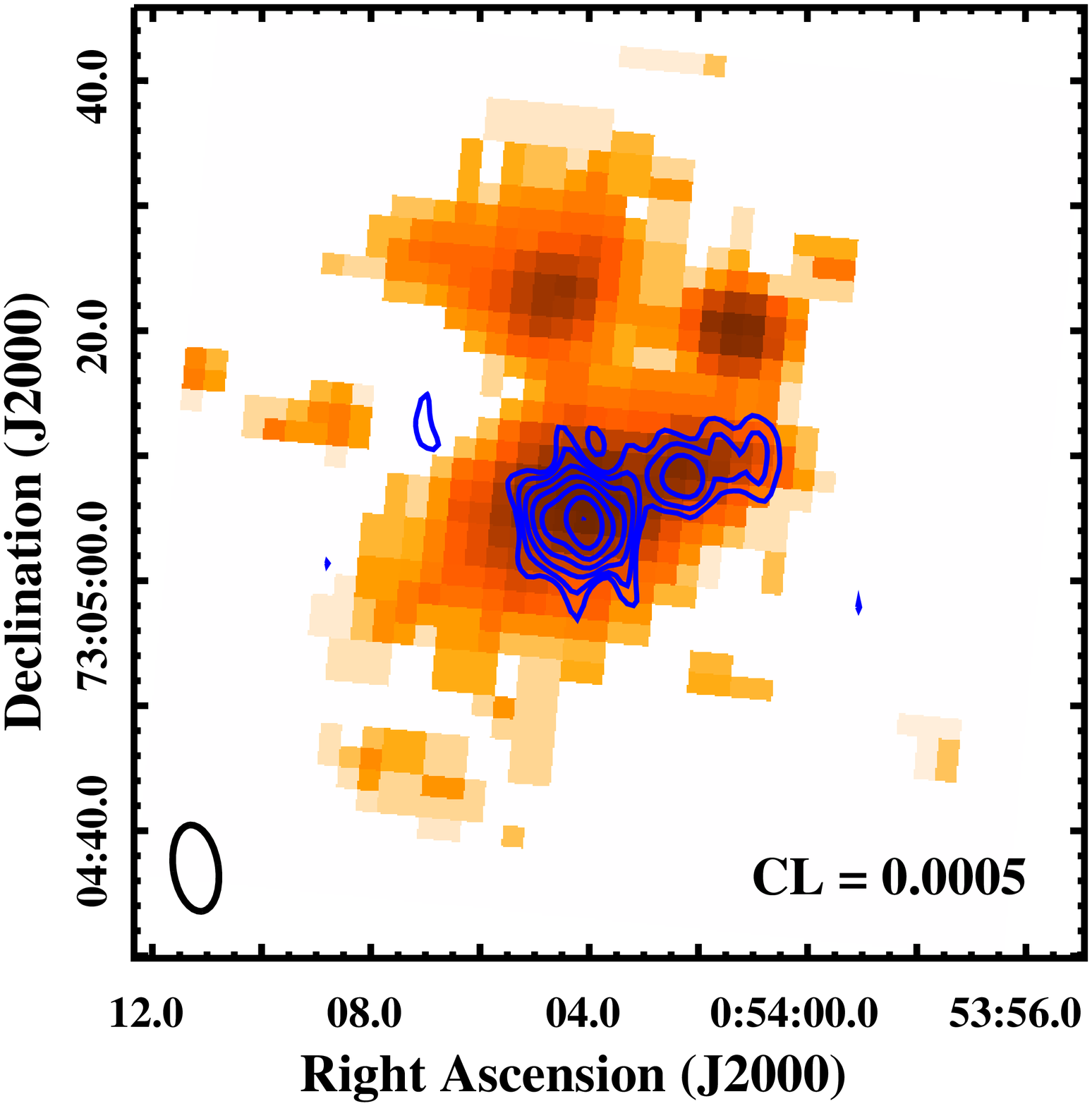} } 
\subfloat[J1036$+$0221]{\includegraphics[width=0.27\textwidth, bb=75 25 740 670, clip=true]{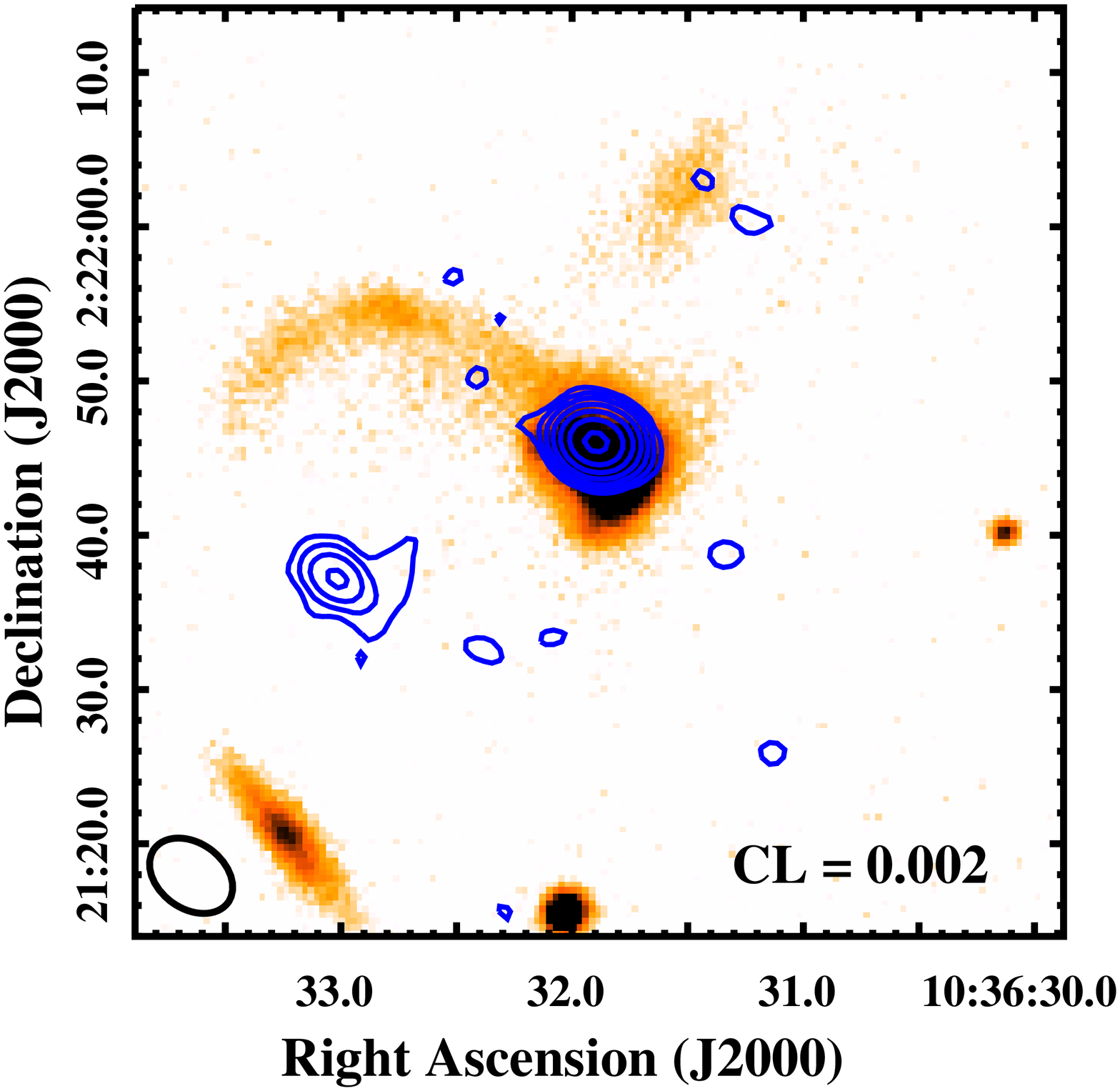} } 
\subfloat[J1100$+$1002]{\includegraphics[width=0.26\textwidth, bb=85 45 755 725, clip=true]{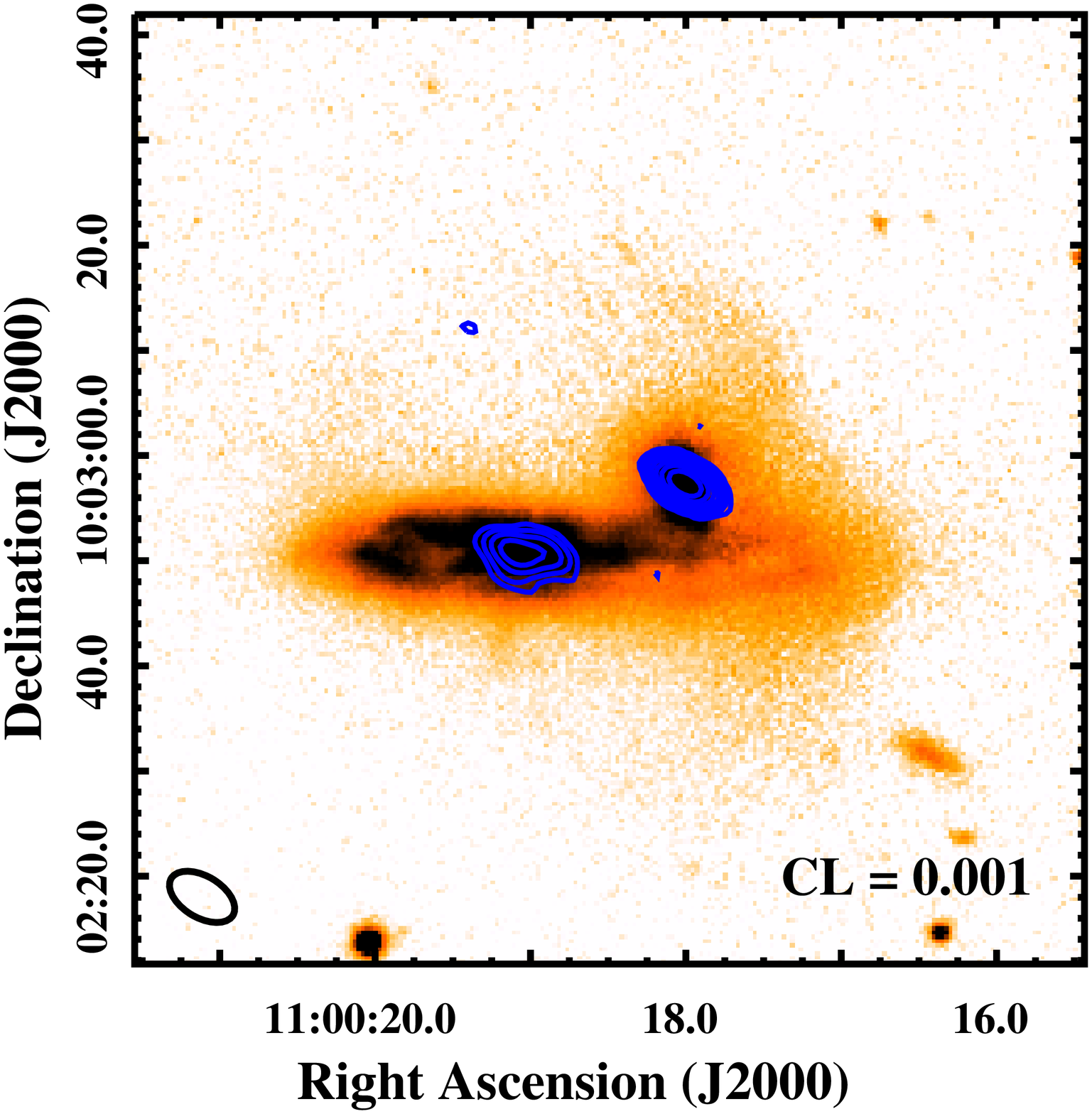} } \hspace{0.01cm}
\subfloat[J1108$-$1015]{\includegraphics[width=0.26\textwidth, bb=95 0 765 690, clip=true]{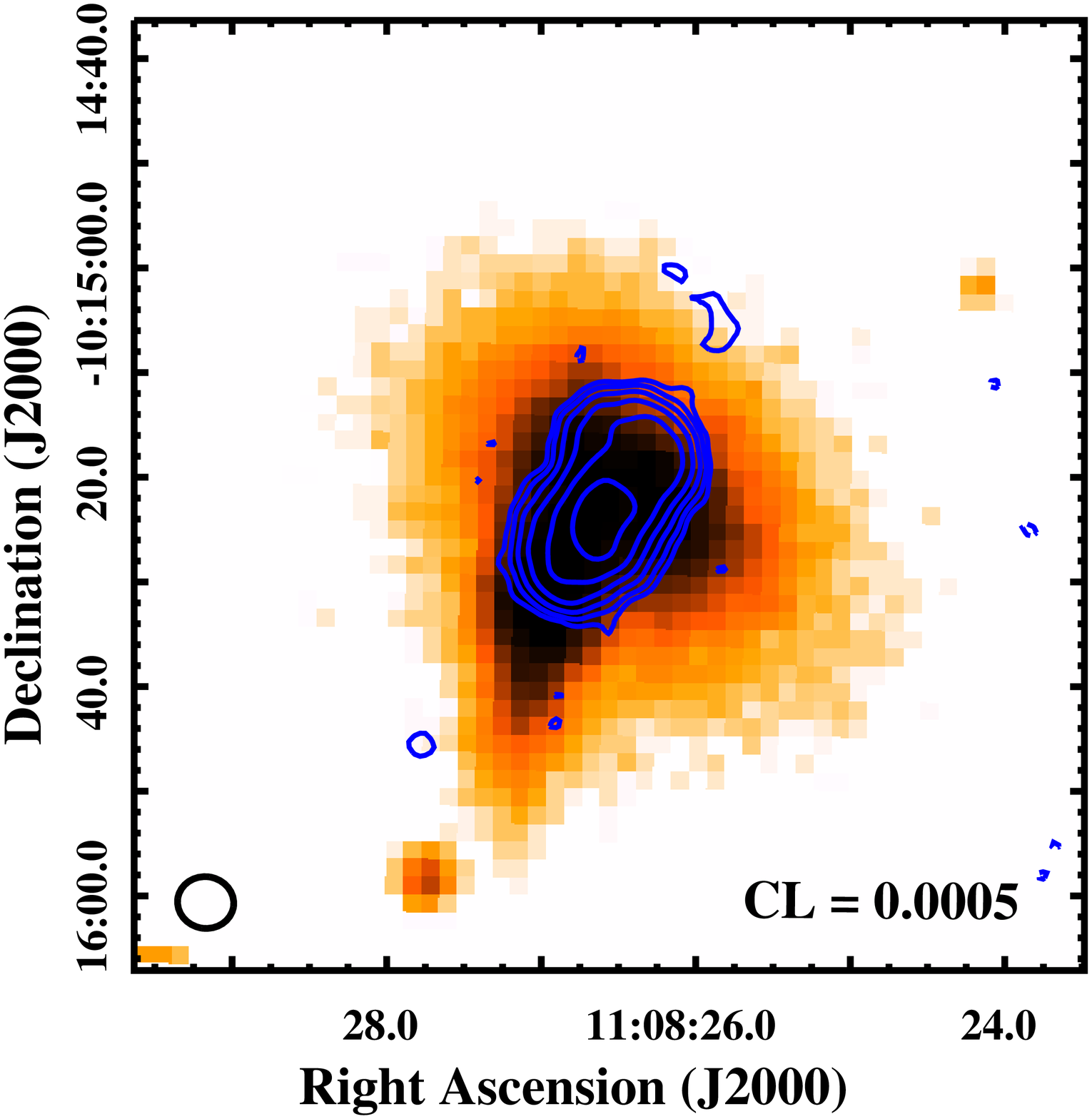} } 
\subfloat[J1214$+$2931]{\includegraphics[width=0.26\textwidth, bb=90 25 780 715, clip=true]{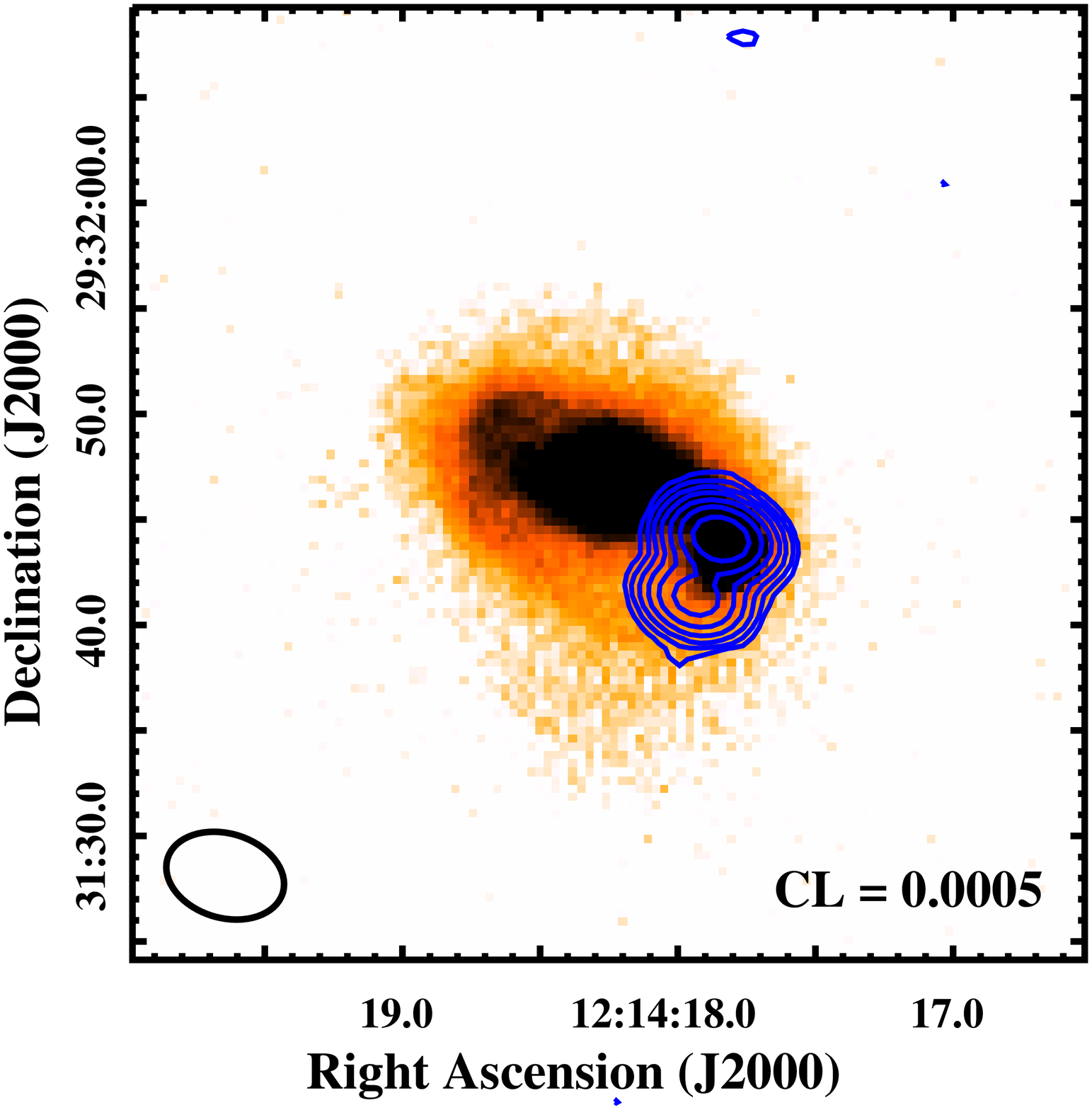} } 
\subfloat[J1315$+$6207]{\includegraphics[width=0.26\textwidth, bb=70 30 765 725, clip=true]{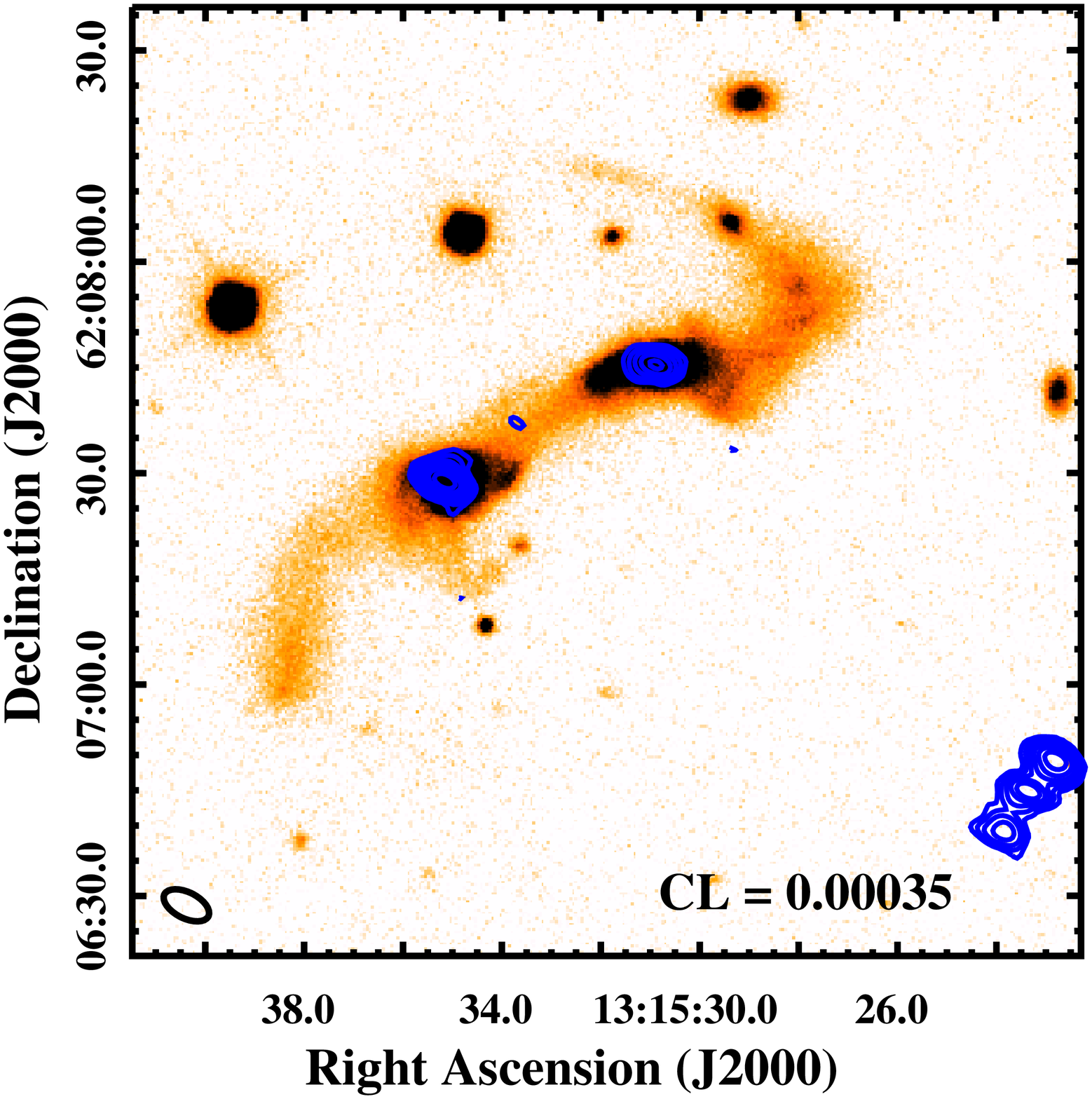} } \hspace{0.01cm}
\subfloat[J1320$+$3408]{\includegraphics[width=0.26\textwidth, bb=85 25 750 700, clip=true]{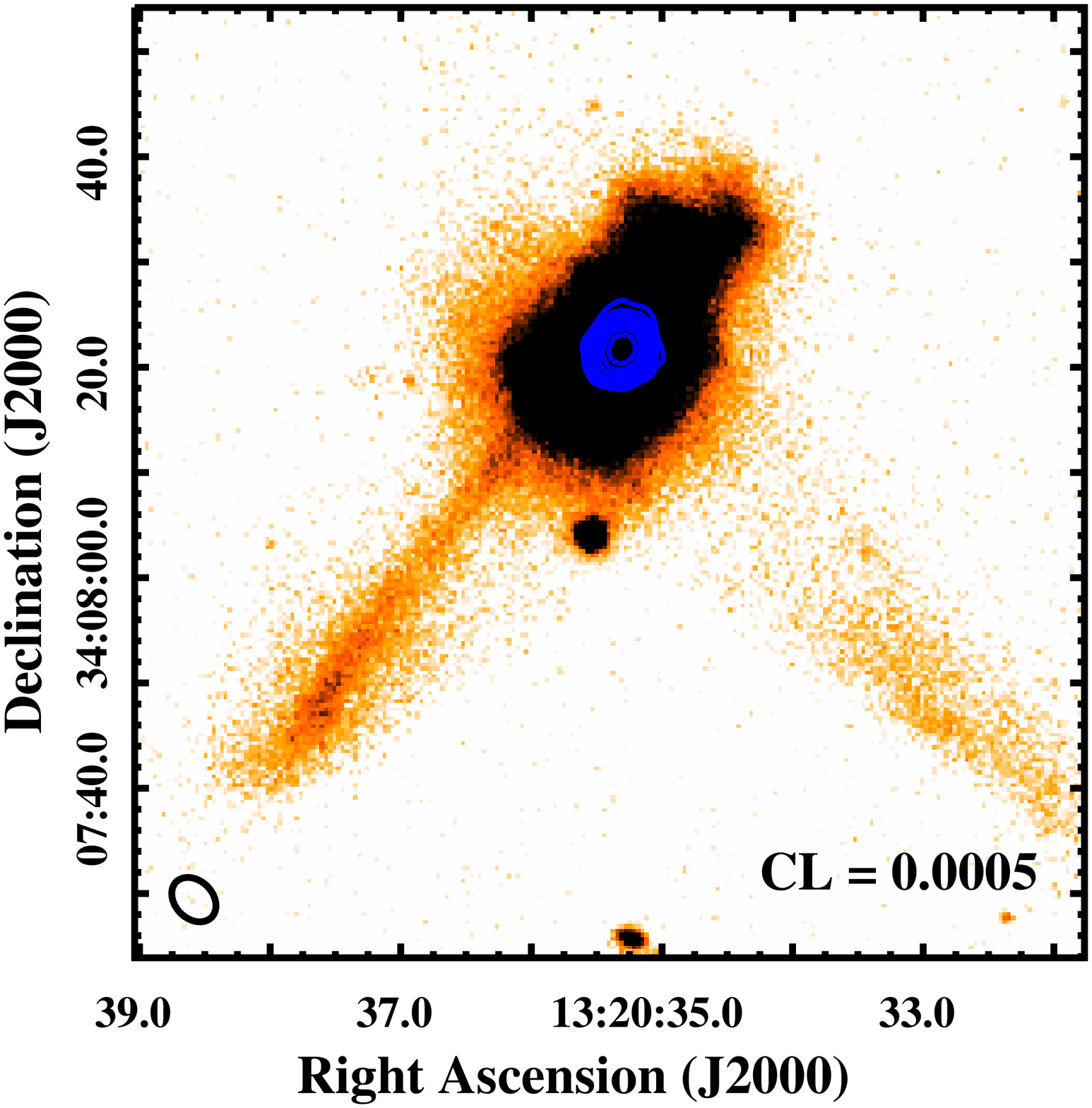} } 
\subfloat[J1356$+$1026]{\includegraphics[width=0.26\textwidth, bb=45 35 710 700, clip=true]{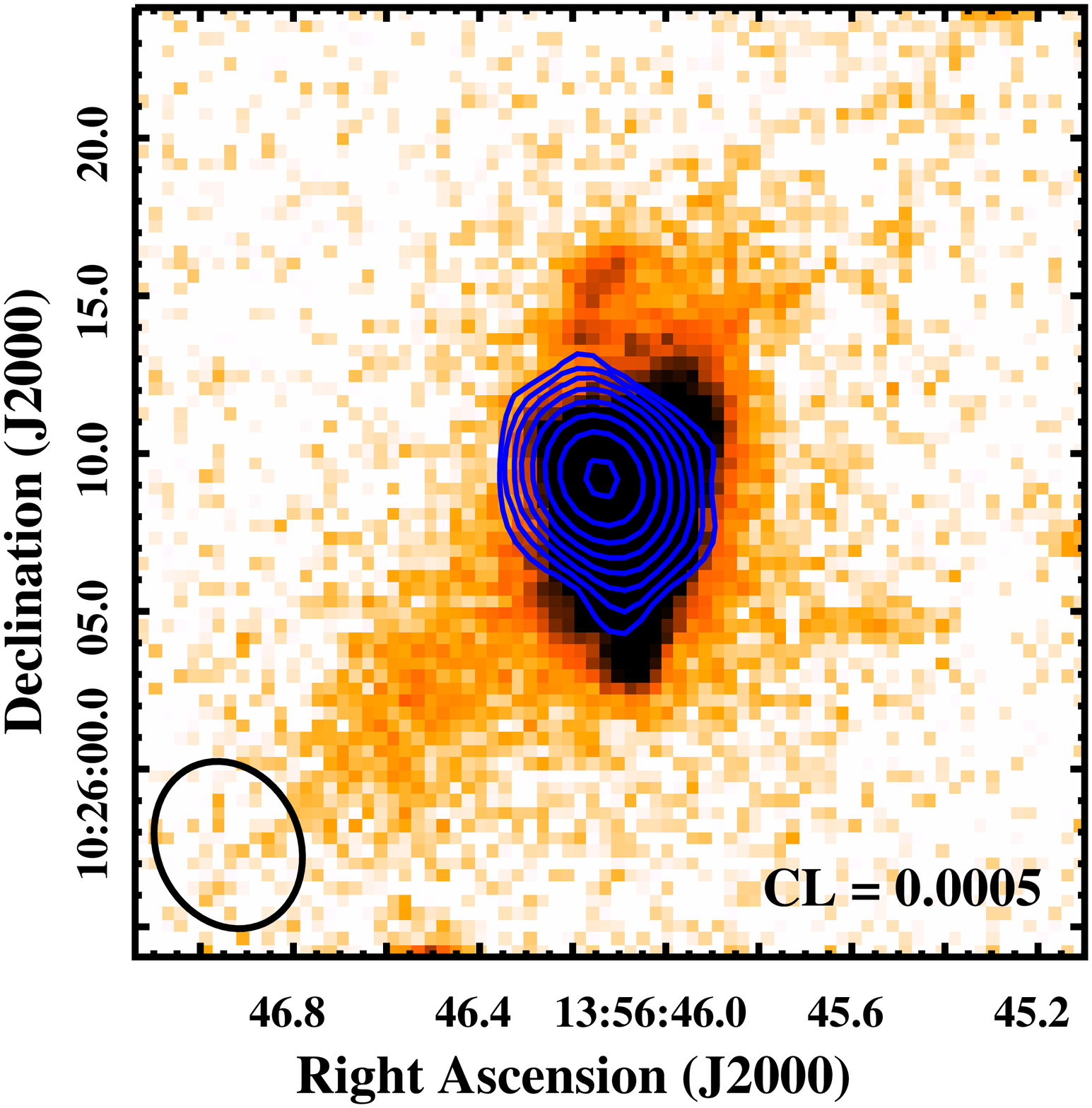} }
\subfloat[J1356$+$1822]{\includegraphics[width=0.26\textwidth, bb=95 50 715 675, clip=true]{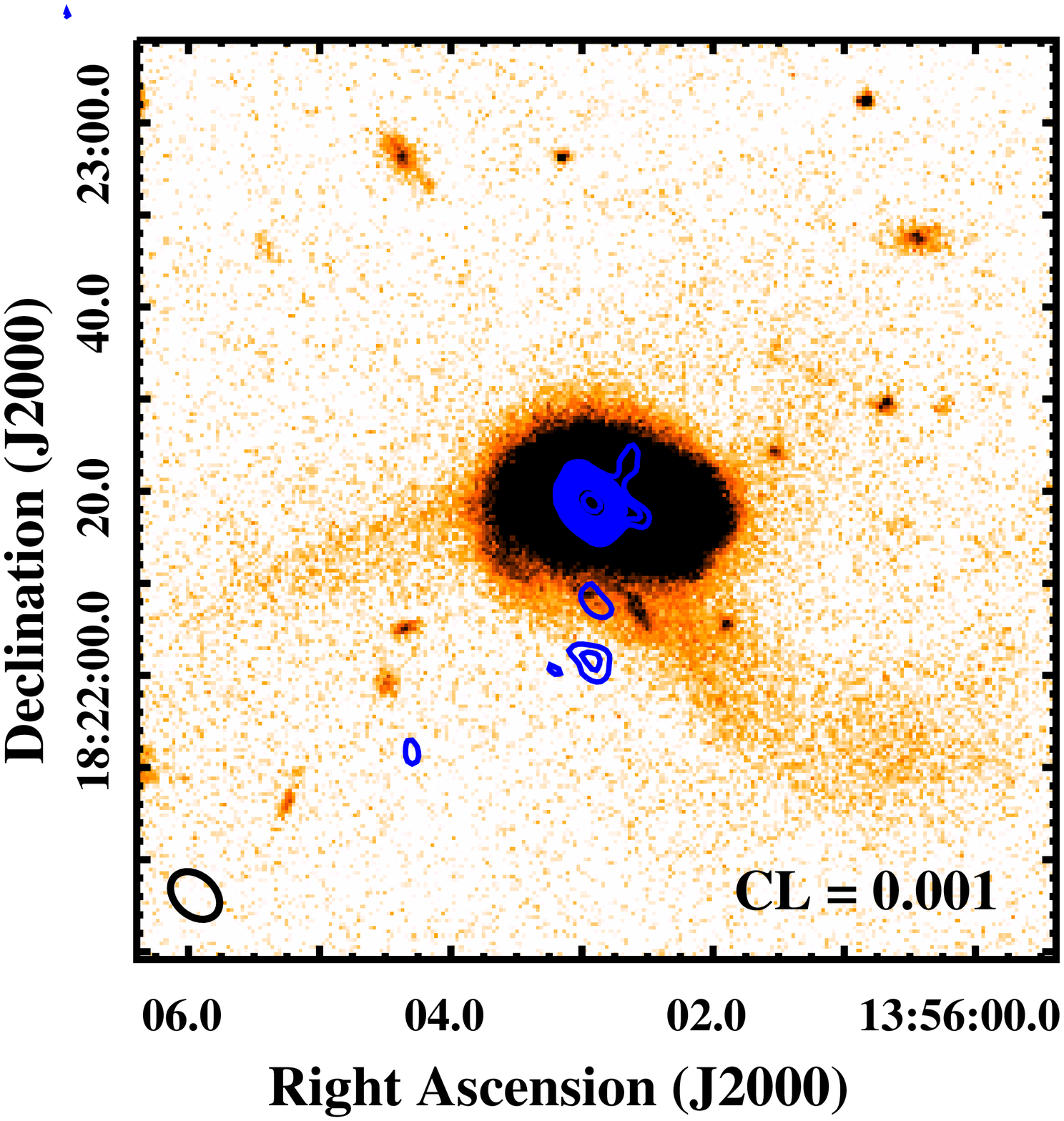} } \hspace{0.01cm} 
\subfloat[J2054$+$0041]{\includegraphics[width=0.26\textwidth, bb=100 40 730 700, clip=true]{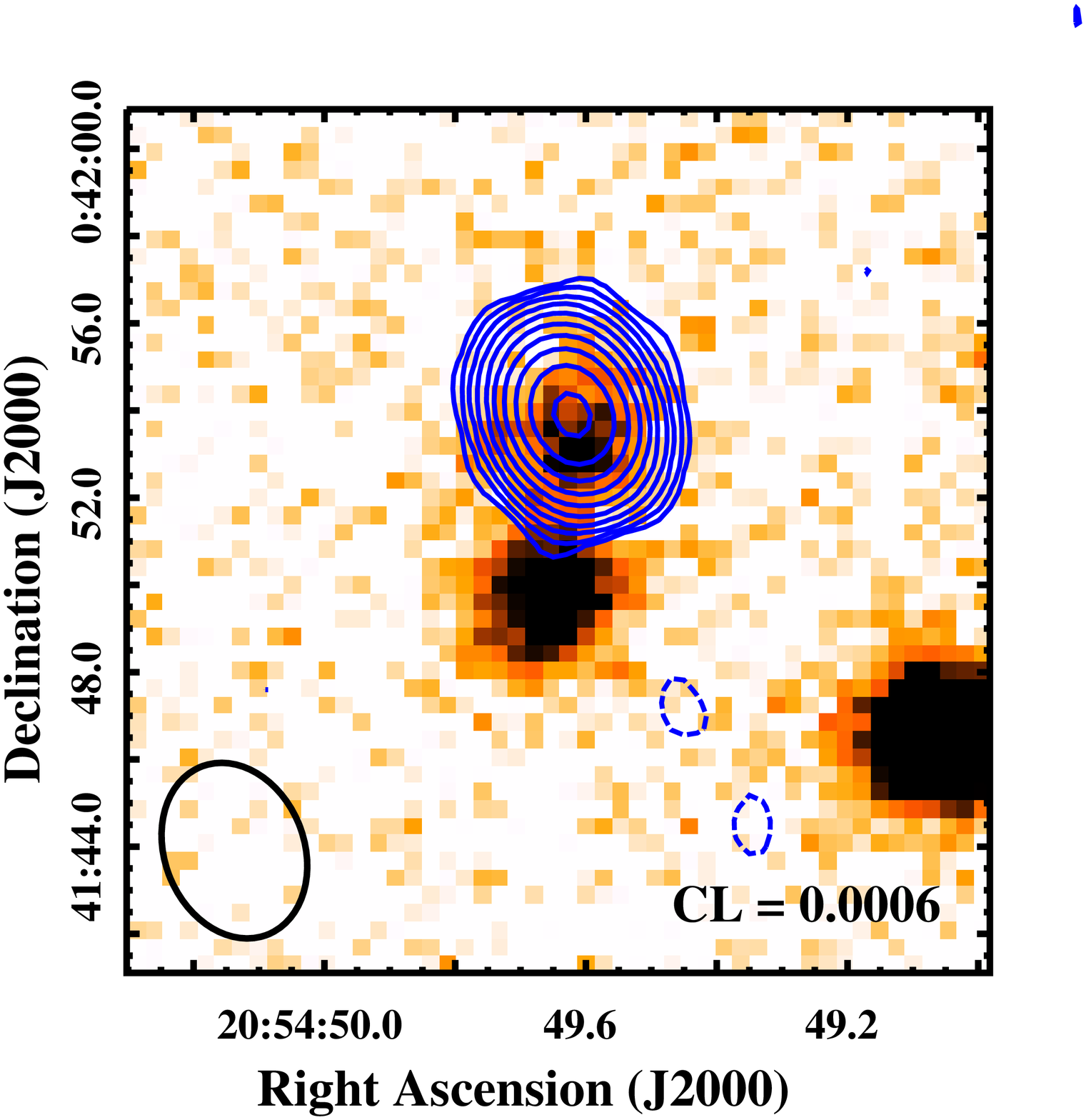} } 
\caption{Optical images (Digitized Sky Survey (DSS) for (a) and (d), and SDSS $r$-band for the rest) overlaid with 1.4 GHz continuum contours of the galaxy mergers.
The restoring beam of the continuum map is shown in the bottom left corner. The contour levels are plotted as CL $\times$ ($-$1,1,2,4,8,...)\,Jy~beam$^{-1}$, where
CL is given in the bottom right corner. Solid lines correspond to positive values while dashed lines correspond to negative values.}
\label{fig:overlay}
\end{figure*}

\begin{figure*}
\subfloat{\includegraphics[height=0.7\textheight, angle=90]{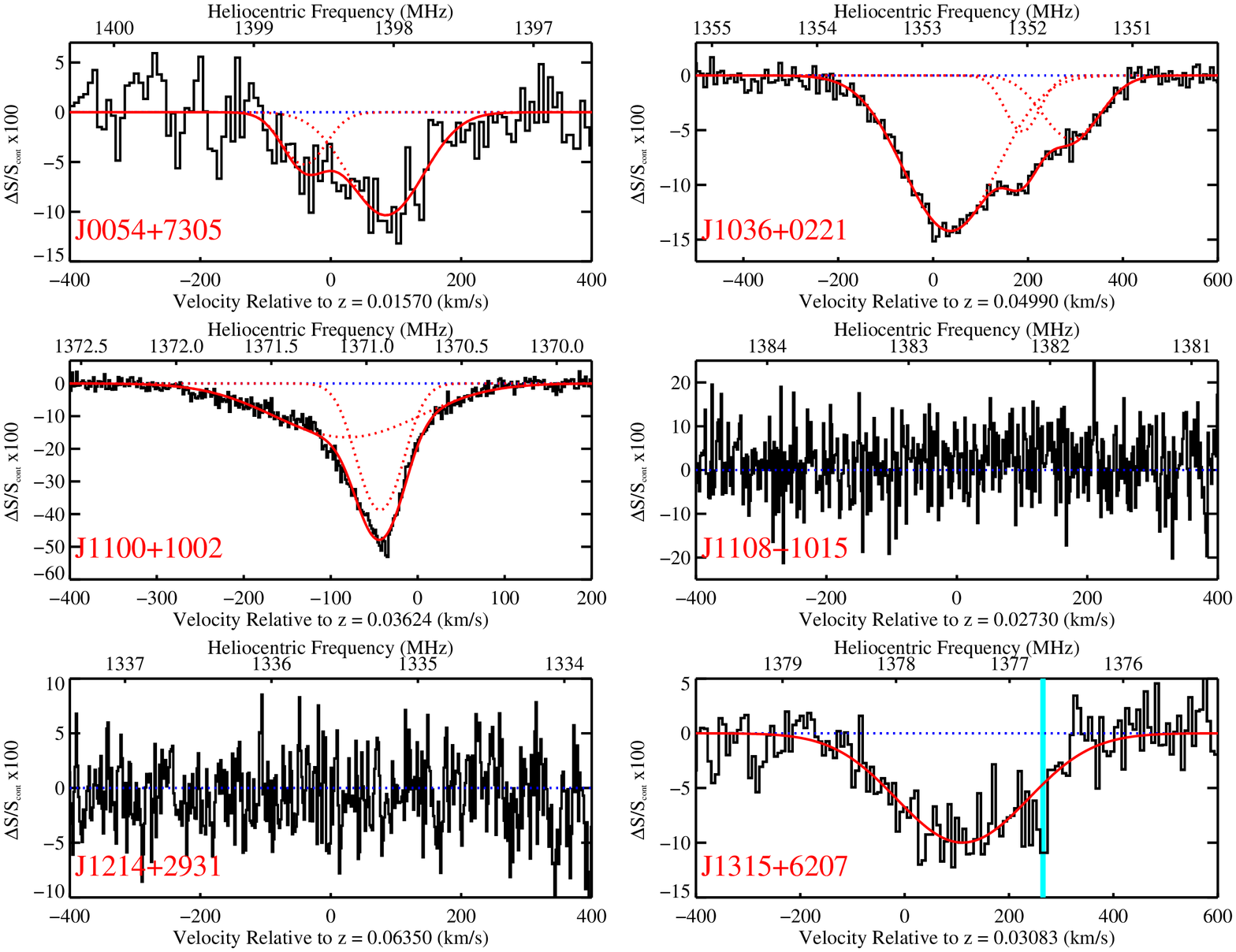} } \hspace{0.1cm}
\subfloat{\includegraphics[height=0.7\textheight, angle=90, trim={6cm 0 0 0}, clip=true]{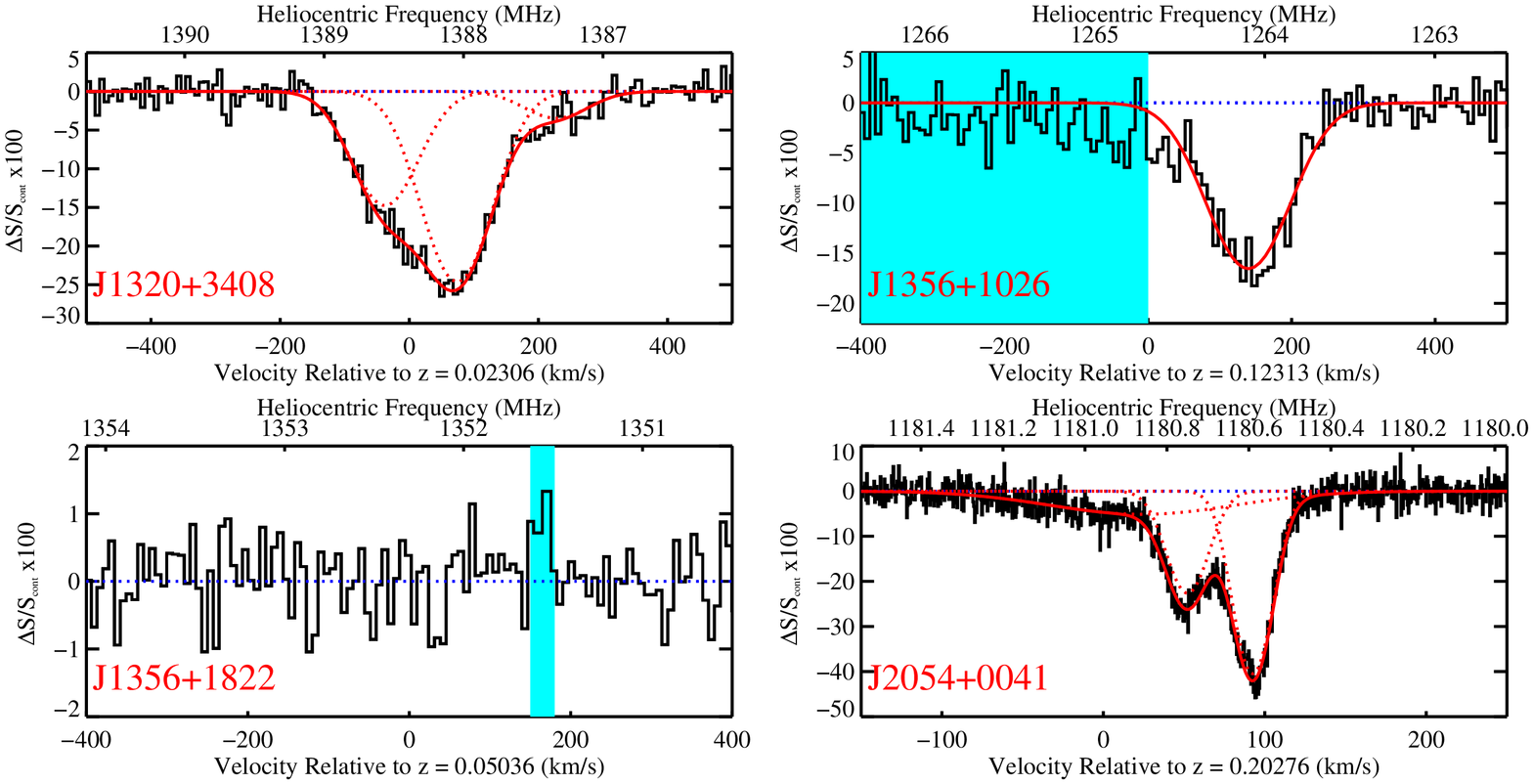} } 
\caption{\hi\ \21\ absorption spectra (shown as percentage change in the continuum flux, $\Delta S/S_{cont}\times100$) 
towards the radio sources undergoing merger in our sample. The velocity scale in each panel is defined with respect 
to the systemic redshift of the radio sources estimated from optical emission lines. Spectra are extracted towards 
the peak of the radio continuum in all the cases. Multi-component Gaussian fits are overplotted in the case of detections
$-$ individual components in dotted lines and resultant fits in solid lines. The shaded region marks the frequency ranges 
affected by radio frequency interference (RFI).}
\label{fig:21cmspectra}
\end{figure*}

We have detected absorption towards seven out of the ten merging systems. The radio continuum maps from our observations overlaid
on the optical images of the mergers are shown in Fig.~\ref{fig:overlay}, and the absorption spectra are shown in Fig.~\ref{fig:21cmspectra}. 
The typical spatial resolution of the radio data is $\sim2''$. The absorption spectra were extracted at the locations of the 
continuum peak flux density in all the cases. The parameters derived from the spectra are listed in Table~\ref{tab:results}. 
In the case of systems where there are two radio continuum peaks, we also list the parameters derived from the spectra towards
the weaker peaks. The optical depth limits for the non-detections have been estimated assuming a velocity width of 100\,\kms\ 
throughout. Further, we perform multi-component Gaussian fitting of the absorption lines. The number of components is decided by 
checking when the \chin\ is closest to unity. The resultant fits are shown overplotted on the spectra in Fig.~\ref{fig:21cmspectra}, 
and the parameters of the fits are given in Table~\ref{tab:gaussfit}. In principle, the full-width-at-half-maximum (FWHM) of the
components can be used to place upper limit on the gas kinetic temperature, assuming that the line width is purely due to thermal 
motions \citep[see section 4 of][]{dutta2017b}. However, that is not applicable here, since the gas is likely to have complex motions 
due to the ongoing merger (see Section \ref{sec_discussion3}). Moreover, the radio structure at sub-arcsecond-scales can be complex
and affect the line widths in some cases (see Section \ref{sec_discussion4}).
\begin{table*}
\caption{Parameters derived from absorption spectra of the galaxy mergers.}
\centering
\begin{tabular}{ccccccccc}
\hline
Source & Peak Flux    & Total Flux & Spectral     & $\tau_{\rm p}$ & \taudv\ & \nhi\                  & $v_{\rm 90}$ & \vshift\ \\
       & Density      & Density    & rms          &                &         & (\ts$/100$ K)($1/$\fc) &              &          \\
       & (\mjb)       & (mJy)      & (\mjb)       &                & (\kms)  & ($10^{20}$\,\cms)      & (\kms)       & (\kms)   \\
(1)    & (2)          & (3)        & (4)          & (5)            & (6)     & (7)                    & (8)          & (9)      \\
\hline
J0054$+$7305 &  32 &  77 & 1.2 & 0.14 $\pm$ 0.03 & 20 $\pm$ 2 & 37 $\pm$ 3  & 296 & 104   \\ 
             &   4 &  13 & 0.8 & $\le$0.195      & $\le$17    & $\le$31     & --- & ---   \\
J1036$+$0221 & 132 & 165 & 0.9 & 0.16 $\pm$ 0.01 & 50 $\pm$ 1 & 92 $\pm$ 1  & 447 & 2     \\
             &  14 &  21 & 0.9 & $\le$0.065      & $\le$4     & $\le$6      & --- & ---   \\
J1100$+$1002 & 125 & 129 & 2.0 & 0.76 $\pm$ 0.02 & 66 $\pm$ 1 & 121 $\pm$ 1 & 224 & $-$35 \\
             &  14 &  25 & 1.8 & $\le$0.137      & $\le$3     & $\le$5      & --- & ---   \\
J1108$-$1015 &  23 & 220 & 1.7 & $\le$0.072      & $\le$3     & $\le$6      & --- & ---   \\
             &  13 &  89 & 1.7 & $\le$0.130      & $\le$6     & $\le$11     & --- & ---   \\
J1214$+$2931 &  59 &  62 & 1.9 & $\le$0.031      & $\le$1     & $\le$2      & --- & ---   \\
             &  23 &  26 & 2.5 & $\le$0.111      & $\le$7     & $\le$13     & --- & ---   \\
J1315$+$6207 &  38 &  48 & 0.8 & 0.13 $\pm$ 0.02 & 30 $\pm$ 1 & 55 $\pm$ 2  & 319 & 93    \\
             &   7 &  10 & 0.9 & $\le$0.133      & $\le$9     & $\le$17     & --- & ---   \\ 
J1320$+$3408 &  62 & 109 & 0.9 & 0.31 $\pm$ 0.01 & 62 $\pm$ 1 & 113 $\pm$ 2 & 302 & 50    \\
J1356$+$1026 &  72 &  79 & 1.7 & 0.20 $\pm$ 0.02 & 26 $\pm$ 1 & 47 $\pm$ 2  & 216 & 148   \\
J1356$+$1822 & 369 & 388 & 2.0 & $\le$0.005      & $\le$0.3   & $\le$1      & --- & ---   \\
J2054$+$0041 & 362 & 376 & 7.8 & 0.61 $\pm$ 0.02 & 30 $\pm$ 1 & 55 $\pm$ 1  & 127 & 94    \\
\hline
\end{tabular}
\label{tab:results}
\begin{flushleft} 
{\it Notes.}
Column 1: galaxy merger name. 
Column 2: peak flux density in \mjb. Results from spectra extracted towards the weaker radio continuum peak, in the case of systems with multiple peaks, are also listed.
Column 3: total flux density in mJy. 
Column 4: spectral rms in \mjb\ (at the spectral resolution specified in Column 6 of Table~\ref{tab:obslog}). 
Column 5: peak optical depth in case of detections; 1$\sigma$ upper limit, i.e. rms, of the optical depth in case of non-detections.
Column 6: integrated optical depth in case of detections; 3$\sigma$ upper limit on the integrated optical depth with data smoothed to 100 \kms\ in case of non-detections.
Column 7: \nhi\ assuming spin temperature, \ts\ = 100 K, and covering factor, \fc\ = 1, in units of $10^{20}$\,\cms\ (3$\sigma$ upper limit in case of non-detections).
Column 8: velocity width which contains 90\% of the total optical depth ($v_{\rm 90}$) in \kms\ in case of detections. 
Column 9: velocity shift between the systemic redshift of the radio source (from optical emission lines) and the peak optical depth (\vshift), in \kms, in case of detections
(with positive sign indicating redshifted absorption and negative sign indicating blueshifted absorption).
\end{flushleft}
\end{table*}
\begin{table} 
\caption{Details of the Gaussian fits to the absorption lines detected from mergers.}
\centering
\begin{tabular}{ccccc}
\hline
Source & No. & \zabs\ & FWHM   & \taup\ \\
       &     &        &        &        \\
       &     &        & (\kms) &        \\
(1)    & (2) & (3)    & (4)    & (5)    \\
\hline
J0054$+$7305 & 1 & 0.01562 $\pm$ 0.00002 &  78 $\pm$ 10 & 0.05 $\pm$ 0.02 \\
             & 2 & 0.01599 $\pm$ 0.00001 & 133 $\pm$  8 & 0.11 $\pm$ 0.01 \\
J1036$+$0221 & 1 & 0.05003 $\pm$ 0.00001 & 224 $\pm$  9 & 0.15 $\pm$ 0.01 \\
             & 2 & 0.05056 $\pm$ 0.00003 &  84 $\pm$ 18 & 0.05 $\pm$ 0.04 \\
             & 3 & 0.05092 $\pm$ 0.00005 & 140 $\pm$ 25 & 0.06 $\pm$ 0.02 \\
J1100$+$1002 & 1 & 0.03596 $\pm$ 0.00001 & 189 $\pm$  2 & 0.18 $\pm$ 0.01 \\
             & 2 & 0.03609 $\pm$ 0.00001 &  58 $\pm$  1 & 0.49 $\pm$ 0.01 \\
J1315$+$6207 & 1 & 0.03121 $\pm$ 0.00001 & 288 $\pm$  7 & 0.11 $\pm$ 0.01 \\            
J1320$+$3408 & 1 & 0.02293 $\pm$ 0.00003 & 118 $\pm$ 11 & 0.16 $\pm$ 0.05 \\
             & 2 & 0.02331 $\pm$ 0.00002 & 124 $\pm$ 11 & 0.28 $\pm$ 0.06 \\
             & 3 & 0.02384 $\pm$ 0.00005 & 115 $\pm$ 26 & 0.04 $\pm$ 0.02 \\
J1356$+$1026 & 1 & 0.12365 $\pm$ 0.00001 & 133 $\pm$  3 & 0.18 $\pm$ 0.01 \\
J2054$+$0041 & 1 & 0.20286 $\pm$ 0.00001 & 133 $\pm$  3 & 0.05 $\pm$ 0.01 \\
             & 2 & 0.20297 $\pm$ 0.00001 &  30 $\pm$  1 & 0.26 $\pm$ 0.01 \\
             & 3 & 0.20313 $\pm$ 0.00001 &  27 $\pm$  1 & 0.52 $\pm$ 0.01 \\
\hline
\end{tabular}
\label{tab:gaussfit}
\begin{flushleft} {\it Notes.}
Column 1: galaxy merger name. Column 2: identification number of absorption component. Column 3: redshift of the component. 
Column 4: full-width-at-half-maximum (FWHM) in \kms\ of the component. 
Column 5: peak optical depth (\taup) of the component. 
\end{flushleft}
\end{table}
\subsection{Individual Systems}
\label{sec_individual}
Below we describe the individual systems from our sample in brief.
\subsubsection{J0054$+$7305}
\label{sec_j0054+7305}
This system consists of the luminous infrared galaxy \citep[LIRG;][]{sanders1996}, MCG~$+$12$-$02$-$001 \citep{sanders2003}. Optical integral field 
spectroscopy of the central region of this LIRG classifies it as a Composite system \citep{alonso2009}. The system has been detected in \hi\ \21\ 
emission \citep[integrated flux density = 6.6 Jy~\kms;][]{courtois2011}, as well as neutral carbon and molecular CO emission \citep{israel2015}. 
{\it Hubble Space Telescope / Advanced Camera for Surveys (HST/ACS)} imaging of this system reveals it to be a major merger, with separation between 
the two interacting galaxies to be 5.3 kpc \citep{kim2013}. We detect resolved radio emission from one of the galaxies in our GMRT 1.4 GHz continuum 
map. We detect broad and shallow absorption towards the peak of the continuum emission. A two-component Gaussian fit to the absorption profile produces 
a residual spectrum nearly consistent with the rms noise, and addition of further components is not warranted by the data. The stronger component is
redshifted by $\sim$84\,\kms\ from the systemic redshift of the radio source, 0.01570 $\pm$ 0.00007 \citep{strauss1992}, while the second component is 
blueshifted by $\sim$40\,\kms.

\begin{figure}
\includegraphics[height=0.36\textwidth, bb=75 10 745 700, clip=true]{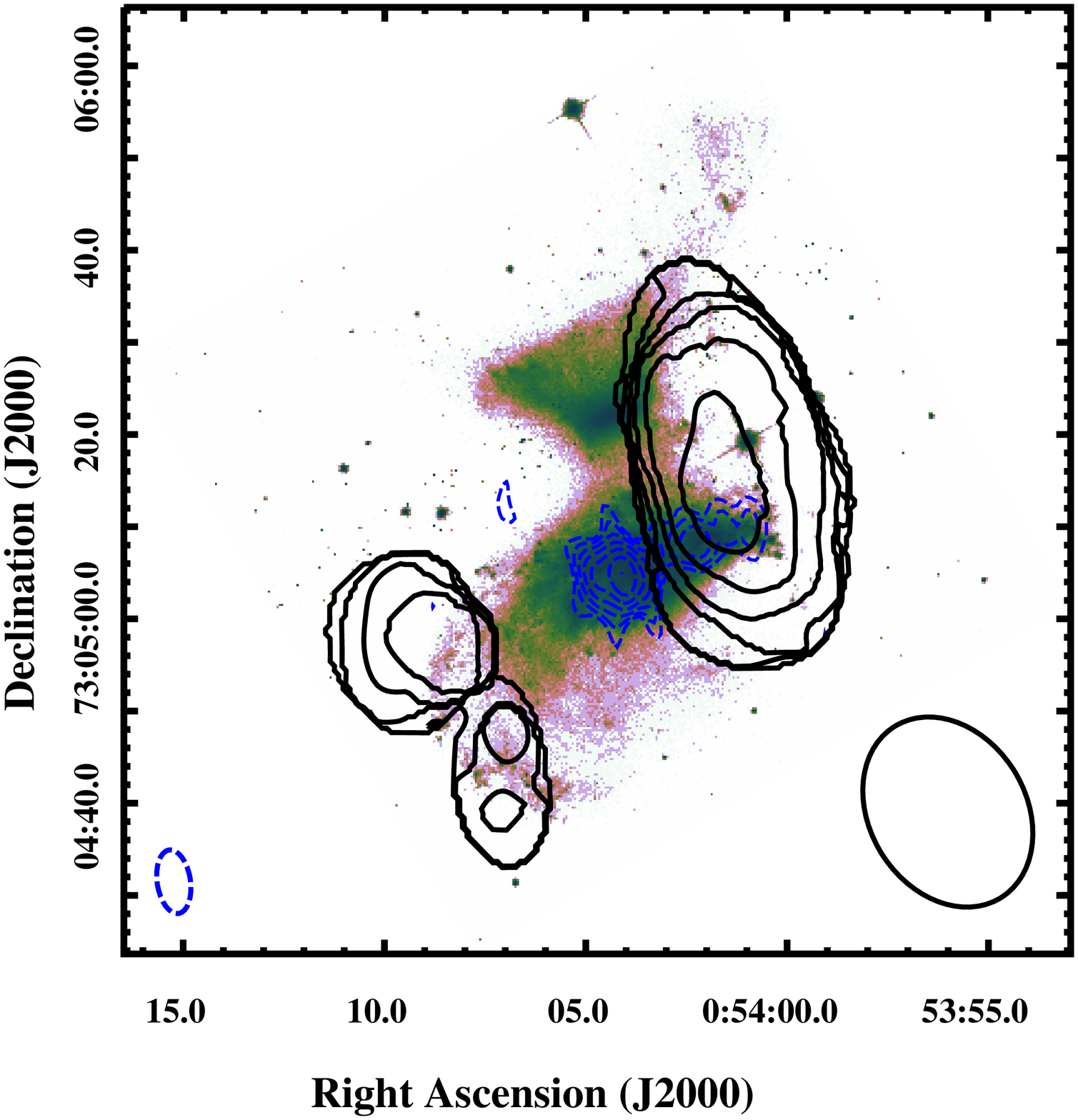} 
\includegraphics[height=0.34\textwidth, bb=50 415 510 750, clip=true]{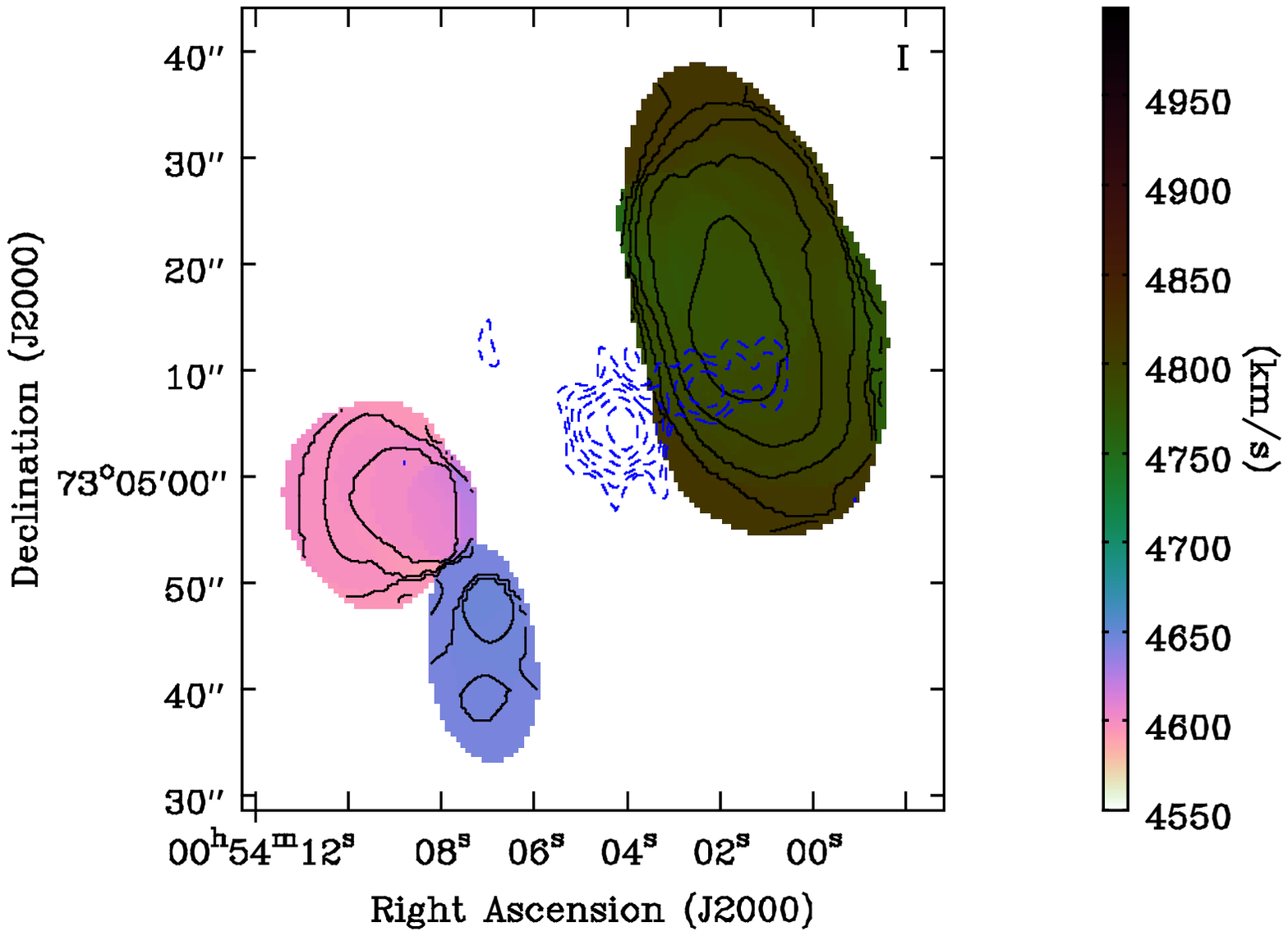}
\includegraphics[height=0.40\textwidth, angle=90]{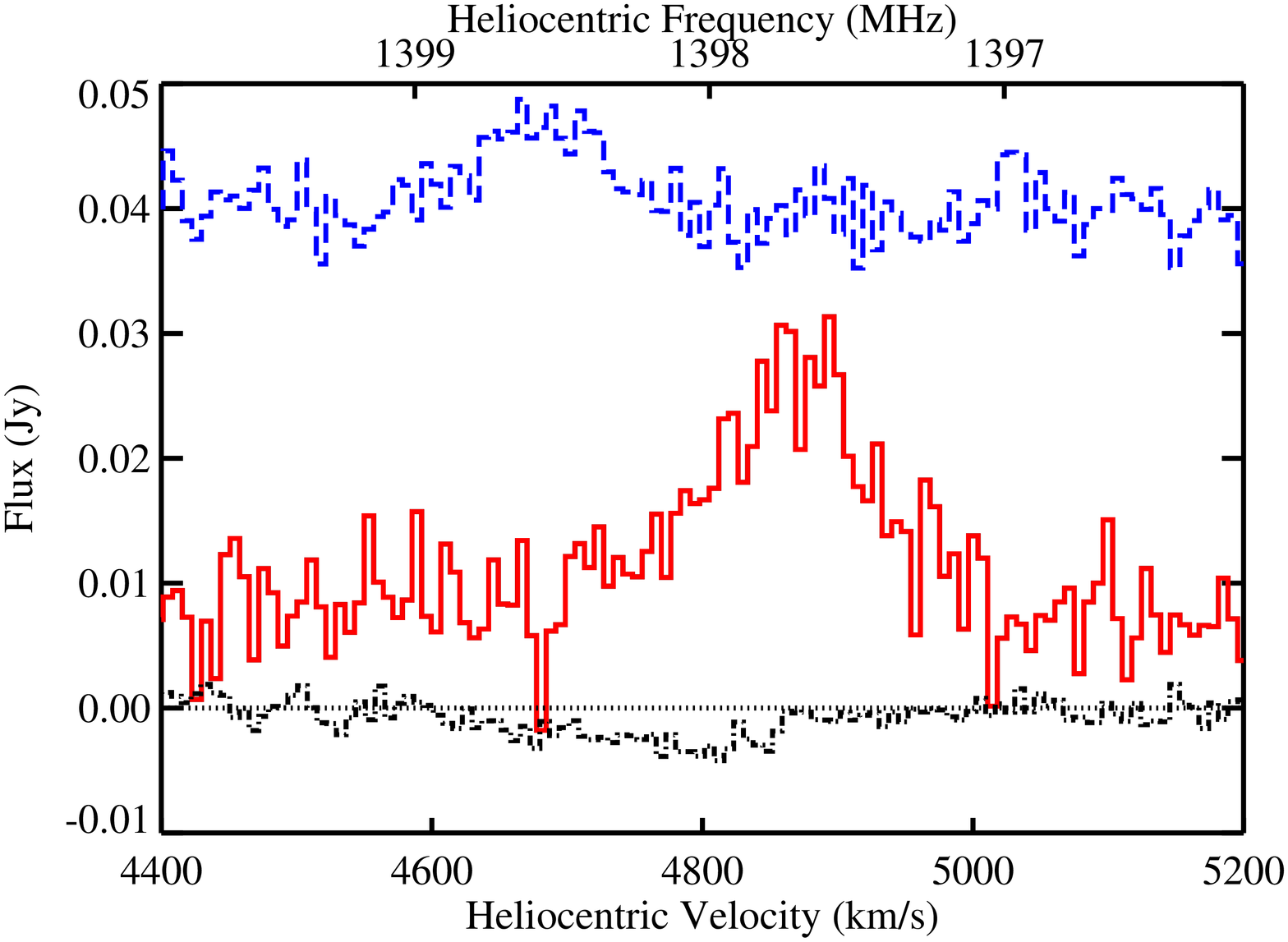}
\caption{{\it Top:} {\it HST/ACS} image (taken with F814W grating) of J0054$+$7305 overlaid with the GMRT \hi\ \21\ emission moment-0 map in solid contours. 
The GMRT 1.4 GHz radio continuum is also shown in dashed contours. The contour levels of the moment-0 map correspond to \nhi\ = 1, 2, 4, 8, 16, 32 $\times$ $10^{20}$\,\cms. 
The restoring beam of the moment-0 map, of size 10.9$''$ $\times$ 8.5$''$, is shown at the bottom right corner. 
The restoring beam of the continuum map is shown at the bottom left corner. 
The contour levels and restoring beam of the continuum map are the same as in panel (a) of Fig.~\ref{fig:overlay}.
{\it Centre:} The GMRT moment-1 map showing the velocity field of the \hi\ emission. 
The moment-0 map contours and the continuum contours, as shown in the top plot, are overlaid. 
{\it Bottom:} The emission spectrum from summing over the emitting region on the right is shown as the solid histogram. 
The emission spectrum from summing over the emitting region on the left in shown as the dotted histogram.
The absorption spectrum is shown as the dashed-dotted histogram. 
The spectra have been shifted arbitrarily on the $y$-axis for display purpose. 
}
\label{fig:j0054_overlays}
\end{figure}

Furthermore, we detect \hi\ \21\ emission in our GMRT data. To map the emission, we imaged the continuum subtracted visibilities with {\tt ROBUST = 1} 
weighting and different {\it uv}-tapers. In Fig.~\ref{fig:j0054_overlays}, we show the moment-0 and moment-1 maps obtained from the spectral cube (imaged with 
an {\it uv}-taper of 20 kilolambda) with spatial resolution of 10.9$''$ $\times$ 8.5$''$ (corresponding to 3.5 $\times$ 2.7~kpc$^2$ at the redshift of the galaxy). 
These maps were created by integrating the emission using a mask true only for pixels with flux greater than three times the single channel noise ($\sim$1.3\,\mjb). 
We recover $\sim$45\% of the total emission that has been detected from this system by the Green Bank Telescope \citep{courtois2011}. The total flux density does
not vary significantly in spectral cubes of different spatial resolutions ($\sim10''-40''$). We checked for the presence of emission by extracting spectra at different
locations around the radio source from the spectral cubes. As can be seen from Fig.~\ref{fig:j0054_overlays}, we do not detect emission near the peak of the radio
continuum, where we detect the absorption. In the spectral cube from which we extract the absorption spectrum (spatial resolution of 3.5$''$ $\times$ 1.8$''$; 
obtained with {\tt ROBUST = 0} weighting and no {\it uv}-taper), we get a 3$\sigma$ upper limit of \nhi\ $\le1.3\times10^{23}$\,\cms\ at the location of the peak
optical depth integrated over the velocity spread of the absorption. Using this and the optical depth, we can place a limit on \ts$/$\fc\ $\le$ 3600~K. Comparatively,
from the spectral cube of lower spatial resolution of 10.9$''$ $\times$ 8.5$''$, we obtain a 3$\sigma$ upper limit of \nhi\ $\le1.4\times10^{22}$\,\cms\ and \ts$/$\fc\ 
$\le$ 380~K. From the velocity map in Fig.~\ref{fig:j0054_overlays}, it can be seen that the emission of the left region is redshifted with respect to that of the right.
Comparing the velocity of the emission with that of the absorption, we find that the absorption occurs at intermediate velocities between the two emission peaks. 
\subsubsection{J1036$+$0221}
\label{sec_j1036+0221}
\begin{figure*}
\includegraphics[height=0.24\textwidth, bb=94 15 745 645, clip=true]{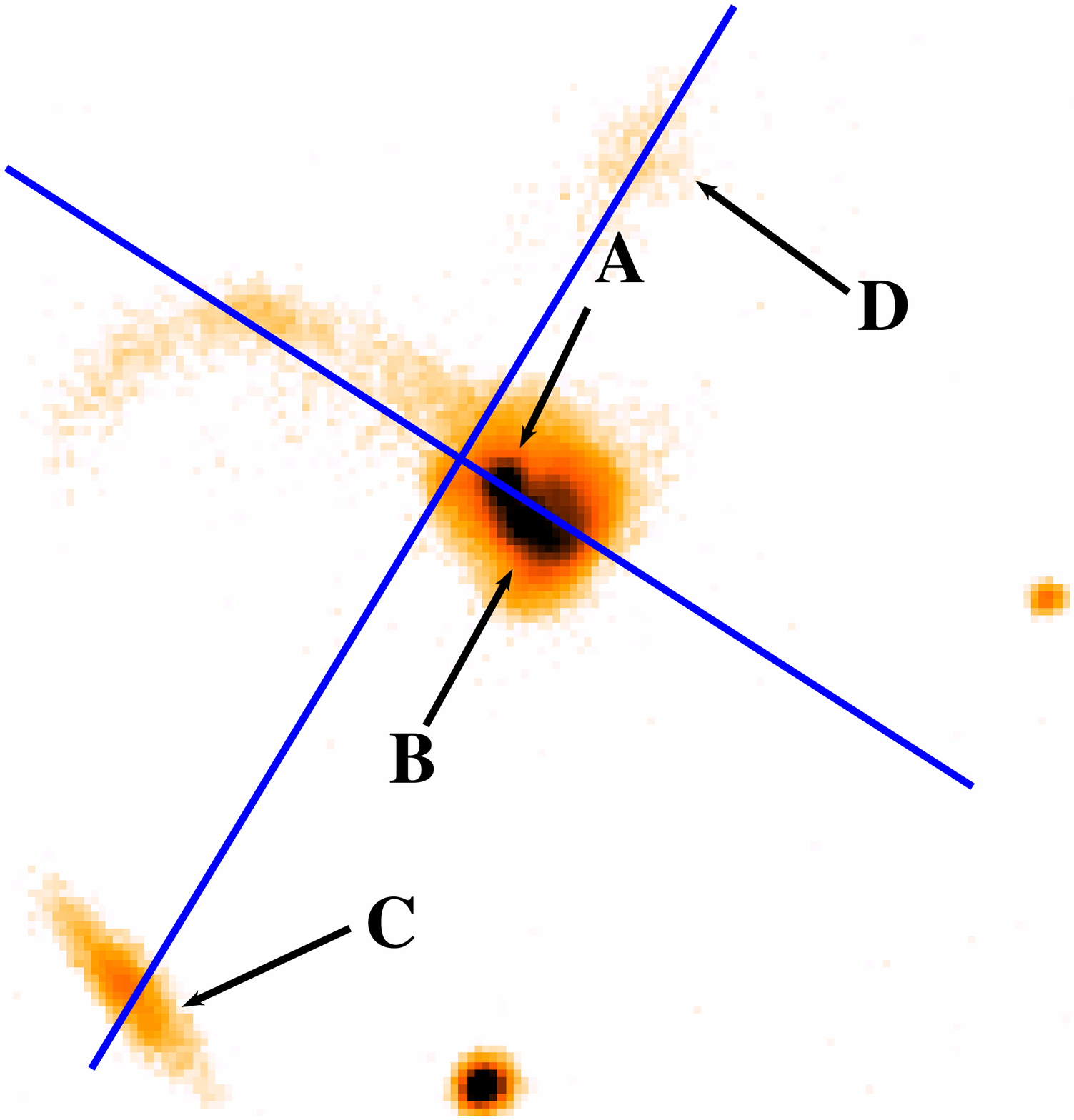} 
\includegraphics[height=0.37\textwidth, angle=90]{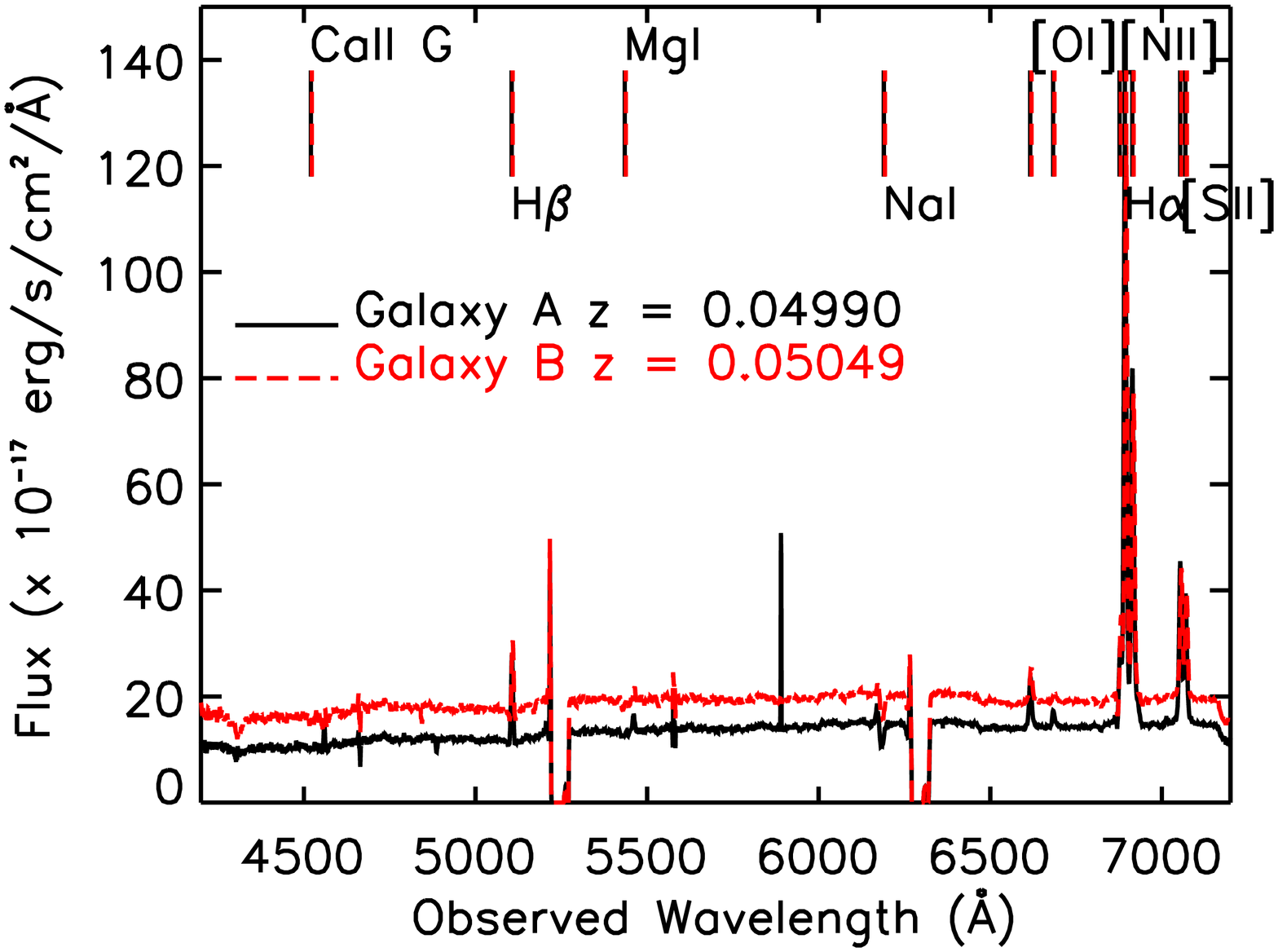} 
\includegraphics[height=0.37\textwidth, angle=90]{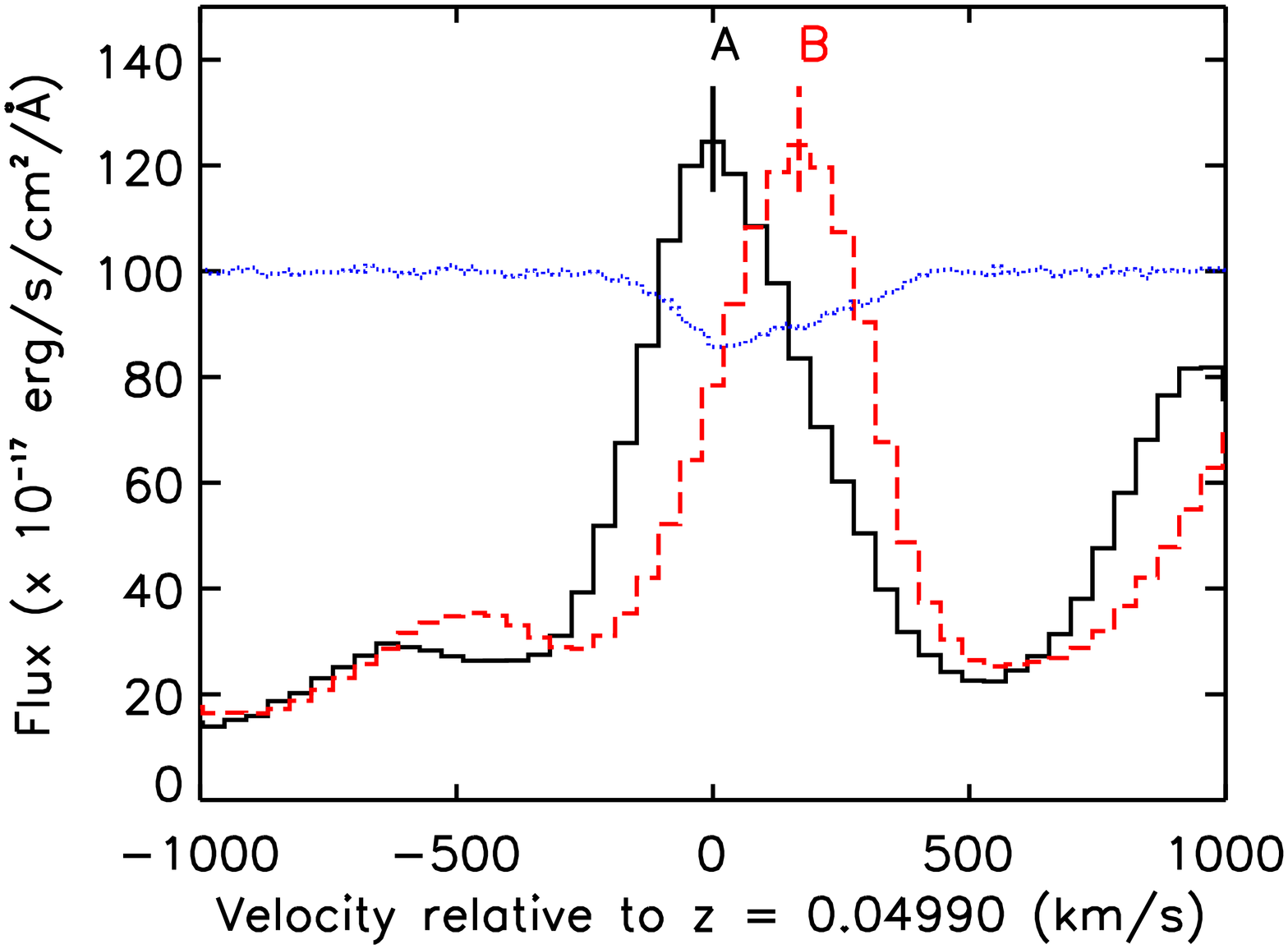}
\caption{{\it Left:} SDSS $r$-band image of J1036$+$0221 showing the slit positions used in our SALT observations.
The radio emission is associated with the nuclei marked `A'.
{\it Centre:} SALT spectra of the nuclear regions, marked `A' and `B' in the left panel, are shown in solid and dashed lines, respectively.
Various emission and absorption lines detected from the two nuclei are marked by vertical ticks.  
{\it Right:} Zoom-in view of the \ha\ emission from the central region of the nuclei, `A' and `B', in solid and dashed lines, respectively. 
The absorption detected towards the radio source `A' is overplotted in dotted lines. 
The absorption spectrum has been smoothed by two pixels and its y-axis has been scaled by an arbitrary value for display purpose.
The peak optical emission from the nuclei of `A' and `B' are marked by solid and dashed vertical ticks, respectively.}
\label{fig:j1036_saltspec}
\end{figure*}

This system (VIII Zw 090) is classified as an ultraluminous infrared galaxy \citep[ULIRG;][]{kilerci2014} and a post-merger \citep{ellison2013}, 
i.e. it consists of galaxies which are in the final stages of their interaction. \hi\ \21\ emission has not been detected from this system 
\citep[log~M(\hi) $<10.10$ M$_\odot$;][]{ellison2015}. This system consists of two interacting nuclei (separated by $\sim$2~kpc) and a tidal tail 
(see left panel of Fig.~\ref{fig:j1036_saltspec}). One of the interacting nuclei shows radio continuum emission in our GMRT 1.4 GHz map. We have 
detected broad ($\sim450$\,\kms) absorption towards this radio source. There is another weaker ($\sim$21~mJy) radio source, $\sim$18~kpc south-east 
of the strong one (see panel (b) of Fig.~\ref{fig:overlay}), towards which we do not detect absorption (\nhi\ $\le$ $6\times10^{20}$\,\cms, for 
\ts\ = 100~K and \fc\ = 1). There is no optical counterpart of the second radio source in the SDSS images, and it is uncertain whether it is 
related to the present system or is a background source. 

From our SALT long-slit data, we extracted spectra along the two slit position angles shown in the left panel of Fig.~\ref{fig:j1036_saltspec}, 
using apertures of sizes $\sim2.5''\times1.5''$ ($\sim2.5\times1.5$~kpc$^2$). The two nuclei in this system are spatially resolved in the long-slit
observations, and nebular emission lines of \ha, \hb, \nii, \sii\ and \oi, and absorption lines of \nai, \mgi\ and \caii\ are detected in their 
spectra (see centre panel of Fig.~\ref{fig:j1036_saltspec}). The \oiii\ emission lines fall in the gap between two detectors unfortunately. We
fit single component Gaussian to the emission lines to estimate the fluxes. We estimate the optical depth at the intrinsic $V$-band, $\tau_V^{Balmer}$, 
using \ha\ and \hb\ emission line ratio and equation 3 of \citet{argence2009} and equation 3 of \citet{wild2007}.\footnote{We use an intrinsic 
ratio of 2.85 as expected for the case B recombination \citep{osterbrock2006}. When we use a higher ratio of 3.1 as maybe applicable for AGN 
\citep{osterbrock2006}, it leads to slightly lower SFRs, but the values are consistent within the uncertainties.} The $\tau_V^{Balmer}$ ranges 
over $\sim2-6$ in the central region, and drops down to $\sim0.1-0.4$ over the tidal tail. Then using the \ha\ flux corrected for dust attenuation, 
we estimate the star formation rate (SFR), following \citet{kennicutt1998}. Note, however, that there is likely contribution from AGN to 
the emission particularly in the central regions, so the SFR values can be treated as upper limits.  The SFR is higher in the central region 
($\sim5-10$~M$_\odot$~yr$^{-1}$) and goes down to $\sim0.001$~M$_\odot$~yr$^{-1}$ in the tail region. Additionally, using \nii$/$\ha\ line 
ratio and the $N2$ index given in \citet{pettini2004}, we estimate the metallicity, $12 + log(O/H)$, which ranges over $8.7-9.0$.

Since the \oiii\ line is not covered, we cannot use the BPT diagnostic \citep{baldwin1981} to classify the emission line region. But the high 
log~\nii$/$\ha\ values ($\ge-0.2$) in the central region suggest that both the nuclei harbour AGNs \citep{osterbrock2006}. In addition, based on 
the nebular emission line ratios (log~(\oiii$/$\hb) $\sim$0.03, log~(\nii$/$\ha) $\sim-$0.25) from SDSS spectrum of the central region of the 
merger, it can be classified as a composite AGN/starburst system \citep[following classification schemes of][]{baldwin1981,kewley2001,kauffmann2003}.
The emission lines originating from the nuclei (`A') co-spatial with the peak of the radio continuum are blueshifted from those of the other 
nuclei (`B') by $\sim170$\,\kms. We measure the redshifts of the nuclear regions, `A' and `B', as 0.04990 $\pm$ 0.00005 and 0.05049 $\pm$ 0.00005, 
respectively. We estimate the SFR to be 7 $\pm$ 2~M$_\odot$~yr$^{-1}$ and 9 $\pm$ 3~M$_\odot$~yr$^{-1}$ in the two nuclei, `A' and `B', respectively. 
The absorption is best fitted with three Gaussian components. The strongest absorption component occurs at the systemic redshift of the nuclei `A'
(see right panel of Fig.~\ref{fig:j1036_saltspec}). The other two components are redshifted by $\sim150$\,\kms\ and $\sim250$\,\kms\ with respect
to the peak absorption. The component that is redshifted by $\sim150$\,\kms\ coincides in velocity with the peak optical emission from the nuclei `B'. 

In the two-dimensional optical spectra, we noticed \ha\ emission extending beyond the stellar continuum. Such extended emission is also seen in the 
{\it Galaxy Evolution Explorer (GALEX)} ultraviolet (UV) images of this merger. The \ha\ filaments could be arising from shock-heated gas resulting 
from the ongoing merger. The \nii$/$\ha\ ratio and the velocity dispersion of the emission lines can be used to separate shocked and star-forming 
regions \citep[see][and references therein]{mortazavi2018}. However, the low velocity resolution of our spectrum is not suitable for studying the 
kinematics of the ionized gas in detail. But the high log~\nii$/$\ha\ value ($\sim-0.2$) in the extended emission seen in the two-dimensional spectrum
suggests that shocks are the likely source of ionization.

Further, from the SALT spectrum, we identify another galaxy to be at the same redshift (0.05061 $\pm$ 0.00005) as that of the merger. This galaxy 
(marked `C' in Fig.~\ref{fig:j1036_saltspec}) is at a separation of $\sim30''$ ($\sim22$~kpc) south-east of the merger in the SDSS image, and could 
be interacting with the merging system. Note, however, that the galaxy seen $\sim17''$ to the north-west of the merger in the SDSS image (marked `D' 
in Fig.~\ref{fig:j1036_saltspec}), is actually not part of the merger, but is found to be a foreground galaxy at $z$ = 0.02189 $\pm$ 0.00005 from
the SALT spectrum.
\subsubsection{J1100$+$1002}
\label{sec_j1100+1002}
This system, UGC 06081, consists of two galaxies (nuclei separated by $\sim$11~kpc) undergoing a major merger. The main OH 18-cm absorption 
lines have not been detected in this system \citep{kazes1985}. However, strong intrinsic \hi\ \21\ absorption has been detected in this system 
previously by \citet{bothun1983}, \citet{williams1983} and \citet{darling2011}. Since this system has two radio sources (separation $\sim16''$),
which were not resolved in the previous single dish spectral line observations, the source of the absorption had remained ambiguous. Using our 
interferometric spectral line observations, we confirm that the absorption arises towards the stronger radio source in the north-west (see panel
(c) of Fig.~\ref{fig:overlay}). No absorption is detected towards the weaker radio source in the south-east (see Table~\ref{tab:results}). The 
absorption consists of two main components $-$ a strong narrow one and a broad shallow one, that is blueshifted with respect to the peak absorption.
The total integrated optical depth obtained by us is consistent with that estimated by \citet{darling2011}, who had assumed that the absorption 
arises in the stronger north-west radio source seen in the FIRST image.

\begin{figure*}
\includegraphics[height=0.20\textwidth, bb=40 100 720 600, clip=true]{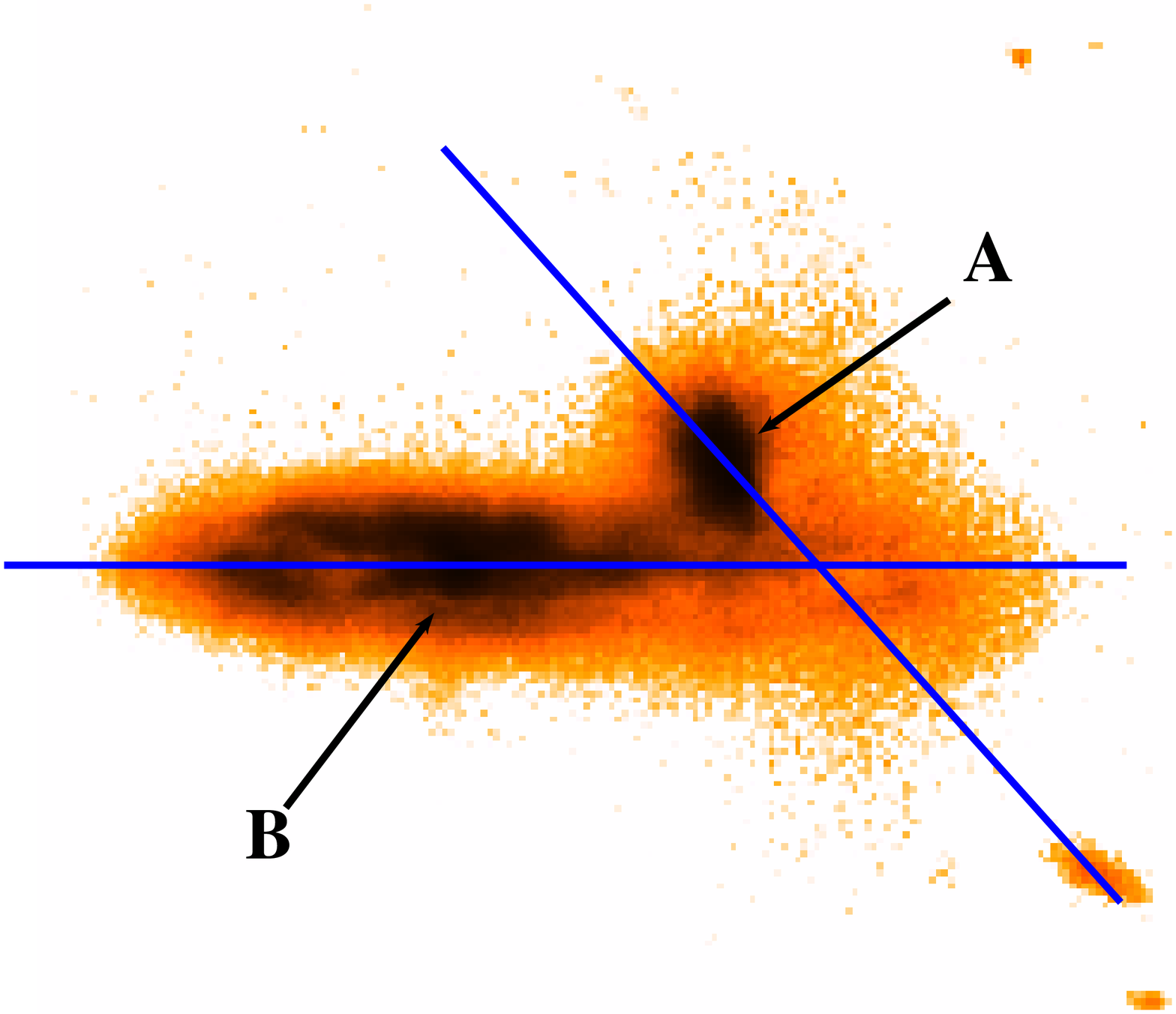} 
\includegraphics[height=0.35\textwidth, angle=90]{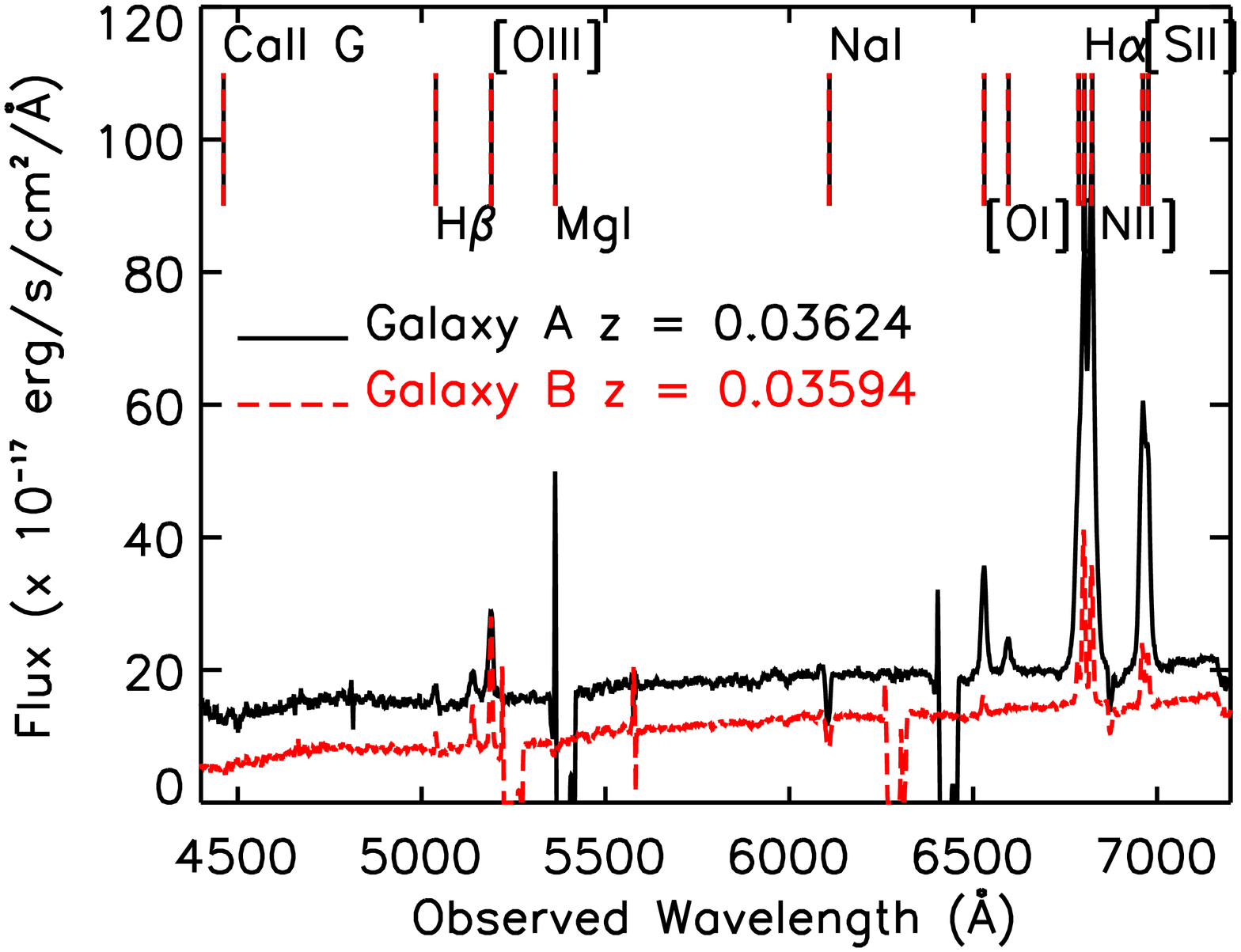} 
\includegraphics[height=0.35\textwidth, angle=90]{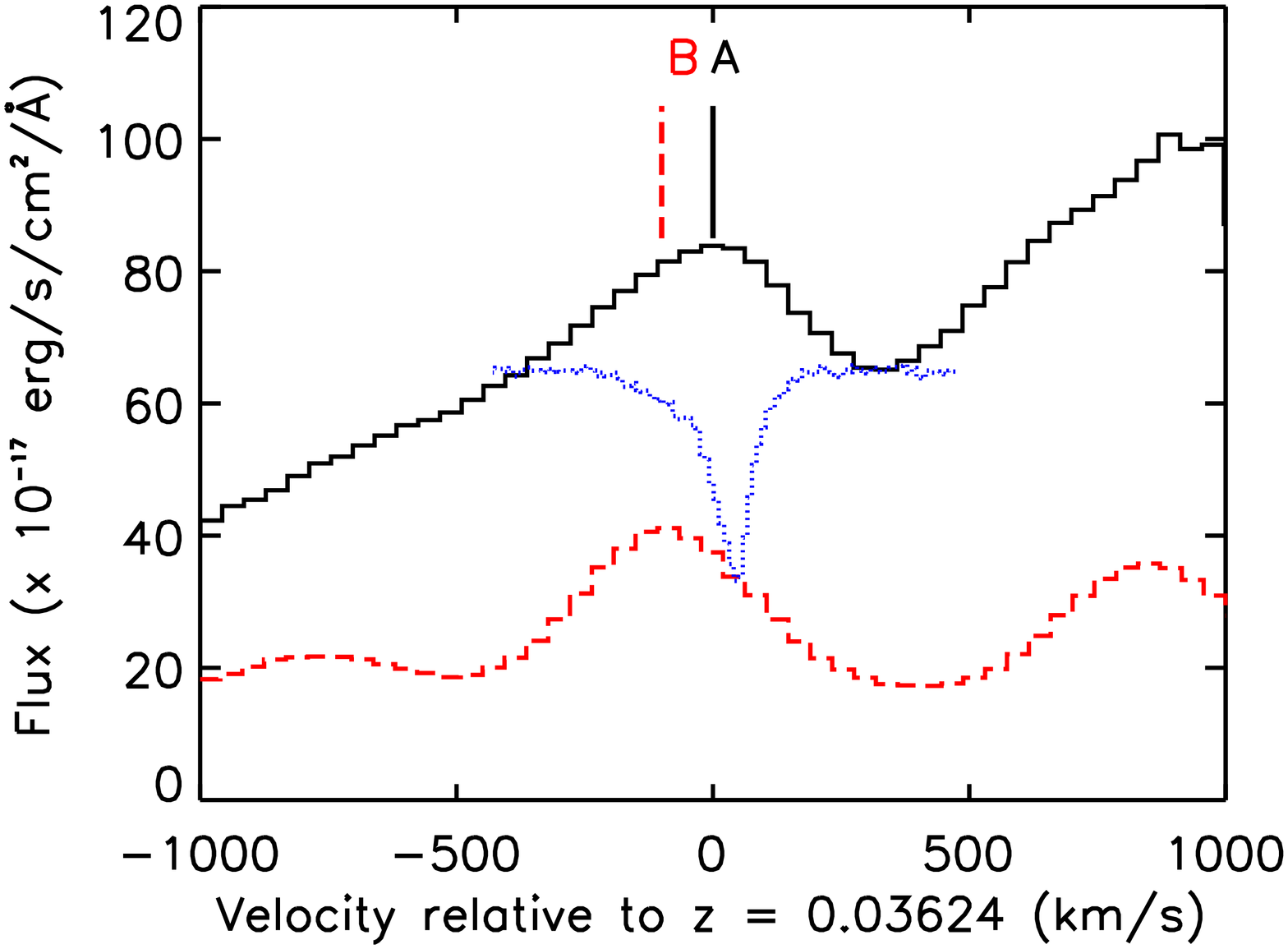}
\caption{{\it Left:} SDSS $r$-band image of J1100$+$1002 showing the slit positions used in our SALT observations.
The stronger radio emission is associated with the nuclei of galaxy marked `A'.
{\it Centre:} SALT spectra of the nuclear region of the two galaxies, marked `A' and `B' in the left panel, are shown in solid and dashed lines, respectively.
Various emission and absorption lines detected from the galaxies are marked by vertical ticks.  
{\it Right:} Zoom-in view of the \ha\ emission from the central region of the galaxies, `A' and `B', in solid and dashed lines, respectively. 
The absorption detected towards the radio source `A' is overplotted in dotted lines. 
The absorption spectrum has been smoothed by five pixels and its y-axis has been scaled by an arbitrary value for display purpose.
The peak optical emission from the nuclei of `A' and `B' are marked by solid and dashed vertical ticks, respectively.}
\label{fig:j1100_saltspec}
\end{figure*}

We extracted spectra from our SALT long-slit observations taken along the two slit position angles shown in the left panel of Fig.~\ref{fig:j1100_saltspec}.
We used apertures of sizes $\sim3.8''\times1.5''$ ($\sim2.7\times1.1$~kpc$^2$). We detect nebular emission lines of \ha, \hb, \nii, \sii, \oi\ and \oiii, 
and absorption lines of \nai, \mgi\ and \caii\ from both the galaxies in our SALT long-slit spectra (see centre panel of Fig.~\ref{fig:j1100_saltspec}). 
We estimate the dust optical depth ($\tau_V^{Balmer}$) and extinction-corrected SFR as described in Section~\ref{sec_j1036+0221}. The $\tau_V^{Balmer}$
values are high in the central parts ($\sim2-7$) of the two galaxies compared to the outer discs ($\sim0.2-0.5$). The SFR is similarly high in the central
parts ($\sim0.5-5$~M$_\odot$~yr$^{-1}$) and lower in the outer parts ($\sim0.001-0.1$~M$_\odot$~yr$^{-1}$). We estimate the metallicity using the $O3N2$ 
index of \citet{pettini2004}. The metallicity ($12 + log(O/H)$) ranges over $8.4-8.8$, with the central regions tending to have lower metallicity compared
to the outer regions. This is consistent with the metallicity dilution observed in the central regions of galaxy pairs due to low-metallicity gas being 
funneled to the centre during a merger \citep{kewley2006,ellison2008,scudder2012,cortijo2017}.

The emission lines originating from the stronger radio source (`A') are very broad ($\sim700$\,\kms), and they are redshifted by $\sim90$\,\kms\ from those 
originating from the weaker radio source (`B'). We estimate the redshifts, based on the emission lines from the nuclear region, to be 0.03624 $\pm$ 0.00008 
and 0.03594 $\pm$ 0.00005 for `A' and `B', respectively. The SFR in the nuclear region of `A' and `B' is 4.5 $\pm$ 1.0 and 2.0 $\pm$ 1.0 M$_\odot$~yr$^{-1}$, 
respectively. Though, we note again that the AGN contributes to the line emission in the central regions. Emission line ratio diagnostic \citep[BPT diagram;][]{baldwin1981} 
indicates that the gas in the central regions of both the galaxies is likely to be ionized by an AGN \citep[following][]{kewley2001,kauffmann2003}. The peak 
of the absorption is blueshifted by $\sim$35\,\kms\ from the peak optical emission from the stronger radio source, `A', whereas the broader absorption component
is blueshifted by $\sim$80\,\kms\ (see right panel of Fig.~\ref{fig:j1100_saltspec}). Though, as noted above, the emission lines from this source are very broad 
and likely to have multiple components.

From the line ratios in the outer regions, we find that the AGN contribution to the gas ionization is likely to extend to $\sim$10~kpc from the central regions.
We also detect \ha\ and \oiii\ emission extending beyond the stellar continuum around the merging system in the two-dimensional spectra, which could indicate
the presence of shocked gas, similar to the system J1036$+$0221. In addition, the GALEX images of this system show a hole or deficit in the UV emission from 
the weaker radio source, `B', that is co-spatial with the peak of the radio continuum, infrared and optical emission, which could indicate the presence of a 
dusty AGN. The lack of absorption towards this radio source is thus interesting. Though the radio source is weak (flux $\sim$25~mJy), we can rule out strong
absorption, i.e. \nhi\ $\le5\times10^{20}$\cms, for \ts\ = 100~K and \fc\ = 1.
\subsubsection{J1108$-$1015}
\label{sec_j1108-1015}
\begin{figure*}
\includegraphics[height=0.25\textwidth, bb=100 30 750 650, clip=true]{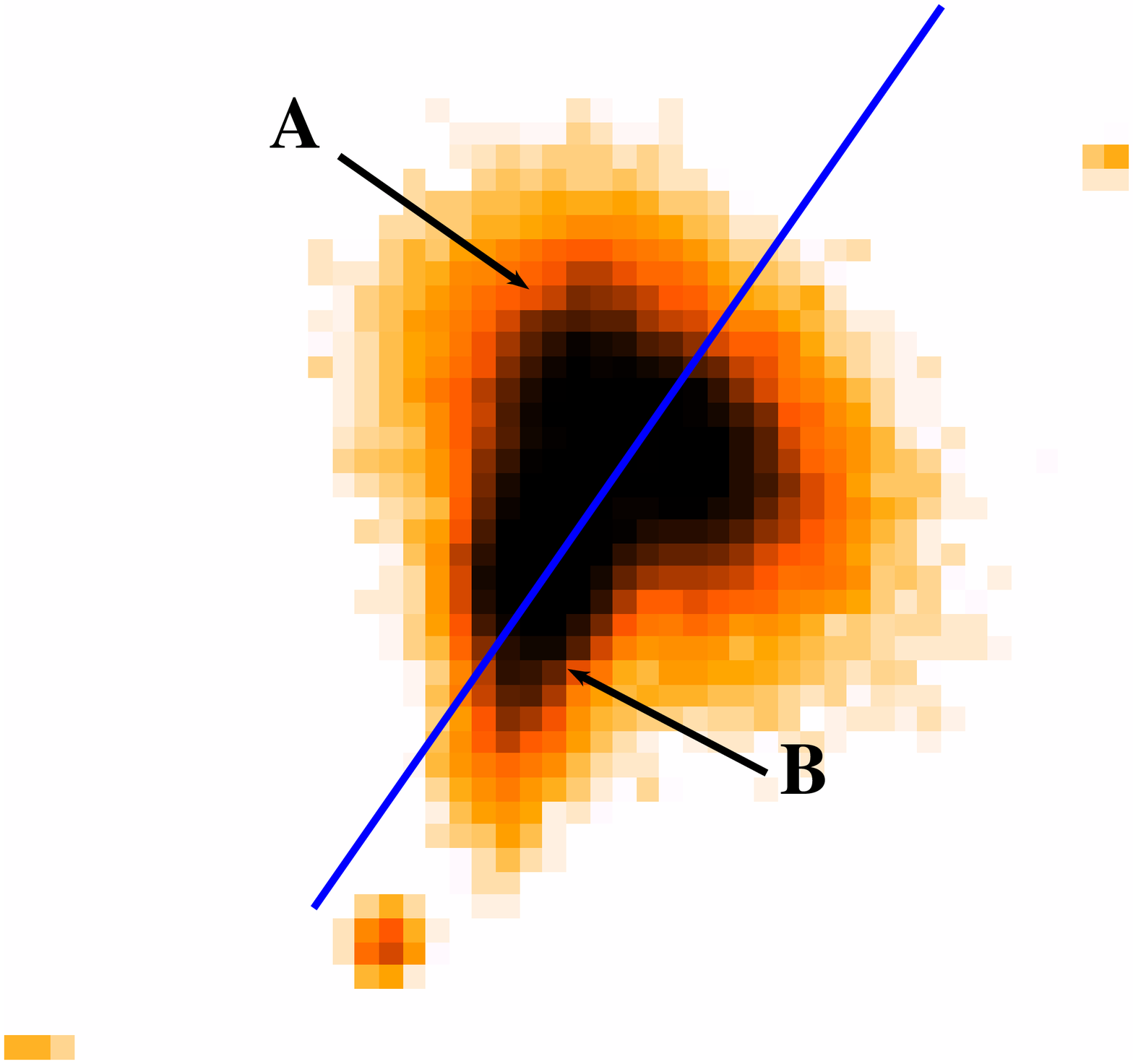} 
\includegraphics[height=0.36\textwidth, angle=90]{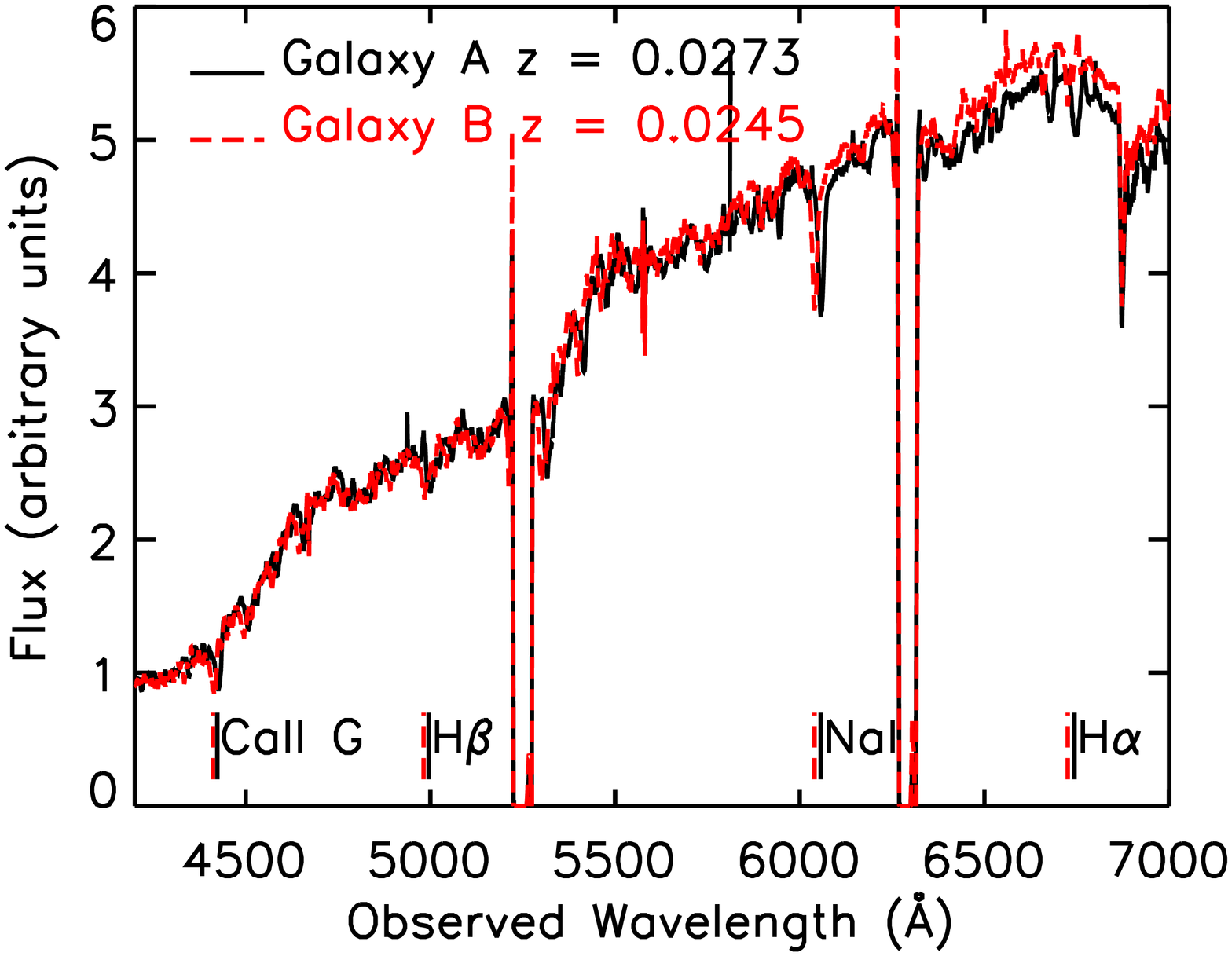}
\caption{{\it Left:} DSS image of J1108$-$1015 showing the position of the slit used in our SALT observations.
{\it Right:} SALT spectra of the two galaxies, marked `A' and `B' in the left panel, are shown in solid and dashed lines, respectively.
Absorption lines of \caii, \nai, \ha\ and \hb\ detected from the galaxies are marked by vertical ticks.}
\label{fig:j1108_saltspec}
\end{figure*}

This system, also known as NGC 3537, consists of two interacting elliptical galaxies, with their nuclei at a projected separation of $\sim$6~kpc. 
Absorption from \ha, \hb, \caii\ and \nai\ are detected in our SALT long-slit spectra of both the galaxies (see Fig.~\ref{fig:j1108_saltspec}),
indicating that these galaxies are non-star-forming. We estimate the redshift of the two galaxies marked `A' and `B' in Fig.~\ref{fig:j1108_saltspec},
as 0.0273 $\pm$ 0.0003 and 0.0245 $\pm$ 0.0003, respectively. The peak of the \ha\ absorption lines from the two galaxies are separated by $\sim$800\,\kms. 
We do not detect any extended emission between the two galaxies in the two-dimensional spectra. The radio emission is resolved in our GMRT 1.4 GHz 
continuum map, with the peak emission accounting for only $\sim$7\% of the total emission. \hi\ \21\ emission has been detected from this system 
with a total flux density of 7.5 $\pm$ 1.1 Jy~\kms\ \citep{richter1987}. However, we do not detect absorption towards any of the continuum 
components, with \nhi\ $\le$ $6\times10^{20}$\,\cms, for \ts\ = 100~K and \fc\ = 1, towards the continuum peak. 
\subsubsection{J1214$+$2931}
\label{sec_j1214+2931}
This system was observed as part of our {\it `Miscellaneous Sample'} of quasar-galaxy-pairs \citep{dutta2017a}. Also known as Was 49, this is an 
isolated, dual AGN system consisting of the disk galaxy Was 49a, with a low luminosity Seyfert 2 nucleus, and the Type 2 AGN Was 49b, corotating 
at a projected radial separation of $\sim$8~kpc from the nucleus of Was 49a \citep{moran1992}. The velocities of the optical emission lines from 
the two nuclei differ by $\sim$70\,\kms. The extreme luminosity of Was 49b is unusual for an AGN in the smaller galaxy of a minor-merger system 
\citep{secrest2017}. Was 49b is also radio-loud, and its radio emission is resolved in our GMRT 1.4 GHz continuum map. We do not detect absorption
towards any of the continuum points, which leads to \nhi\ $\le$ $2\times10^{20}$\,\cms, for \ts\ = 100~K and \fc\ = 1, towards the continuum peak.
\subsubsection{J1315$+$6207}
\label{sec_j1315+6207}
This is a major merger (also known as UGC 8335, Arp 238) between two galaxies at a projected separation of $\sim$21~kpc. \hi\ \21\ and CO emission 
has been detected from this LIRG \citep{huchtmeier1989,leech2010}, however no OH 18-cm absorption has been detected \citep{henkel1986,staveley1987}.
Based on {\it Herschel} far-infrared and sub-millimetre imaging observations, the two interacting galaxies have SFRs of 8 and 55 M$_\odot$~yr$^{-1}$
and dust masses of 6 and 7 $\times10^{7}$ M$_\odot$ \citep{cao2016}. The nuclei of both the interacting galaxies show radio emission in our GMRT map. 
We detect broad absorption towards the stronger radio source. Single component Gaussian gives best fit to the absorption. The peak optical depth is
redshifted by $\sim$90\,\kms\ with respect to the systemic velocity measured from optical emission lines. The second radio source is too weak to 
obtain a strong constraint on the absorption (\nhi\ $\le2\times10^{21}$\,\cms, for \ts\ = 100~K and \fc\ = 1). 

Absorption has been previously reported from this system in \citet{maccagni2017}. The total optical depth reported by them is three times 
less than that obtained by us, though the FWHM of the absorption is similar. Note also that the total continuum flux densities in both the
data match within 10\%. The beam of their observations ($\sim$45$''\times$12$''$) using the Westerbork Synthesis Radio Telescope (WSRT) has 
$\sim$90 times larger area compared to that of our observations. The difference in the optical depth is not expected if the absorbing gas 
uniformly covers the radio source. However, in the presence of partial coverage, the optical depth measured using data obtained with WSRT 
tend to be lower than that measured using higher spatial resolution data obtained with GMRT. Thus, the higher optical depth we measure points
towards the clumpy nature of the absorbing gas and partial coverage of the radio source. 

We also detect another source in our GMRT radio continuum image at $\sim$1.5$'$ (or $\sim$60~kpc) south-west of J1315$+$6207 (see panel (f) of Fig.~\ref{fig:overlay}). 
There is no redshift information available for this source in the literature, so we do not know whether it is a background source or not. We do not detect 
absorption against this source. If the radio source happens to be a background source, we get \nhi\ $\le$ $5\times10^{20}$\,\cms, for \ts\ = 100~K 
and \fc\ = 1, towards the strongest emission ($\sim$19\,\mjb) of the continuum.
\subsubsection{J1320$+$3408}
\label{sec_j1320+3408}
This major merger (also known as IC 883, UGC 8387, Arp 193) consists of a LIRG that is classified as a starburst-AGN composite. The radio source 
has a core-jet morphology at sub-parsec scales \citep{romero2017}. Detection of \hi\ \21\ absorption towards the radio source has been reported 
previously \citep{heckman1983,clemens2004,gereb2015}. \citet{clemens2004}, using sub-arcsecond-scale \hi\ absorption and CO emission data, found 
differences in the velocity structure of the atomic and molecular gas components in the central regions of this merger. Our GMRT absorption 
spectrum is consistent with the results reported in the literature. We find that the absorption profile is best fitted with three Gaussian 
components. The SDSS spectra of the central region of this system measures the redshift from nebular emission lines as 0.02306 $\pm$ 0.00001. 
The strongest absorption is redshifted by $\sim$70\,\kms\ with respect to this. Among the other two components, one is redshifted by $\sim$200\,\kms\ 
and the other is blueshifted by $\sim$40\,\kms.
\subsubsection{J1356$+$1026}
\label{sec_j1356+1026}
This system is classified as an ULIRG and consists of two merging nuclei separated by $\sim$2.5~kpc. The northern nucleus hosts a luminous obscured 
Type 2 double-peaked \oiii\ AGN \citep{greene2009,liu2010,fu2011}. This system has not been detected in OH 18-cm \citep{darling2000}. CO molecular 
emission has been mapped from this system \citep{sun2014}. Both molecular and ionized gas outflows have been observed emerging from the obscured 
quasar \citep{sun2014,greene2012}. Extended X-ray emission has also been observed around this quasar, that is co-spatial with the ionized gas outflow. 
This quasar shows compact radio emission in our GMRT 1.4 GHz map. Absorption is detected against the radio emission, and is best fit with a single 
Gaussian. The peak of the absorption is redshifted by $\sim$150\,\kms\ from the systemic redshift of $z = 0.1231$ measured from SDSS spectra. We are
unable to securely confirm any outflowing neutral gas components due to strong radio frequency interference (RFI) over the corresponding frequency
ranges. The total optical depth obtained by us is a factor of 2.5 times higher than that obtained by \citet{maccagni2017} using WSRT, while the velocity
width is similar. The total continuum flux densities in the two data are consistent within a factor of 1.3. We note that the signal-to-noise ratio 
(SNR) of the WSRT absorption spectrum is low, with the peak optical depth being detected at only $\sim$2$\sigma$ significance (though the total optical 
depth is detected at $\sim$8$\sigma$ significance). The higher optical depth measured using GMRT data compared to that measured using WSRT data could 
be related to the larger beam size of the latter and partial coverage of the radio source as discussed in Section~\ref{sec_j1315+6207}.
\subsubsection{J1356$+$1822}
\label{sec_j1356+1822}
This is an ULIRG (known as Mrk 463 or UGC 8850) with double nuclei (separated by $\sim$4~kpc) and tidal tails \citep{mazzarella1991}. The eastern 
nucleus is classified as a Seyfert 2 galaxy and shows strong radio emission. The western nucleus is also an AGN, making this a binary AGN system 
\citep{bianchi2008}. It has been detected in CO emission \citep{pearson2016}, but not in \hi\ \21\ emission \citep{bieging1983}. Recently, 
\citet{treister2018} have conducted a detailed study of this system using optical, infrared and sub-mm integral field unit (IFU) spectroscopy.
They detect both outflowing ionized gas and inflowing molecular gas around the eastern nucleus. Interestingly, we do not detect any absorption
(\nhi\ $\le$ $10^{20}$\,\cms, for \ts\ = 100~K and \fc\ = 1) towards the radio source, which is co-spatial with the eastern nucleus. 
This could be because most of the cold gas in the circumnuclear region is in the molecular form. The non-detection could also be related to the 
complex radio structure and small covering factor of the cold gas clumps. While the radio source is compact in our GMRT 1.4 GHz continuum map, 
it is resolved in the 18-cm sub-arcsecond-scale ($\sim$18~pc resolution) map presented in \citet{kukula1999}. The total flux is $\sim$25\% of 
the arcsecond-scale flux and there are multiple radio components spread over $\sim$1~kpc in this map, with the peak flux density being 38\,\mjb. 
\subsubsection{J2054$+$0041}
\label{sec_j2054+0041}
\begin{figure}
\includegraphics[height=0.5\textwidth, angle=90]{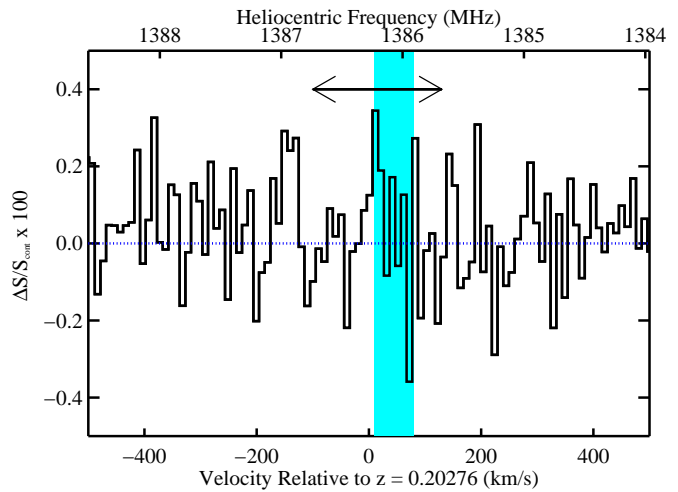} 
\caption{VLA OH 18-cm absorption spectrum towards J2054$+$0041. The spectrum has been smoothed to $\sim$10\,\kms\ for display purpose. 
The shaded region marks the frequency ranges affected by RFI. The arrow demarcates the frequency range over which \hi\ \21\ absorption is detected.}
\label{fig:j2054_oh}
\end{figure}

This system was initially selected as a quasar-galaxy-pair to study the cold gas around low-$z$ galaxies \citep[{\it `Miscellaneous Sample'} of][]{dutta2017a}.
However, our SALT long-slit observations revealed it to be a merging system. It consists of two interacting galaxies at $z\sim0.2$, at a projected separation 
of $\sim$11~kpc. One of them is the radio source PKS 2052$+$005, which is classified as a flat-spectrum radio source \citep{healey2007}. We detect strong and 
broad multi-component absorption towards the radio source using VLA. This is the only system in our sample whose redshifted OH 18-cm absorption lines fall in
frequency ranges that are relatively unaffected by strong RFI. So we searched for OH absorption, i.e. the main 1665 and 1667 MHz lines, using VLA. We do not
detect any OH absorption, though part of the frequency range, over which \hi\ absorption has been detected, is affected by RFI (see Fig.~\ref{fig:j2054_oh}). 
The radio source, similar as in the \hi\ data, is compact with a peak flux density of 378~\mjb. We reach a $3\sigma$ optical depth sensitivity of, $\int\tau_{1667 MHz}dv$ $\le$0.07\,\kms, 
for a velocity width of 30\,\kms\ (the FWHM of the two narrow Gaussian components detected in \hi\ absorption; see Table~\ref{tab:gaussfit}). From this we 
estimate $N$(OH) $\le 1.6 \times 10^{14}$\,\cms, assuming an excitation temperature of 10 K \citep[following equation 1 of][]{liszt1996}, i.e. $N$(OH)$/$\nhi\ $\le10^{-7}$ 
for the individual \hi\ absorption components. Thus, the \hi\ absorption is most likely tracing the diffuse gas in the circumnuclear region of this merger. 

\begin{figure*}
\includegraphics[height=0.45\textwidth, angle=90]{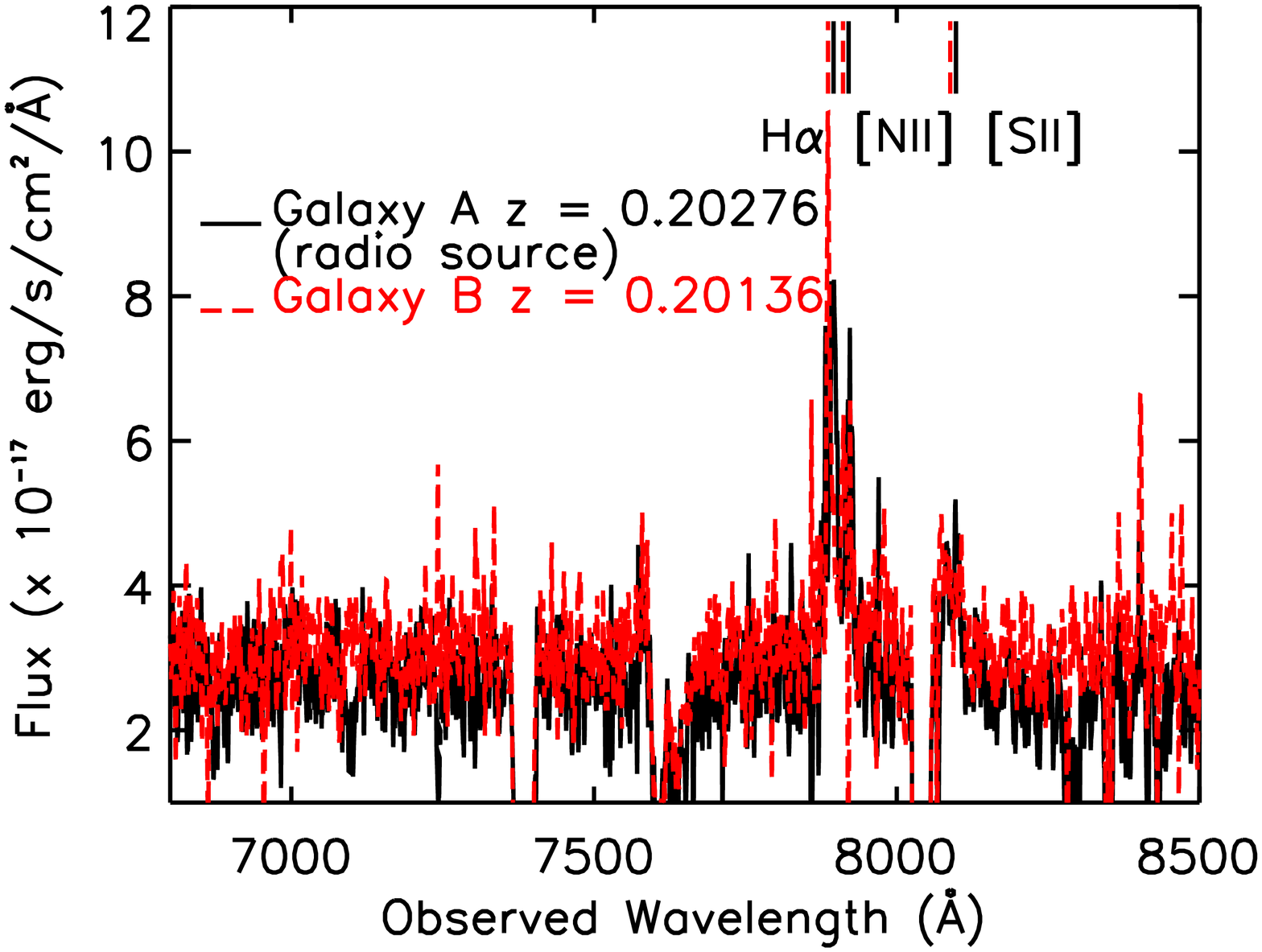} 
\includegraphics[height=0.45\textwidth, angle=90]{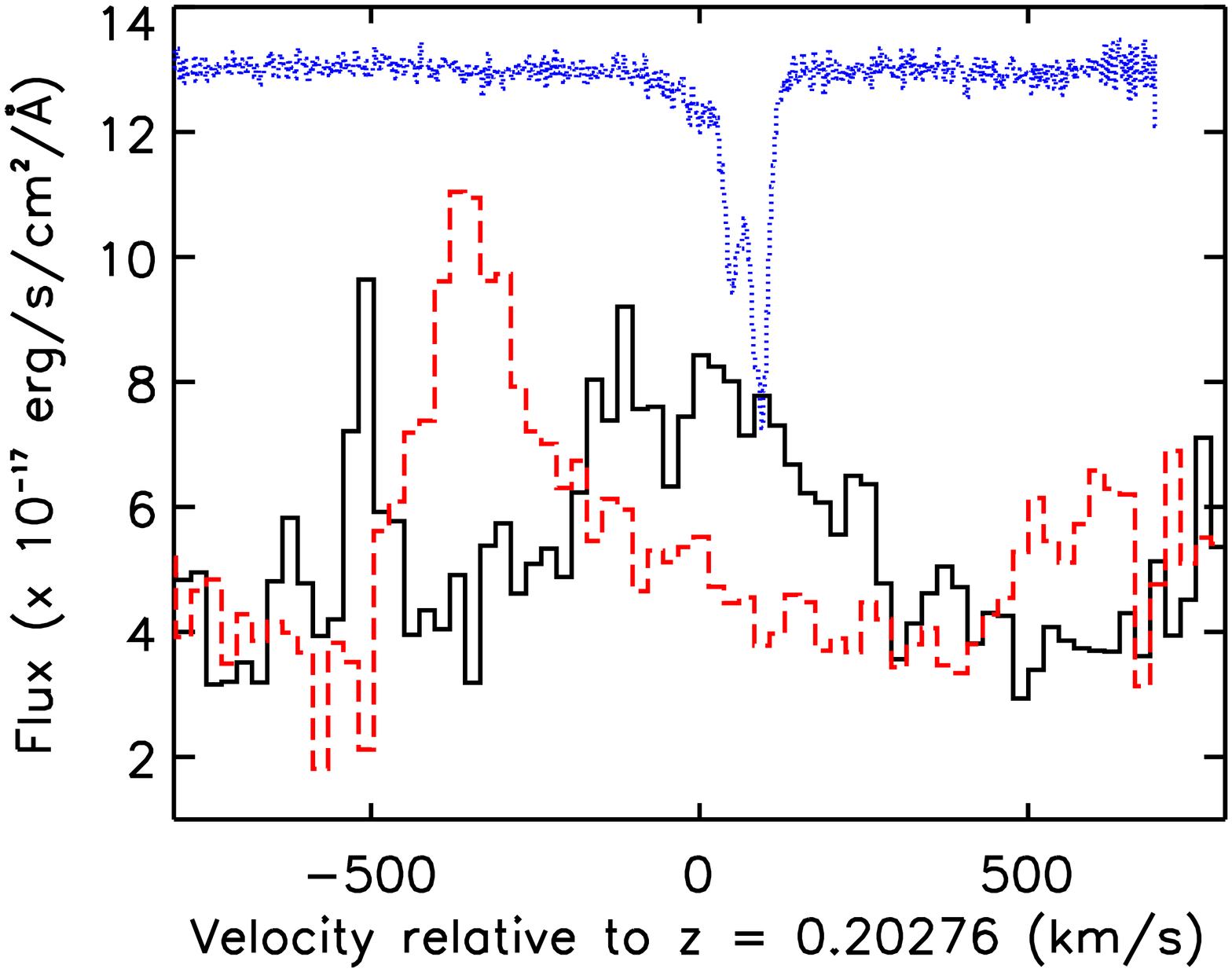}
\caption{{\it Left:} SALT spectra of the two galaxies, `A' and `B', that constitute the merging system J2054$+$0041 are shown in solid and dashed lines, respectively.
Galaxy `A' shows radio emission. Emission lines of \ha, \nii\ and \sii\ detected from the galaxies are marked by vertical ticks. The spectra have been smoothed
by three pixels. {\it Right:} Zoom-in view of the \ha\ emission from the galaxies, `A' and `B', in solid and dashed lines, respectively. The absorption detected
towards radio source `A' is overplotted in dotted lines. The absorption spectrum has been smoothed by five pixels and its y-axis has been scaled by an arbitrary
value for display purpose.}
\label{fig:j2054_saltspec}
\end{figure*}

\ha\ emission is detected from both the merging galaxies in our SALT long-slit spectra, separated by $\sim$350\,\kms\ (see left panel of Fig.~\ref{fig:j2054_saltspec}). 
\nii\ and \sii\ emission are also detected, but these wavelength ranges are affected by sky emission lines and have poor SNR. The \hi\ \21\ absorption consists of three 
components $-$ two narrow ones and a broad one. The velocity range of the absorption coincides with that of the \ha\ emission from the radio galaxy (`A', $z$ = 0.2028 $\pm$ 0.0005; 
see right panel of Fig.~\ref{fig:j2054_saltspec}). No absorption is detected at the velocity of the other galaxy (`B', $z$ = 0.2014 $\pm$ 0.0005), whose nebular emission
is blueshifted with respect to that of the radio galaxy (`A') by $\sim350$\,\kms. From the \ha\ emission, we estimate the SFR (uncorrected for extinction) to be $0.59 \pm 0.02$ 
and $0.49 \pm 0.02$ M$_\odot$~yr$^{-1}$ for the galaxies `A' and `B', respectively \citep[following][]{kennicutt1998}.
%
%
\section{Discussion} 
\label{sec_discussion}
\subsection{Detection rate of neutral gas in mergers}
\label{sec_discussion1}
The detection rate of absorption in our sample of mergers is 70 $\pm$ 26\%. Absorption is not detected in three systems, 
J1108$-$1015, J1214$+$2931 and J1356$+$1822. The system J1108$-$1015, consisting of two red non-star-forming galaxies, 
is likely to be a gas-poor, dry merger. \hi\ \21\ emission has been detected from this system using single dish observation
(Section~\ref{sec_j1108-1015}), which has spatial resolution of few arcmins and does not resolve the system. Though this
indicates the presence of neutral gas, the non-detection of absorption towards the central radio source suggests the lack
of cold neutral gas in the central region (few kiloparsecs) of this system. J1214$+$2931 and J1356$+$1822 are double nuclei
systems, where one of the nuclei shows radio emission. While J1356$+$1822 shows tidal tails, J1214$+$2931 is a minor merger
without any significant morphological disturbances visible in the SDSS image. As mentioned in Section~\ref{sec_j1356+1822},
J1356$+$1822 does show the presence of molecular gas around both its nuclei. Our GMRT spectra indicates that \nhi\ $\lesssim10^{20}$
towards these radio sources (Table~\ref{tab:results}).

Next, to get a more statistically robust estimate of the detection rate of absorption in mergers, we have compiled 
a list of all low-$z$ ($z\le0.2$) strong radio sources that are associated with galaxy mergers and have been searched 
for absorption in the literature (see Table~\ref{tab:litsample}). We visually inspected the available optical images 
(mostly from SDSS, in some cases from DSS) of the $z\le0.2$ radio sources that have been searched for absorption in 
the literature\footnote{Using compilations given in $-$ \citet{gupta2006,chandola2011,chandola2013,gereb2015,maccagni2017}.
Note that in these studies, the targets were selected purely based on their radio emission and are unbiased with respect to
their optical morphology.}. We also cross-matched samples of local galaxy mergers from the literature (see Section~\ref{sec_observations}) 
with the above compilations of absorption searches in low-$z$ radio sources. For identification of mergers we followed 
the same criteria as adopted for selecting our sample (see Section~\ref{sec_observations}). We find 19 merging systems 
associated with radio sources at $z\le0.2$, and 17 of them have been detected in absorption (see Table~\ref{tab:litsample}).
In the case of multiple radio components in a system, we have considered the absorption detected towards the strongest radio 
component. Combining the above mergers with our sample gives a detection rate of $83 \pm 17$\% in merging systems. This is 
considerably higher (by $\sim3-4$ times) than the typical detection rates ($\sim20-30$\%) in samples of low-$z$ radio sources
\citep[see for example table 1 of][]{maccagni2017}. \citet{maccagni2017} also find a similar high rate ($\sim78$\%) of incidence
of absorption in a smaller sample (9) of interacting systems compared to samples of non-interacting radio galaxies. The high 
detection rate in merging systems indicates large covering factor of neutral gas in their circumnuclear regions.
\begin{table*}
\caption{Details of $z\le0.2$ galaxy mergers from the literature that have been searched for \hi\ \21\ absorption.}
\centering
\begin{tabular}{cccccccc}
\hline
Name & Coordinates & $z$ & $S$(1.4~GHz)  & \nhi\                  & FWHM   & \vshift\ & Ref. \\
     & (J2000)     &     & (mJy)         & (\ts$/100$ K)($1/$\fc) & (\kms) & (\kms)   &      \\
     & RA~~~~~~Dec &     &               & ($10^{20}$\,\cms)      &        &          &      \\
(1)  & (2)         & (3) & (4)           & (5)                    & (6)    & (7)      & (8)  \\
\hline
NGC~2623                   & 08:38:24 $+$25:45:17 & 0.0185 & 95   & 67     & 254 & 64     & 1  \\
UGC~5101                   & 09:35:52 $+$61:21:12 & 0.0394 & 147  & 54     & 536 & $-$78  & 2  \\
SDSS~J094221.98$+$062335.2 & 09:42:22 $+$06:23:35 & 0.1232 & 108  & 91     & 75  & 0      & 3  \\
NGC~2992                   & 09:45:42 $-$14:19:35 & 0.0077 & 227  & 57     & 95  & $-$18  & 4  \\
SDSS~J105327.25$+$205835.9 & 10:53:27 $+$20:58:36 & 0.0526 & 84   & 5      & 156 & $-$58  & 2  \\
NGC~3690                   & 11:28:32 $+$58:33:43 & 0.0104 & 282  & 94     & 207 & 165    & 1  \\
MRK~231                    & 12:56:14 $+$56:52:25 & 0.0422 & 243  & 29     & 179 & 0      & 5  \\  
NGC~4922                   & 13:01:25 $+$29:18:50 & 0.0234 & 40   & 15     & 148 & 54     & 6  \\
NGC~5222                   & 13:34:56 $+$13:44:32 & 0.0231 & 34   & 29     & 130 & $-$101 & 6  \\
NGC~5256                   & 13:38:17 $+$48:16:32 & 0.0276 & 55   & 13     & 140 & 176    & 6  \\
Mrk~273                    & 13:44:42 $+$55:53:13 & 0.0373 & 132  & 86     & 570 & 85     & 2  \\
4C$+$12.50                 & 13:47:33 $+$12:17:24 & 0.1217 & 4860 & 3      & 135 & 45     & 7  \\
IRASF~14394$+$5332         & 14:41:04 $+$53:20:09 & 0.1045 & 40   & $\le$5 & --- & ---    & 6  \\
VV~059                     & 15:08:05 $+$34:23:23 & 0.0456 & 130  & 125    & 102 & $-$178 & 8  \\
3C~321                     & 15:31:43 $+$24:04:19 & 0.0961 & 81   & 92     & 33  & 235    & 9  \\          
Arp~220                    & 15:34:57 $+$23:30:11 & 0.0181 & 316  & 132    & 238 & $-$32  & 10  \\
NGC~6040                   & 16:04:26 $+$17:44:31 & 0.0409 & 72   & $\le$3 & --- & ---    & 2  \\
SDSS~J163804.02$+$264329.0 & 16:38:04 $+$26:43:29 & 0.0652 & 36   & 25     & 111 & 96     & 6  \\
NGC~6240                   & 16:52:58 $+$02:24:03 & 0.0245 & 427  & 128    & 348 & $-$44  & 11 \\
\hline
\end{tabular}
\label{tab:litsample}
\begin{flushleft} {\it Notes.}
Column 1: galaxy merger name. Column 2: J2000 coordinates of the galaxies undergoing merger. Column 3: redshift of the galaxy merger. 
Column 4: total flux density at 1.4~GHz in mJy of the radio source from FIRST (from NVSS in the case of NGC~2992 and NGC~6240). 
Column 5: \nhi\ assuming \ts\ = 100 K and \fc\ = 1, in units of $10^{20}$\,\cms. 
Column 6: full-width-at-half-maximum (FWHM) in \kms\ of the absorption.
Column 7: velocity shift (\vshift) in \kms\ of the peak absorption and the galaxy systemic optical redshift.
Column 8: reference for the \hi\ \21\ absorption data $-$
1: \citet{dickey1986}; 2: \citet{gereb2015}; 3: \citet{srianand2015}; 4: \citet{gallimore1999}; 5: \citet{carilli1998b}; 
6: \citet{maccagni2017}; 7: \citet{gupta2006}; 8: \citet{chandola2011}; 9: \citet{chandola2012}; 10: \citet{mundell2001}; 11: \citet{baan2007} \\
\end{flushleft}
\end{table*}
\subsection{Association of high column density neutral gas with mergers}
\label{sec_discussion2}
\begin{figure}
\includegraphics[height=0.5\textwidth, angle=90]{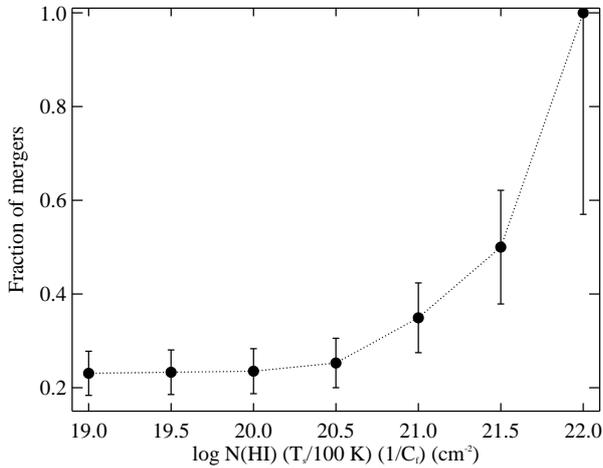} 
\caption{The fraction of merging systems among the radio sources showing \hi\ \21\ absorption at $z\lesssim0.2$ with increasing \nhi\ cut-off.
Here we have converted the total measured optical depth to \nhi\ assuming \ts\ = 100 K and \fc\ = 1.}
\label{fig:merger_frac}
\end{figure}

\citet{srianand2015} have noted that high column density absorbing gas, i.e. \nhi\ $\sim10^{22}$\,\cms, appear to be
associated with galaxy mergers. We plot the fraction of merging systems among low-$z$ radio sources showing absorption
as a function of different \nhi\ cut-offs in Fig.~\ref{fig:merger_frac}. For this we have used the literature compilations 
of $z\le0.2$ intrinsic absorbers as mentioned in Section~\ref{sec_discussion1}, and our sample of mergers and the mergers
listed in Table~\ref{tab:litsample}. We have considered \ts\ = 100 K and \fc\ = 1 to convert the optical depth of the 
absorption to \nhi. It can be seen that the fraction of mergers increases as the \nhi\ cut-off increases. About 50\% 
of the absorbers with \nhi\ $> 5 \times 10^{21}$\,\cms\ arise from radio sources that are associated with mergers, and 
all the five sources having \nhi\ $>10^{22}$\,\cms\ are associated with mergers. Hence, we confirm the point made by 
\citet{srianand2015} that extremely strong associated absorbers have very high probability of arising from merging systems. 
Though we note the caveat that visual identification of mergers is subjective and in some cases we are limited by the 
available optical data. However, visual classification is still considered the most reliable method for identification 
of merging/interacting systems \citep[e.g.][]{weston2017}, and majority of the sources considered here have optical 
images and spectra available in SDSS. 

\begin{figure*}
\includegraphics[height=0.45\textwidth, angle=90]{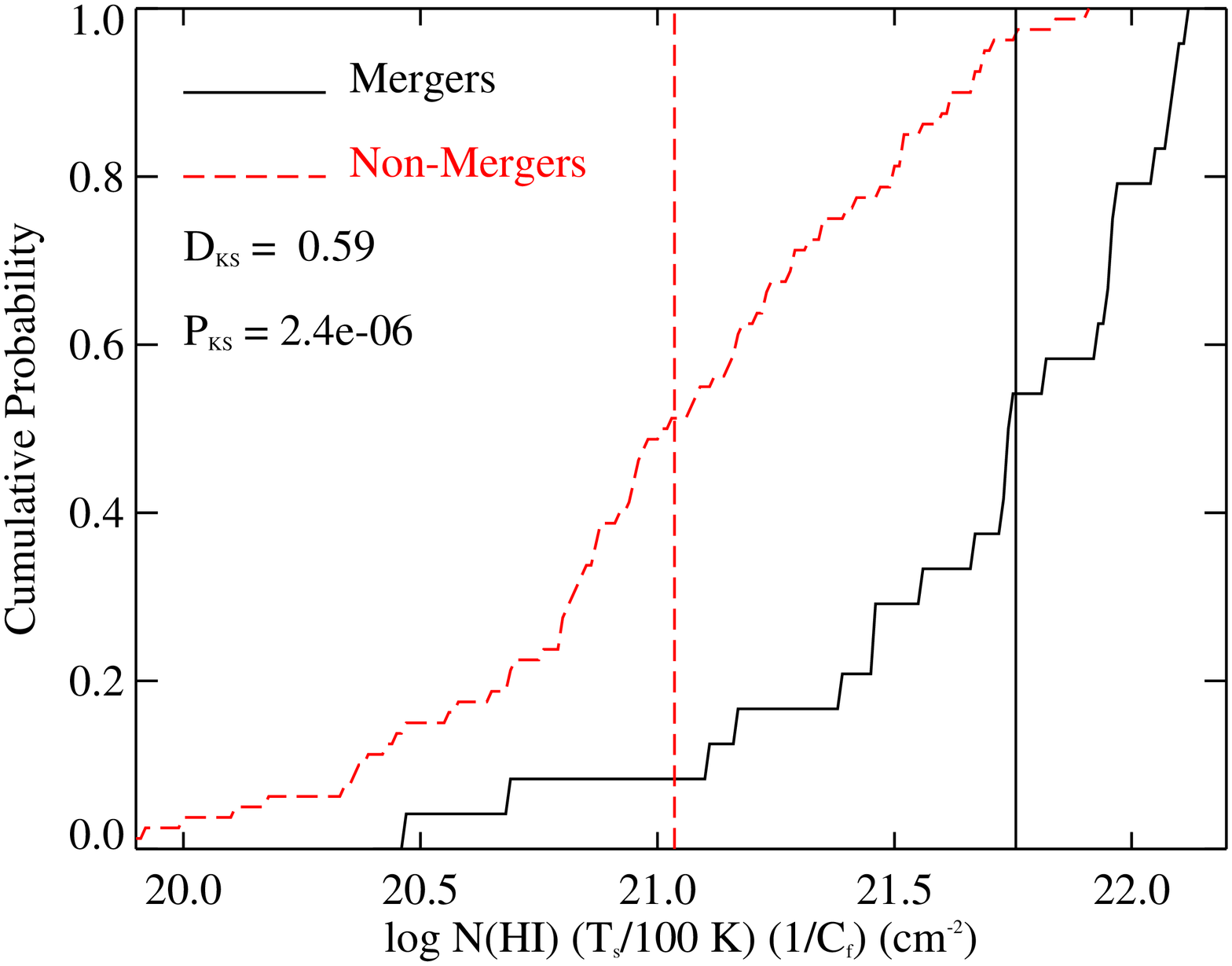}
\includegraphics[height=0.45\textwidth, angle=90]{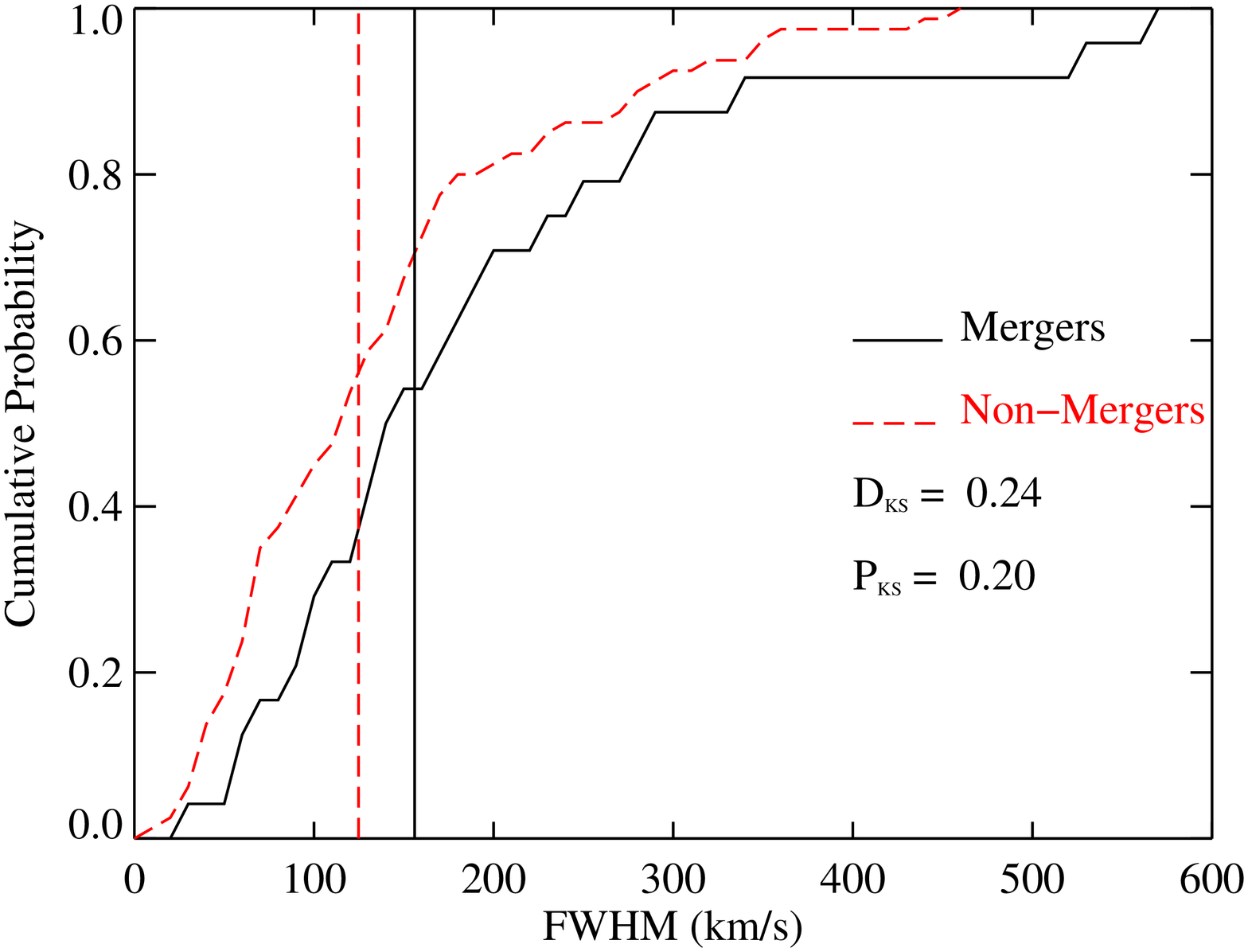}
\caption{{\it Left:} The cumulative distribution of \nhi\ (assuming \ts\ = 100 K and \fc\ = 1) for \hi\ \21\ absorption detected
from radio sources associated with mergers (solid line) compared to that for non-merging radio sources (dashed line). The vertical
solid and dashed lines mark the median \nhi\ of the merger and non-merger samples, respectively.
{\it Right:} The cumulative distribution of FWHM of the absorption lines detected from mergers (solid line) compared to that from
non-mergers (dashed line). The median FWHM of the merger and non-merger samples are marked by vertical solid and dashed lines,
respectively. The results from two-sided KS tests between the two distributions are indicated in both the panels.}
\label{fig:cumdist}
\end{figure*}

The association of high \nhi\ gas with mergers can be further seen from the left panel of Fig.~\ref{fig:cumdist}, which shows
the cumulative distribution of \nhi\ of absorption detected from radio sources associated with mergers compared to that of 
intrinsic absorption from non-merging radio sources. The merging systems show stronger absorption with a median \nhi\ = 
$6\times10^{21}$\,\cms, which is six times higher compared to the median \nhi\ in the non-merging sources. A two-sided 
Kolmogorov-Smirnov (KS) test between the two \nhi\ distributions indicates that the maximum deviation between the two 
cumulative distribution functions is $D_{\rm KS}$ = 0.59, with a probability of $P_{\rm KS}$ = $2.4 \times 10^{-6}$ (where 
$P_{\rm KS}$ is the probability of finding this $D_{\rm KS}$ value or lower by chance). Therefore, the \nhi\ distribution in 
samples of radio-loud AGNs can be used to statistically identify merger populations at high-$z$ and in future/ongoing blind 
\hi\ \21\ absorption surveys with the Square Kilometre Array pathfinders/precursors \citep{morganti2015,gupta2016}.
\subsection{Kinematics of neutral gas in mergers}
\label{sec_discussion3}
The absorption lines detected in our sample are broad ($\sim100-400$\,\kms), and we typically resolve two to three components
in each absorption, with the FWHM of individual components ranging over $30-200$\,\kms. The absorption profiles of the mergers 
compiled from the literature are also broad ($\gtrsim100$\,\kms), although the individual components may not be distinguished in 
some cases due to low spectral resolution data. We compare the widths of the absorption lines detected from mergers and non-mergers 
in the right panel of Fig.~\ref{fig:cumdist}. For the sources from the samples of \citet{gereb2015} and \citet{maccagni2017}, we use 
the Busy function FWHM provided by them. For the other sources from the literature, we measure the FWHM of the absorption profile as 
the full width at half the peak optical depth, using the Gaussian parameters provided. While the velocity widths of absorption lines 
arising from mergers tend to be slightly larger on average, the difference between the two cumulative distribution functions is not 
very significant (with $D_{\rm KS}$ = 0.24 and $P_{\rm KS}$ = 0.20). The slightly larger widths of the absorption lines in merging 
sources can be attributed to complex gas flow processes due to the ongoing merger. 

\begin{figure*}
\includegraphics[width=0.36\textwidth, angle=90]{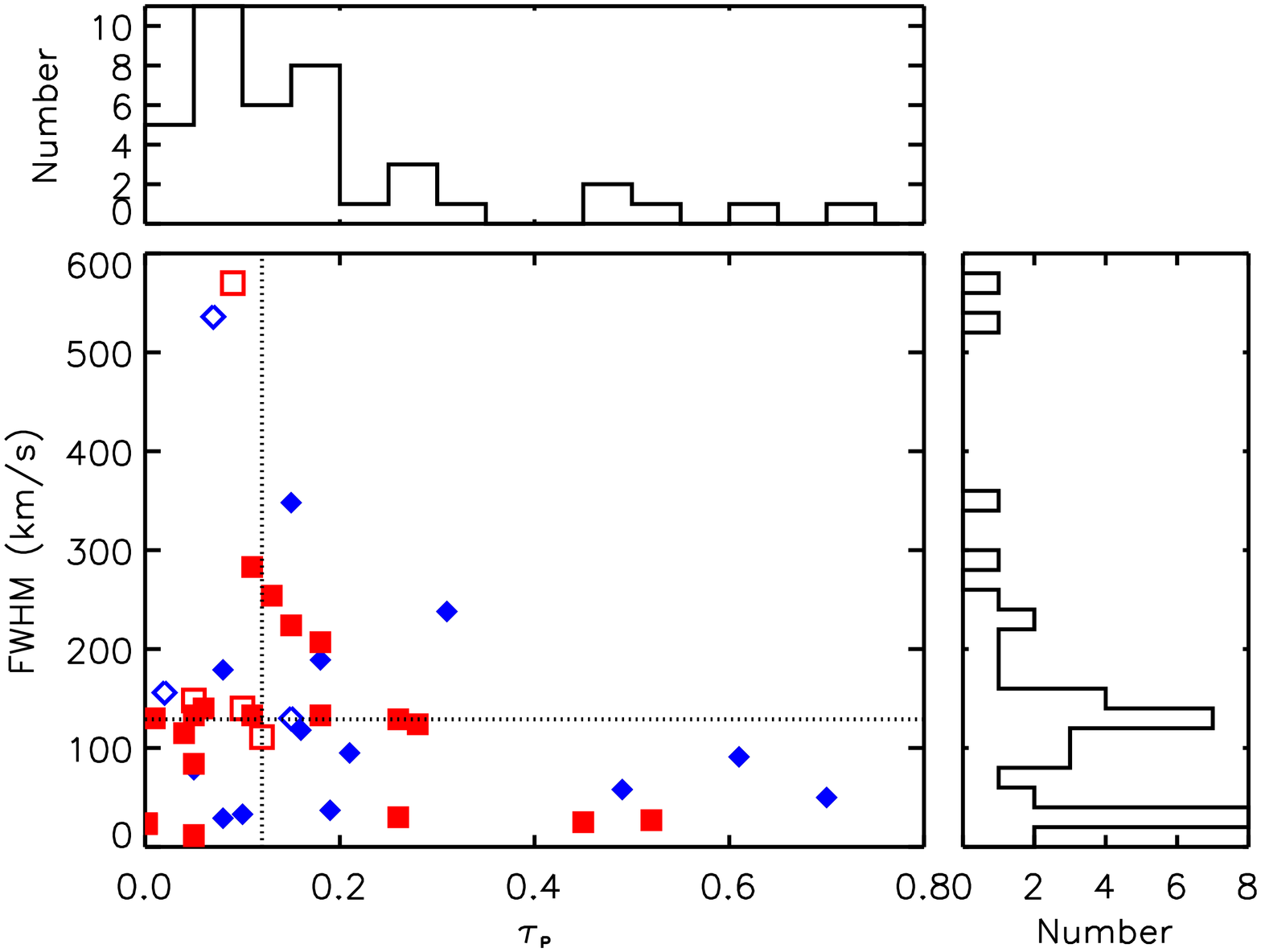}
\includegraphics[width=0.36\textwidth, angle=90]{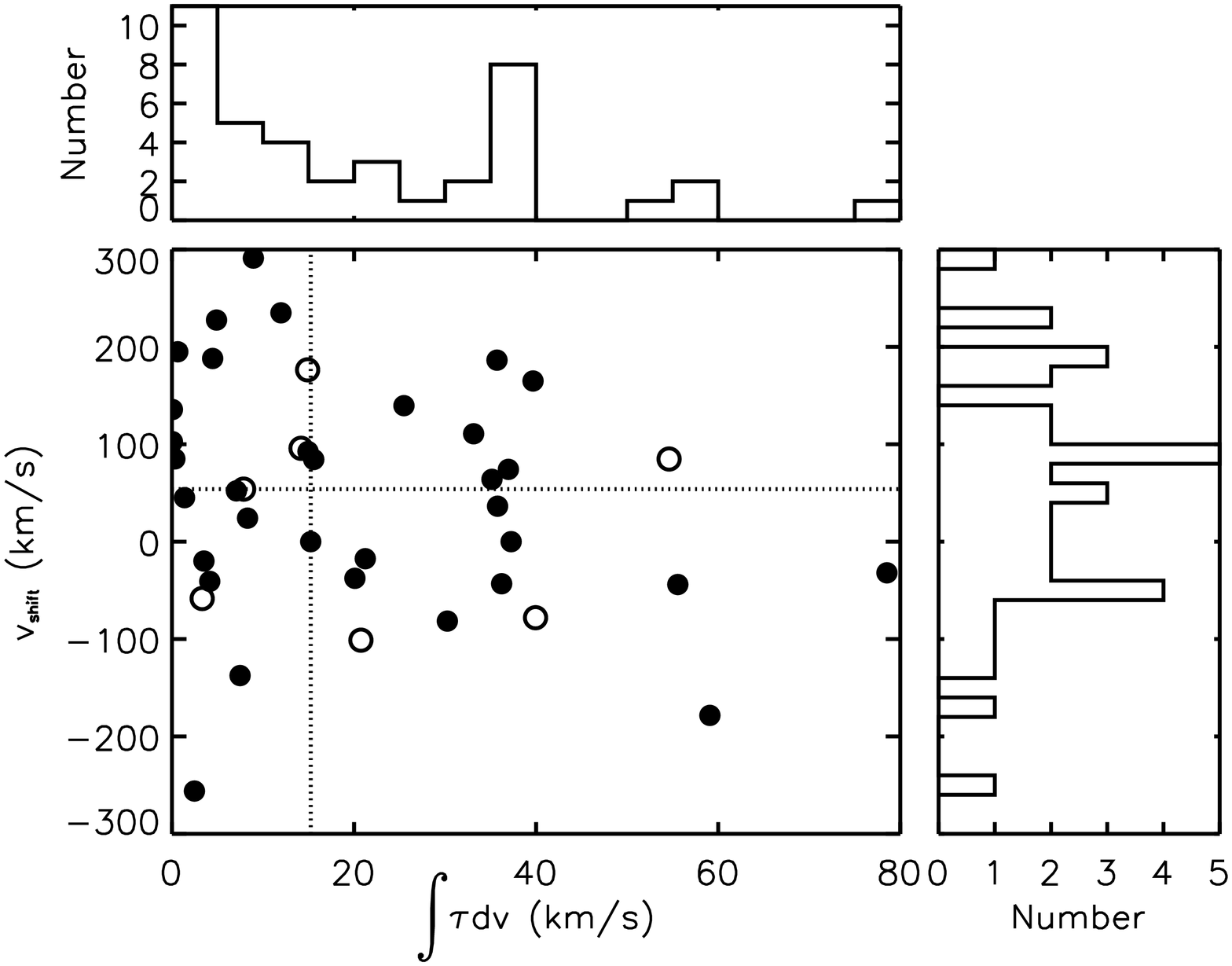}
\caption{{\it Left:} The distribution of FWHM and peak optical depth (\taup) of the individual absorption components. 
The diamonds denote blueshifted components (\vshift\ $\le$0\,\kms), and the squares denote redshifted components 
(\vshift\ $>$0\,\kms). The open symbols denote the measurements from \citet{gereb2015} and \citet{maccagni2017}, for which 
individual absorption component parameters have not been provided. The horizontal and vertical lines demarcate the median FWHM 
and \taup, respectively. The histograms of the distribution of FWHM and \taup\ are shown to the right and top of the plot, respectively.
{\it Right:} The distribution of velocity shift between the peak absorption and the galaxy systemic redshift (\vshift)
and the total optical depth (\taudv) of the individual absorption components. The open symbols are the same as in the left
panel. The horizontal and vertical lines demarcate the median \vshift\ and \taudv, respectively. The histograms of the distribution 
of \vshift\ and \taudv\ are shown to the right and top of the plot, respectively.
}
\label{fig:vshift}
\end{figure*}

The left panel of Fig.~\ref{fig:vshift} shows the distribution of FWHM and peak optical depth (\taup) of the individual absorption
components detected from mergers. Since individual component parameters are not provided for the absorbers listed in \citet{gereb2015}
and \citet{maccagni2017}, we plot these measurements as open symbols. It can be seen that there exist absorption components that are 
either narrow and strong, or broad and shallow. In addition to the velocity widths, the velocity shift between the absorption lines 
and the emission lines can be used to infer about the kinematics and origin of the absorbing gas. The right panel of Fig.~\ref{fig:vshift}
shows the distribution of this velocity shift (\vshift) for the different absorption components as a function of their total optical 
depth (\taudv). We have considered the systemic velocity of the stronger radio source as reference in case of mergers consisting of
multiple sources. The stronger radio source is the one showing absorption and is associated with the AGN activity. Hence, the choice 
of the stronger radio source as the velocity reference is motivated by our main interest of checking for connection of the absorbing 
gas with the AGN activity. The typical uncertainty in the \vshift\ measurements is $\sim$20-30\,\kms. The median \vshift\ is $\sim$54\,\kms, 
with $\sim$60\% of the components being redshifted (\vshift\ $\ge0$\,\kms). For comparison, the median \vshift\ among low-$z$ non-mergers 
in the sample of \citet{maccagni2017} is $\sim$ $-$20\,\kms, with $\sim$40\% of the absorption being redshifted. Further, if we consider 
blueshifted and redshifted absorption components to have \vshift\ $\le$ $-$100\,\kms\ and $\ge100$\,\kms, respectively, then $\sim$10\% 
of the absorption components in mergers are blueshifted and $\sim$30\% of them are redshifted. Compared to this, $\sim$28\% and $\sim$9\% 
of the absorption are blueshifted and redshifted, respectively, among the non-mergers \citep[in the sample presented in][]{maccagni2017}. 
The fraction of redshifted absorption among merging sources hence appears to be higher by about three times than what is typically found 
in low-$z$ radio sources. Note that in general, studies of intrinsic absorption in radio sources observe excess of blueshifted absorption 
\citep[e.g.][]{vermeulen2003,gupta2006}.

Further from Fig.~\ref{fig:vshift} (right panel), we see that the absorption components with smaller \taudv\ ($\le$15\,\kms, corresponding to 
\nhi\ $\lesssim3\times10^{19}$\,\cms, for \ts\ = 100~K and \fc\ = 1) show a range in \vshift\ ($-$300 to 300\,\kms), while those with larger 
\taudv\ are usually within $\sim$200\,\kms\ of the systemic velocity. Similarly, we find that the narrower (FWHM $\lesssim$100\,\kms) absorption 
components span the whole range of \vshift, while the broader ones are centred within $\sim$100\,\kms. This implies that the stronger and broader 
absorption components are likely to be tracing gas rotating in the circumnuclear discs, while the weaker components could be tracing both co-rotating 
and inflowing/outflowing gas. 

Note that the kinematics of the absorbers from \citet{gereb2015} and \citet{maccagni2017} are measured by fitting the Busy function, while those of ours 
and the other absorbers from the literature are measured by fitting Gaussian functions. We have checked that excluding the former measurements does not 
affect the statistical results regarding the kinematics of the absorbers in mergers as obtained above.
\subsection{Radio morphology of the mergers}
\label{sec_discussion4}
Among our sample, six of the radio sources are compact in the 1.4 GHz arcsecond-scale continuum maps, while four show resolved emission (see Fig.~\ref{fig:overlay}). 
Out of the 19 mergers in the literature sample, four sources show resolved continuum emission at arcsecond-scales, and the rest are compact (i.e., 
deconvolved sizes $\le2''$ in FIRST images). The merger, NGC~3690/IC~694, consists of three main radio components in the arcsecond-scale 1.4 GHz 
image, and absorption has been detected towards all of them \citep{baan1990}. Similarly, absorption has been detected across the complex radio 
structure extending over $\sim$8~kpc in the merger NGC~6240 \citep{baan2007}. Note that two out of three non-detections in our sample are towards 
radio sources that have resolved emission at arcsecond-scales, and the radio emission is resolved at sub-arcsecond-scales for the third non-detection
(see Section~\ref{sec_j1356+1822}). In case of the two non-detections from the literature, the radio sources are compact at arcsecond-scales, and 
one of them is also compact in the available 5 GHz sub-arcsecond-scale image \citep{helmboldt2007}. The non-detections in these two cases do not 
seem to be directly related to the radio morphology. Note that the radio power at 1.4 GHz of the sources with non-detections (log~$P_{1.4GHz} 
\sim 23.4-24.4$~W~Hz$^{-1}$) is not different compared to that of the sources with detections (log~$P_{1.4GHz} \sim 22.5-26.3$~W~Hz$^{-1}$). 

Sub-arcsecond-scale spectroscopy is not available for the sources in our sample, except for J1320$+$3408 (see Section~\ref{sec_j1320+3408}). 
The radio continuum emission at 1.4 GHz in this case has an elongated structure over $\sim$1~kpc, and absorption is detected against 
this entire region \citep{clemens2004}. Sub-arcsecond-scale absorption spectroscopy have been carried out in the case of few mergers from the 
literature as well. \citet{srianand2015} have resolved the radio source SDSS~J094221.98$+$062335.2 into a compact symmetric object of extent 89~pc 
and detected absorption towards both the lobes, with the absorption being consistent with arising from a circumnuclear disk. \citet{carilli2000} 
have studied the 1.4 GHz radio continuum and absorption in Mrk~273 at parsec-scales, and found evidence for a rotating gas disk of diameter 
350~pc. \citet{morganti2004} have found that the radio source in 4C$+$12.50 has a structure extending over $\sim$325~pc and the absorption 
is offset from the nucleus or core by $\sim$50~pc, which could be tracing the interstellar medium around the radio source. \citet{mundell2001} 
have detected spatially resolved absorption against the extended radio continuum emission in the central $\sim$900~pc of Arp~220, and found the 
extended absorption to be consistent with two counter-rotating disks of neutral hydrogen connected by a bridge. The above studies highlight the 
importance of sub-arcsecond-scale follow-up spectroscopy of the absorption detected in mergers, and the need to increase such observations, 
in order to understand the true origin of the absorbing gas in these complex systems.
\subsection{Optical properties and AGN activity of the mergers}
\label{sec_discussion5}
By definition of our sample (Section~\ref{sec_sample}), the galaxies typically show disturbed morphologies and tidal features in the available optical images. 
In the cases of J1100$-$1015, J1214$+$2931 and J2054$+$0041, no tidal features are visible in the optical images (Fig.~\ref{fig:overlay}), but they consist of
two very nearby ($\lesssim$10~kpc) galaxies/nuclei. Similarly, all the galaxies in the literature sample except three exhibit clear tidal features or disturbed
morphology. The remaining three consist of close pairs of galaxies. The galaxies both in our sample and in the literature appear to be star-forming based on the
available optical spectra, except for J1108$-$1015 which does not show any nebular emission (see Section~\ref{sec_j1108-1015}). Based on SDSS magnitude-dependent
colour cut \citep[as defined in][]{blanton2007,weinmann2006}, about 60\% of the galaxies undergoing merger are blue and rest are red. Thirteen of the mergers consist
of two interacting galaxies that have SDSS photometry available. If we define wet, dry and mixed mergers to comprise of blue-blue, red-red and blue-red galaxies, 
respectively, then three, four and six among these thirteen would be wet, dry and mixed mergers, respectively. Further, we can define mergers, where individual
galaxies cannot be distinguished, as wet or dry based on their colours blue or red, respectively. Then overall, $\sim$50\%, $\sim$27\% and $\sim$23\% of the 
mergers would be wet, dry and mixed, respectively. We do not find any difference in the optical colours and morphology of the galaxies which host the strong radio
sources and those that do not. The presence and strength of the absorption, moreover, does not seem to be related to the optical properties like morphology and 
colour, and the projected separation between the galaxies.

All the mergers in our sample (except for J1108$-$1015; see above) can be classified as AGNs or AGN-starburst composites based on optical nebular emission line 
ratio diagnostics. The AGN nature is further indicated by the presence of compact strong radio emission in these systems. Similarly, the mergers from the literature 
are classified optically as AGNs/Composites as well. In most of the cases, the presence of AGNs in the mergers is further confirmed with multi-wavelength 
(radio, infrared, and X-ray) data. Further, most of the mergers are also classified as LIRGs or ULIRGs (infrared luminosity $\sim10^{11-12}$~L$_\odot$ for 
LIRGs; $>10^{12}$~L$_\odot$ for ULIRGs). This is expected since the strong infrared emission of LIRGs and ULIRGs usually appears to be instigated by interactions 
\citep{sanders1996}. The mergers are dust-rich, with $\sim$91\% of the mergers which show absorption having WISE colour, $W2-W3>2$, consistent with the results 
of \citet{chandola2017}. Although, the presence of large amount of dust in the central regions of the mergers can obscure the central powering mechanism, i.e
the contribution of AGN and starburst is difficult to determine based on optical data. For example, the relative contribution of AGN and starburst to the extreme 
infrared luminosity of Arp~220 is still a matter of debate \citep[e.g.][]{martin2016}. However, it is found that the AGN activity in mergers can start before the 
final coalescence and be present along with the star formation \citep{ellison2011}. \citet{muller2018}, for example, have observed both AGN-driven and starburst-driven
outflows in the dual black hole merger, NGC~6240.

The large amount of neutral gas detected in majority of the mergers (Sections~\ref{sec_discussion1} and \ref{sec_discussion2}), and the excess of 
redshifted absorption, indicated by the velocity shift of the absorbing gas with respect to the emission lines of the radio sources (Section~\ref{sec_discussion3}), 
could maybe point towards a connection between the neutral gas near the centre of the mergers and the AGN activity of the radio sources. For instance,
in the mergers J1036$+$0221 and J1100$+$1002, the strongest absorption component occurs close to the systemic velocity of the radio source, implying 
that the absorbing gas could be arising from the circumnuclear disc or torus associated with the AGN. Additionally, we find redshifted (e.g. J1036$+$0221) 
and blueshifted (e.g. J1100$+$1002) absorption components among the mergers, which could represent infalling and outflowing gas clumps, respectively. 
We plan to conduct more in-depth studies of these systems in the future, in order to connect the absorbing gas with the AGN activity of the mergers.
%
%
\section{Summary}  
\label{sec_summary} 
We have presented in this work a search for neutral gas towards radio sources at $z\le0.2$ that are associated with galaxy mergers,
in an attempt to understand the connection between the presence of neutral gas in their circumnuclear regions and the radio activity.
Using GMRT and VLA observations, we have detected \hi\ \21\ absorption in seven out of ten merging systems. The absorption are broad
($\sim100-400$\,\kms) and strong [\nhi\ $\sim10^{21-22}$\,\cms\ for \ts\ = 100 K and \fc\ = 1]. We have detected and studied the 
\hi\ \21\ emission from one of the systems. We have also searched for OH 18-cm absorption towards one of these systems and reported
its non-detection. In addition, we have presented SALT long-slit optical spectra of four systems in our sample. In two of these systems, 
where we were able to carry out spatially resolved optical spectroscopy, we have studied the nebular emission line properties over 
$\sim25-35$~kpc across the systems. The strongest absorption component in these systems coincides with the peak of the emission from 
the radio sources. 

We have compiled a sample of low-$z$ ($z\le0.2$) mergers with radio sources that have been searched for absorption in the literature.
Combining this with our sample we obtain the following results:
\begin{itemize}
 \item The detection rate of absorption in low-$z$ radio sources that are part of merging systems is $83 \pm 17$\%. 
       This is $\sim3-4$ times higher than the detection rates of intrinsic absorption in samples of low-$z$ radio galaxies.
 \item Mergers account for most of the strong absorption detected in low-$z$ radio sources, and the fraction of mergers
       increases with increasing \nhi\ cut-off. Approximately 40\% of the absorbers with \nhi\ $>10^{21}$\,\cms\ arise from mergers, 
       and 100\% of the absorbers with \nhi\ $>10^{22}$\,\cms\ arise from mergers. The distribution of \nhi\ of the absorption 
       arising from mergers is significantly different from that of non-mergers, with mergers giving rise to six times stronger absorption
       on average. 
 \item The absorption from mergers tend to be broad and multi-component. The line widths could be affected by the complex gas kinematics
       in the mergers as well as the complex radio continuum morphology. However, we do not find any significant difference 
       in the FWHM distribution of the absorbers between mergers and non-mergers. 
 \item About 60\% and 30\% of the absorption components have velocity shift from the systemic velocity of $\ge$0\,\kms\ and $\ge+$100\,\kms, 
       respectively. This is 1.5 and 3 times higher, respectively, than what is found for intrinsic absorption in non-interacting 
       low-$z$ radio sources. The stronger and broader absorption components among mergers are usually within $\sim\pm$100\,\kms\ of 
       the systemic velocities. 
\end{itemize}

We have found statistical evidence for the presence of large amount of neutral gas in the central regions of merging galaxies showing 
strong radio emission. While the neutral gas could be fueling the AGN activity of these mergers, direct connection of the absorbing gas 
with the triggering of the AGN is more challenging to establish. For example, it is not straightforward to interpret the kinematics of 
the absorbing gas. The systemic velocities used in the literature and in our sample are estimated using optical nuclear spectra (fibre-spectra 
or long-slit spectra) obtained with different spatial resolutions and different wavelength coverage. The absorption data are also not of 
uniform spectral resolution, coverage and sensitivity. So all of the available spectra may not be sensitive to detect narrow weak infalling
absorption as seen in the case of SDSS~J094221.98$+$062335.2 \citep{srianand2015}, or shallow highly blueshifted absorption as seen in the
case of Mrk~231 \citep{morganti2016}. Furthermore, the current absorption spectra provide information about the neutral gas present in the
merging systems along our line of sight to the radio continuum emission. Information about the gas geometry and kinematics from high spatial 
resolution ionized and molecular emission line mapping is required to associate the different absorbing components with the emitting regions. 
For example, multi-wavelength IFU observations of the merger Mrk~463 presented in \citet{treister2018} suggests that while there exists 
significant reservoir of molecular gas around the double nuclei, only a small fraction of it is feeding the SMBH. More such studies are 
required to understand the relationship between neutral gas in mergers and AGN activity. The systems presented here are ideal for follow-up
observations to investigate in more detail such a connection. \\
  
%
%

\noindent \textbf{ACKNOWLEDGEMENTS} \newline
\noindent 
We thank the anonymous reviewer for useful comments that helped improve the paper.
RD acknowledges support from the Alexander von Humboldt Foundation.
We thank the staff at GMRT, VLA, and SALT for their help during the observations. 
GMRT is run by the National Centre for Radio Astrophysics of the Tata Institute of Fundamental Research. 
The VLA is run by the National Radio Astronomy Observatory (NRAO). 
The NRAO is a facility of the National Science Foundation operated under cooperative agreement by Associated Universities, Inc. 
Some of the observations reported in this paper were obtained with the Southern African Large Telescope (SALT).

This research has made use of the NASA/IPAC Extragalactic Database (NED) which is operated by the Jet Propulsion Laboratory, 
California Institute of Technology, under contract with the National Aeronautics and Space Administration.
Based on photographic data obtained using The UK Schmidt Telescope. The UK Schmidt Telescope was operated by the Royal Observatory Edinburgh, 
with funding from the UK Science and Engineering Research Council, until 1988 June, and thereafter by the Anglo-Australian Observatory. 
Original plate material is copyright (c) the Royal Observatory Edinburgh and the Anglo-Australian Observatory. 
The plates were processed into the present compressed digital form with their permission. 
The Digitized Sky Survey was produced at the Space Telescope Science Institute under US Government grant NAG W-2166.
Based on observations made with the NASA/ESA Hubble Space Telescope, obtained from the data archive at the Space Telescope Science Institute. 
STScI is operated by the Association of Universities for Research in Astronomy, Inc. under NASA contract NAS 5-26555.

Funding for SDSS-III has been provided by the Alfred P. Sloan Foundation, the Participating Institutions, the National Science Foundation, 
and the U.S. Department of Energy Office of Science. The SDSS-III web site is http://www.sdss3.org/.
SDSS-III is managed by the Astrophysical Research Consortium for the Participating Institutions of the SDSS-III Collaboration 
including the University of Arizona, the Brazilian Participation Group, Brookhaven National Laboratory, Carnegie Mellon University, 
University of Florida, the French Participation Group, the German Participation Group, Harvard University, the Instituto de Astrofisica 
de Canarias, the Michigan State/Notre Dame/JINA Participation Group, Johns Hopkins University, Lawrence Berkeley National Laboratory, 
Max Planck Institute for Astrophysics, Max Planck Institute for Extraterrestrial Physics, New Mexico State University, New York University, 
Ohio State University, Pennsylvania State University, University of Portsmouth, Princeton University, the Spanish Participation Group, 
University of Tokyo, University of Utah, Vanderbilt University, University of Virginia, University of Washington, and Yale University. 

%
%
\def\aj{AJ}%
\def\actaa{Acta Astron.}%
\def\araa{ARA\&A}%
\def\apj{ApJ}%
\def\apjl{ApJ}%
\def\apjs{ApJS}%
\def\ao{Appl.~Opt.}%
\def\apss{Ap\&SS}%
\def\aap{A\&A}%
\def\aapr{A\&A~Rev.}%
\def\aaps{A\&AS}%
\def\azh{A$Z$h}%
\def\baas{BAAS}%
\def\bac{Bull. astr. Inst. Czechosl.}%
\def\caa{Chinese Astron. Astrophys.}%
\def\cjaa{Chinese J. Astron. Astrophys.}%
\def\icarus{Icarus}%
\def\jcap{J. Cosmology Astropart. Phys.}%
\def\jrasc{JRASC}%
\def\mnras{MNRAS}%
\def\memras{MmRAS}%
\def\na{New A}%
\def\nar{New A Rev.}%
\def\pasa{PASA}%
\def\pra{Phys.~Rev.~A}%
\def\prb{Phys.~Rev.~B}%
\def\prc{Phys.~Rev.~C}%
\def\prd{Phys.~Rev.~D}%
\def\pre{Phys.~Rev.~E}%
\def\prl{Phys.~Rev.~Lett.}%
\def\pasp{PASP}%
\def\pasj{PASJ}%
\def\qjras{QJRAS}%
\def\rmxaa{Rev. Mexicana Astron. Astrofis.}%
\def\skytel{S\&T}%
\def\solphys{Sol.~Phys.}%
\def\sovast{Soviet~Ast.}%
\def\ssr{Space~Sci.~Rev.}%
\def\zap{$Z$Ap}%
\def\nat{Nature}%
\def\iaucirc{IAU~Circ.}%
\def\aplett{Astrophys.~Lett.}%
\def\apspr{Astrophys.~Space~Phys.~Res.}%
\def\bain{Bull.~Astron.~Inst.~Netherlands}%
\def\fcp{Fund.~Cosmic~Phys.}%
\def\gca{Geochim.~Cosmochim.~Acta}%
\def\grl{Geophys.~Res.~Lett.}%
\def\jcp{J.~Chem.~Phys.}%
\def\jgr{J.~Geophys.~Res.}%
\def\jqsrt{J.~Quant.~Spec.~Radiat.~Transf.}%
\def\memsai{Mem.~Soc.~Astron.~Italiana}%
\def\nphysa{Nucl.~Phys.~A}%
\def\physrep{Phys.~Rep.}%
\def\physscr{Phys.~Scr}%
\def\planss{Planet.~Space~Sci.}%
\def\procspie{Proc.~SPIE}%
\let\astap=\aap
\let\apjlett=\apjl
\let\apjsupp=\apjs
\let\applopt=\ao
\bibliographystyle{mnras}
\bibliography{mybib}

\begin{thebibliography}{}
\makeatletter
\relax
\def\mn@urlcharsother{\let\do\@makeother \do\$\do\&\do\#\do\^\do\_\do\%\do\~}
\def\mn@doi{\begingroup\mn@urlcharsother \@ifnextchar [ {\mn@doi@}
  {\mn@doi@[]}}
\def\mn@doi@[#1]#2{\def\@tempa{#1}\ifx\@tempa\@empty \href
  {http://dx.doi.org/#2} {doi:#2}\else \href {http://dx.doi.org/#2} {#1}\fi
  \endgroup}
\def\mn@eprint#1#2{\mn@eprint@#1:#2::\@nil}
\def\mn@eprint@arXiv#1{\href {http://arxiv.org/abs/#1} {{\tt arXiv:#1}}}
\def\mn@eprint@dblp#1{\href {http://dblp.uni-trier.de/rec/bibtex/#1.xml}
  {dblp:#1}}
\def\mn@eprint@#1:#2:#3:#4\@nil{\def\@tempa {#1}\def\@tempb {#2}\def\@tempc
  {#3}\ifx \@tempc \@empty \let \@tempc \@tempb \let \@tempb \@tempa \fi \ifx
  \@tempb \@empty \def\@tempb {arXiv}\fi \@ifundefined
  {mn@eprint@\@tempb}{\@tempb:\@tempc}{\expandafter \expandafter \csname
  mn@eprint@\@tempb\endcsname \expandafter{\@tempc}}}

\bibitem[\protect\citeauthoryear{{Aditya}, {Kanekar}  \& {Kurapati}}{{Aditya}
  et~al.}{2016}]{aditya2016}
{Aditya} J.~N.~H.~S.,  {Kanekar} N.,   {Kurapati} S.,  2016, \mn@doi [\mnras]
  {10.1093/mnras/stv2563}, \href
  {http://adsabs.harvard.edu/abs/2016MNRAS.455.4000A} {455, 4000}

\bibitem[\protect\citeauthoryear{{Alexander} \& {Hickox}}{{Alexander} \&
  {Hickox}}{2012}]{alexander2012}
{Alexander} D.~M.,  {Hickox} R.~C.,  2012, \mn@doi [\nar]
  {10.1016/j.newar.2011.11.003}, \href
  {http://adsabs.harvard.edu/abs/2012NewAR..56...93A} {56, 93}

\bibitem[\protect\citeauthoryear{{Allison} et~al.,}{{Allison}
  et~al.}{2012}]{allison2012}
{Allison} J.~R.,  et~al., 2012, \mn@doi [\mnras]
  {10.1111/j.1365-2966.2012.21062.x}, \href
  {http://adsabs.harvard.edu/abs/2012MNRAS.423.2601A} {423, 2601}

\bibitem[\protect\citeauthoryear{{Alonso-Herrero},
  {Garc{\'{\i}}a-Mar{\'{\i}}n}, {Monreal-Ibero}, {Colina}, {Arribas},
  {Alfonso-Garz{\'o}n}  \& {Labiano}}{{Alonso-Herrero}
  et~al.}{2009}]{alonso2009}
{Alonso-Herrero} A.,  {Garc{\'{\i}}a-Mar{\'{\i}}n} M.,  {Monreal-Ibero} A.,
  {Colina} L.,  {Arribas} S.,  {Alfonso-Garz{\'o}n} J.,   {Labiano} A.,  2009,
  \mn@doi [\aap] {10.1051/0004-6361/200911813}, \href
  {http://adsabs.harvard.edu/abs/2009A%26A...506.1541A} {506, 1541}

\bibitem[\protect\citeauthoryear{{Argence} \& {Lamareille}}{{Argence} \&
  {Lamareille}}{2009}]{argence2009}
{Argence} B.,  {Lamareille} F.,  2009, \mn@doi [\aap]
  {10.1051/0004-6361:20066998}, \href
  {http://adsabs.harvard.edu/abs/2009A%26A...495..759A} {495, 759}

\bibitem[\protect\citeauthoryear{{Baan} \& {Haschick}}{{Baan} \&
  {Haschick}}{1990}]{baan1990}
{Baan} W.~A.,  {Haschick} A.,  1990, \mn@doi [\apj] {10.1086/169385}, \href
  {http://adsabs.harvard.edu/abs/1990ApJ...364...65B} {364, 65}

\bibitem[\protect\citeauthoryear{{Baan}, {Hagiwara}  \& {Hofner}}{{Baan}
  et~al.}{2007}]{baan2007}
{Baan} W.~A.,  {Hagiwara} Y.,   {Hofner} P.,  2007, \mn@doi [\apj]
  {10.1086/513593}, \href {http://adsabs.harvard.edu/abs/2007ApJ...661..173B}
  {661, 173}

\bibitem[\protect\citeauthoryear{{Baldwin}, {Phillips}  \&
  {Terlevich}}{{Baldwin} et~al.}{1981}]{baldwin1981}
{Baldwin} J.~A.,  {Phillips} M.~M.,   {Terlevich} R.,  1981, \mn@doi [\pasp]
  {10.1086/130766}, \href {http://adsabs.harvard.edu/abs/1981PASP...93....5B}
  {93, 5}

\bibitem[\protect\citeauthoryear{{Barcos-Mu{\~n}oz} et~al.,}{{Barcos-Mu{\~n}oz}
  et~al.}{2017}]{barcos2017}
{Barcos-Mu{\~n}oz} L.,  et~al., 2017, \mn@doi [\apj]
  {10.3847/1538-4357/aa789a}, \href
  {http://adsabs.harvard.edu/abs/2017ApJ...843..117B} {843, 117}

\bibitem[\protect\citeauthoryear{{Barnes} \& {Hernquist}}{{Barnes} \&
  {Hernquist}}{1996}]{barnes1996}
{Barnes} J.~E.,  {Hernquist} L.,  1996, \mn@doi [\apj] {10.1086/177957}, \href
  {http://adsabs.harvard.edu/abs/1996ApJ...471..115B} {471, 115}

\bibitem[\protect\citeauthoryear{{Bianchi}, {Chiaberge}, {Piconcelli},
  {Guainazzi}  \& {Matt}}{{Bianchi} et~al.}{2008}]{bianchi2008}
{Bianchi} S.,  {Chiaberge} M.,  {Piconcelli} E.,  {Guainazzi} M.,   {Matt} G.,
  2008, \mn@doi [\mnras] {10.1111/j.1365-2966.2008.13078.x}, \href
  {http://adsabs.harvard.edu/abs/2008MNRAS.386..105B} {386, 105}

\bibitem[\protect\citeauthoryear{{Bieging} \& {Biermann}}{{Bieging} \&
  {Biermann}}{1983}]{bieging1983}
{Bieging} J.~H.,  {Biermann} P.,  1983, \mn@doi [\aj] {10.1086/113301}, \href
  {http://adsabs.harvard.edu/abs/1983AJ.....88..161B} {88, 161}

\bibitem[\protect\citeauthoryear{{Blanton} \& {Berlind}}{{Blanton} \&
  {Berlind}}{2007}]{blanton2007}
{Blanton} M.~R.,  {Berlind} A.~A.,  2007, \mn@doi [\apj] {10.1086/512478},
  \href {http://adsabs.harvard.edu/abs/2007ApJ...664..791B} {664, 791}

\bibitem[\protect\citeauthoryear{{Bothun} \& {Schommer}}{{Bothun} \&
  {Schommer}}{1983}]{bothun1983}
{Bothun} G.~D.,  {Schommer} R.~A.,  1983, \mn@doi [\apjl] {10.1086/183994},
  \href {http://adsabs.harvard.edu/abs/1983ApJ...267L..15B} {267, L15}

\bibitem[\protect\citeauthoryear{{Cao} et~al.,}{{Cao} et~al.}{2016}]{cao2016}
{Cao} C.,  et~al., 2016, \mn@doi [\apjs] {10.3847/0067-0049/222/2/16}, \href
  {http://adsabs.harvard.edu/abs/2016ApJS..222...16C} {222, 16}

\bibitem[\protect\citeauthoryear{{Carilli} \& {Taylor}}{{Carilli} \&
  {Taylor}}{2000}]{carilli2000}
{Carilli} C.~L.,  {Taylor} G.~B.,  2000, \mn@doi [\apjl] {10.1086/312584},
  \href {http://adsabs.harvard.edu/abs/2000ApJ...532L..95C} {532, L95}

\bibitem[\protect\citeauthoryear{{Carilli}, {Wrobel}  \& {Ulvestad}}{{Carilli}
  et~al.}{1998a}]{carilli1998b}
{Carilli} C.~L.,  {Wrobel} J.~M.,   {Ulvestad} J.~S.,  1998a, \mn@doi [\aj]
  {10.1086/300253}, \href {http://adsabs.harvard.edu/abs/1998AJ....115..928C}
  {115, 928}

\bibitem[\protect\citeauthoryear{{Carilli}, {Menten}, {Reid}, {Rupen}  \&
  {Yun}}{{Carilli} et~al.}{1998b}]{carilli1998a}
{Carilli} C.~L.,  {Menten} K.~M.,  {Reid} M.~J.,  {Rupen} M.~P.,   {Yun} M.~S.,
   1998b, \mn@doi [\apj] {10.1086/305191}, \href
  {http://adsabs.harvard.edu/abs/1998ApJ...494..175C} {494, 175}

\bibitem[\protect\citeauthoryear{{Chambers}, {Miley}  \& {van
  Breugel}}{{Chambers} et~al.}{1987}]{chambers1987}
{Chambers} K.~C.,  {Miley} G.~K.,   {van Breugel} W.,  1987, \mn@doi [\nat]
  {10.1038/329604a0}, \href {http://adsabs.harvard.edu/abs/1987Natur.329..604C}
  {329, 604}

\bibitem[\protect\citeauthoryear{{Chandola} \& {Saikia}}{{Chandola} \&
  {Saikia}}{2017}]{chandola2017}
{Chandola} Y.,  {Saikia} D.~J.,  2017, \mn@doi [\mnras]
  {10.1093/mnras/stw2705}, \href
  {http://adsabs.harvard.edu/abs/2017MNRAS.465..997C} {465, 997}

\bibitem[\protect\citeauthoryear{{Chandola}, {Sirothia}  \&
  {Saikia}}{{Chandola} et~al.}{2011}]{chandola2011}
{Chandola} Y.,  {Sirothia} S.~K.,   {Saikia} D.~J.,  2011, \mn@doi [\mnras]
  {10.1111/j.1365-2966.2011.19607.x}, \href
  {http://adsabs.harvard.edu/abs/2011MNRAS.418.1787C} {418, 1787}

\bibitem[\protect\citeauthoryear{{Chandola}, {Sirothia}, {Saikia}  \&
  {Gupta}}{{Chandola} et~al.}{2012}]{chandola2012}
{Chandola} Y.,  {Sirothia} S.~K.,  {Saikia} D.~J.,   {Gupta} N.,  2012,
  Bulletin of the Astronomical Society of India, \href
  {http://adsabs.harvard.edu/abs/2012BASI...40..139C} {40, 139}

\bibitem[\protect\citeauthoryear{{Chandola}, {Gupta}  \& {Saikia}}{{Chandola}
  et~al.}{2013}]{chandola2013}
{Chandola} Y.,  {Gupta} N.,   {Saikia} D.~J.,  2013, \mn@doi [\mnras]
  {10.1093/mnras/sts499}, \href
  {http://adsabs.harvard.edu/abs/2013MNRAS.429.2380C} {429, 2380}

\bibitem[\protect\citeauthoryear{{Cisternas} et~al.,}{{Cisternas}
  et~al.}{2011}]{cisternas2011}
{Cisternas} M.,  et~al., 2011, \mn@doi [\apj] {10.1088/0004-637X/726/2/57},
  \href {http://adsabs.harvard.edu/abs/2011ApJ...726...57C} {726, 57}

\bibitem[\protect\citeauthoryear{{Clemens} \& {Alexander}}{{Clemens} \&
  {Alexander}}{2004}]{clemens2004}
{Clemens} M.~S.,  {Alexander} P.,  2004, \mn@doi [\mnras]
  {10.1111/j.1365-2966.2004.07647.x}, \href
  {http://adsabs.harvard.edu/abs/2004MNRAS.350...66C} {350, 66}

\bibitem[\protect\citeauthoryear{{Combes} et~al.,}{{Combes}
  et~al.}{2009}]{combes2009}
{Combes} F.,  et~al., 2009, \mn@doi [\aap] {10.1051/0004-6361/200912181}, \href
  {http://adsabs.harvard.edu/abs/2009A%26A...503...73C} {503, 73}

\bibitem[\protect\citeauthoryear{{Comerford}, {Pooley}, {Barrows}, {Greene},
  {Zakamska}, {Madejski}  \& {Cooper}}{{Comerford}
  et~al.}{2015}]{comerford2015}
{Comerford} J.~M.,  {Pooley} D.,  {Barrows} R.~S.,  {Greene} J.~E.,  {Zakamska}
  N.~L.,  {Madejski} G.~M.,   {Cooper} M.~C.,  2015, \mn@doi [\apj]
  {10.1088/0004-637X/806/2/219}, \href
  {http://adsabs.harvard.edu/abs/2015ApJ...806..219C} {806, 219}

\bibitem[\protect\citeauthoryear{{Condon}, {Cotton}, {Greisen}, {Yin},
  {Perley}, {Taylor}  \& {Broderick}}{{Condon} et~al.}{1998}]{condon1998}
{Condon} J.~J.,  {Cotton} W.~D.,  {Greisen} E.~W.,  {Yin} Q.~F.,  {Perley}
  R.~A.,  {Taylor} G.~B.,   {Broderick} J.~J.,  1998, \mn@doi [\aj]
  {10.1086/300337}, \href {http://adsabs.harvard.edu/abs/1998AJ....115.1693C}
  {115, 1693}

\bibitem[\protect\citeauthoryear{{Cortijo-Ferrero} et~al.,}{{Cortijo-Ferrero}
  et~al.}{2017}]{cortijo2017}
{Cortijo-Ferrero} C.,  et~al., 2017, \mn@doi [\mnras] {10.1093/mnras/stx383},
  \href {http://adsabs.harvard.edu/abs/2017MNRAS.467.3898C} {467, 3898}

\bibitem[\protect\citeauthoryear{{Courtois}, {Tully}, {Makarov}, {Mitronova},
  {Koribalski}, {Karachentsev}  \& {Fisher}}{{Courtois}
  et~al.}{2011}]{courtois2011}
{Courtois} H.~M.,  {Tully} R.~B.,  {Makarov} D.~I.,  {Mitronova} S.,
  {Koribalski} B.,  {Karachentsev} I.~D.,   {Fisher} J.~R.,  2011, \mn@doi
  [\mnras] {10.1111/j.1365-2966.2011.18515.x}, \href
  {http://adsabs.harvard.edu/abs/2011MNRAS.414.2005C} {414, 2005}

\bibitem[\protect\citeauthoryear{{Cox}, {Primack}, {Jonsson}  \&
  {Somerville}}{{Cox} et~al.}{2004}]{cox2004}
{Cox} T.~J.,  {Primack} J.,  {Jonsson} P.,   {Somerville} R.~S.,  2004, \mn@doi
  [\apjl] {10.1086/421905}, \href
  {http://adsabs.harvard.edu/abs/2004ApJ...607L..87C} {607, L87}

\bibitem[\protect\citeauthoryear{{Cox}, {Jonsson}, {Somerville}, {Primack}  \&
  {Dekel}}{{Cox} et~al.}{2008}]{cox2008}
{Cox} T.~J.,  {Jonsson} P.,  {Somerville} R.~S.,  {Primack} J.~R.,   {Dekel}
  A.,  2008, \mn@doi [\mnras] {10.1111/j.1365-2966.2007.12730.x}, \href
  {http://adsabs.harvard.edu/abs/2008MNRAS.384..386C} {384, 386}

\bibitem[\protect\citeauthoryear{{Crawford} et~al.,}{{Crawford}
  et~al.}{2010}]{crawford2010}
{Crawford} S.~M.,  et~al., 2010, in Observatory Operations: Strategies,
  Processes, and Systems III. p. 773725, \mn@doi{10.1117/12.857000}

\bibitem[\protect\citeauthoryear{{Croton} et~al.,}{{Croton}
  et~al.}{2006}]{croton2006}
{Croton} D.~J.,  et~al., 2006, \mn@doi [\mnras]
  {10.1111/j.1365-2966.2005.09675.x}, \href
  {http://adsabs.harvard.edu/abs/2006MNRAS.365...11C} {365, 11}

\bibitem[\protect\citeauthoryear{{Darling} \& {Giovanelli}}{{Darling} \&
  {Giovanelli}}{2000}]{darling2000}
{Darling} J.,  {Giovanelli} R.,  2000, \mn@doi [\aj] {10.1086/301403}, \href
  {http://adsabs.harvard.edu/abs/2000AJ....119.3003D} {119, 3003}

\bibitem[\protect\citeauthoryear{{Darling}, {Macdonald}, {Haynes}  \&
  {Giovanelli}}{{Darling} et~al.}{2011}]{darling2011}
{Darling} J.,  {Macdonald} E.~P.,  {Haynes} M.~P.,   {Giovanelli} R.,  2011,
  \mn@doi [\apj] {10.1088/0004-637X/742/1/60}, \href
  {http://adsabs.harvard.edu/abs/2011ApJ...742...60D} {742, 60}

\bibitem[\protect\citeauthoryear{{Dickey}}{{Dickey}}{1986}]{dickey1986}
{Dickey} J.~M.,  1986, \mn@doi [\apj] {10.1086/163793}, \href
  {http://adsabs.harvard.edu/abs/1986ApJ...300..190D} {300, 190}

\bibitem[\protect\citeauthoryear{{Dutta}, {Gupta}, {Srianand}  \&
  {O'Meara}}{{Dutta} et~al.}{2016}]{dutta2016}
{Dutta} R.,  {Gupta} N.,  {Srianand} R.,   {O'Meara} J.~M.,  2016, \mn@doi
  [\mnras] {10.1093/mnras/stv2980}, \href
  {http://adsabs.harvard.edu/abs/2016MNRAS.456.4209D} {456, 4209}

\bibitem[\protect\citeauthoryear{{Dutta}, {Srianand}, {Gupta}, {Momjian},
  {Noterdaeme}, {Petitjean}  \& {Rahmani}}{{Dutta} et~al.}{2017a}]{dutta2017a}
{Dutta} R.,  {Srianand} R.,  {Gupta} N.,  {Momjian} E.,  {Noterdaeme} P.,
  {Petitjean} P.,   {Rahmani} H.,  2017a, \mn@doi [\mnras]
  {10.1093/mnras/stw2689}, \href
  {http://adsabs.harvard.edu/abs/2017MNRAS.465..588D} {465, 588}

\bibitem[\protect\citeauthoryear{{Dutta}, {Srianand}, {Gupta}, {Joshi},
  {Petitjean}, {Noterdaeme}, {Ge}  \& {Krogager}}{{Dutta}
  et~al.}{2017b}]{dutta2017b}
{Dutta} R.,  {Srianand} R.,  {Gupta} N.,  {Joshi} R.,  {Petitjean} P.,
  {Noterdaeme} P.,  {Ge} J.,   {Krogager} J.-K.,  2017b, \mn@doi [\mnras]
  {10.1093/mnras/stw3040}, \href
  {http://adsabs.harvard.edu/abs/2017MNRAS.465.4249D} {465, 4249}

\bibitem[\protect\citeauthoryear{{Ellison}, {Patton}, {Simard}  \&
  {McConnachie}}{{Ellison} et~al.}{2008}]{ellison2008}
{Ellison} S.~L.,  {Patton} D.~R.,  {Simard} L.,   {McConnachie} A.~W.,  2008,
  \mn@doi [\aj] {10.1088/0004-6256/135/5/1877}, \href
  {http://adsabs.harvard.edu/abs/2008AJ....135.1877E} {135, 1877}

\bibitem[\protect\citeauthoryear{{Ellison}, {Patton}, {Mendel}  \&
  {Scudder}}{{Ellison} et~al.}{2011}]{ellison2011}
{Ellison} S.~L.,  {Patton} D.~R.,  {Mendel} J.~T.,   {Scudder} J.~M.,  2011,
  \mn@doi [\mnras] {10.1111/j.1365-2966.2011.19624.x}, \href
  {http://adsabs.harvard.edu/abs/2011MNRAS.418.2043E} {418, 2043}

\bibitem[\protect\citeauthoryear{{Ellison}, {Mendel}, {Patton}  \&
  {Scudder}}{{Ellison} et~al.}{2013}]{ellison2013}
{Ellison} S.~L.,  {Mendel} J.~T.,  {Patton} D.~R.,   {Scudder} J.~M.,  2013,
  \mn@doi [\mnras] {10.1093/mnras/stt1562}, \href
  {http://adsabs.harvard.edu/abs/2013MNRAS.435.3627E} {435, 3627}

\bibitem[\protect\citeauthoryear{{Ellison}, {Fertig}, {Rosenberg}, {Nair},
  {Simard}, {Torrey}  \& {Patton}}{{Ellison} et~al.}{2015}]{ellison2015}
{Ellison} S.~L.,  {Fertig} D.,  {Rosenberg} J.~L.,  {Nair} P.,  {Simard} L.,
  {Torrey} P.,   {Patton} D.~R.,  2015, \mn@doi [\mnras]
  {10.1093/mnras/stu2744}, \href
  {http://adsabs.harvard.edu/abs/2015MNRAS.448..221E} {448, 221}

\bibitem[\protect\citeauthoryear{{Fathi} et~al.,}{{Fathi}
  et~al.}{2013}]{fathi2013}
{Fathi} K.,  et~al., 2013, \mn@doi [\apjl] {10.1088/2041-8205/770/2/L27}, \href
  {http://adsabs.harvard.edu/abs/2013ApJ...770L..27F} {770, L27}

\bibitem[\protect\citeauthoryear{{Ferrarese} \& {Merritt}}{{Ferrarese} \&
  {Merritt}}{2000}]{ferrarese2000}
{Ferrarese} L.,  {Merritt} D.,  2000, \mn@doi [\apjl] {10.1086/312838}, \href
  {http://adsabs.harvard.edu/abs/2000ApJ...539L...9F} {539, L9}

\bibitem[\protect\citeauthoryear{{Fu}, {Myers}, {Djorgovski}  \& {Yan}}{{Fu}
  et~al.}{2011}]{fu2011}
{Fu} H.,  {Myers} A.~D.,  {Djorgovski} S.~G.,   {Yan} L.,  2011, \mn@doi [\apj]
  {10.1088/0004-637X/733/2/103}, \href
  {http://adsabs.harvard.edu/abs/2011ApJ...733..103F} {733, 103}

\bibitem[\protect\citeauthoryear{{Gaibler}, {Khochfar}, {Krause}  \&
  {Silk}}{{Gaibler} et~al.}{2012}]{gaibler2012}
{Gaibler} V.,  {Khochfar} S.,  {Krause} M.,   {Silk} J.,  2012, \mn@doi
  [\mnras] {10.1111/j.1365-2966.2012.21479.x}, \href
  {http://adsabs.harvard.edu/abs/2012MNRAS.425..438G} {425, 438}

\bibitem[\protect\citeauthoryear{{Gallimore}, {Baum}, {O'Dea}, {Pedlar}  \&
  {Brinks}}{{Gallimore} et~al.}{1999}]{gallimore1999}
{Gallimore} J.~F.,  {Baum} S.~A.,  {O'Dea} C.~P.,  {Pedlar} A.,   {Brinks} E.,
  1999, \mn@doi [\apj] {10.1086/307853}, \href
  {http://adsabs.harvard.edu/abs/1999ApJ...524..684G} {524, 684}

\bibitem[\protect\citeauthoryear{{Gebhardt} et~al.,}{{Gebhardt}
  et~al.}{2000}]{gebhardt2000}
{Gebhardt} K.,  et~al., 2000, \mn@doi [\apjl] {10.1086/312840}, \href
  {http://adsabs.harvard.edu/abs/2000ApJ...539L..13G} {539, L13}

\bibitem[\protect\citeauthoryear{{Ger{\'e}b}, {Maccagni}, {Morganti}  \&
  {Oosterloo}}{{Ger{\'e}b} et~al.}{2015}]{gereb2015}
{Ger{\'e}b} K.,  {Maccagni} F.~M.,  {Morganti} R.,   {Oosterloo} T.~A.,  2015,
  \mn@doi [\aap] {10.1051/0004-6361/201424655}, \href
  {http://adsabs.harvard.edu/abs/2015A%26A...575A..44G} {575, A44}

\bibitem[\protect\citeauthoryear{{Greene}, {Zakamska}, {Liu}, {Barth}  \&
  {Ho}}{{Greene} et~al.}{2009}]{greene2009}
{Greene} J.~E.,  {Zakamska} N.~L.,  {Liu} X.,  {Barth} A.~J.,   {Ho} L.~C.,
  2009, \mn@doi [\apj] {10.1088/0004-637X/702/1/441}, \href
  {http://adsabs.harvard.edu/abs/2009ApJ...702..441G} {702, 441}

\bibitem[\protect\citeauthoryear{{Greene}, {Zakamska}  \& {Smith}}{{Greene}
  et~al.}{2012}]{greene2012}
{Greene} J.~E.,  {Zakamska} N.~L.,   {Smith} P.~S.,  2012, \mn@doi [\apj]
  {10.1088/0004-637X/746/1/86}, \href
  {http://adsabs.harvard.edu/abs/2012ApJ...746...86G} {746, 86}

\bibitem[\protect\citeauthoryear{{Gupta}, {Salter}, {Saikia}, {Ghosh}  \&
  {Jeyakumar}}{{Gupta} et~al.}{2006}]{gupta2006}
{Gupta} N.,  {Salter} C.~J.,  {Saikia} D.~J.,  {Ghosh} T.,   {Jeyakumar} S.,
  2006, \mn@doi [\mnras] {10.1111/j.1365-2966.2006.11064.x}, \href
  {http://adsabs.harvard.edu/abs/2006MNRAS.373..972G} {373, 972}

\bibitem[\protect\citeauthoryear{{Gupta} et~al.,}{{Gupta}
  et~al.}{2016}]{gupta2016}
{Gupta} N.,  et~al., 2016, in Proceedings of MeerKAT Science: On the Pathway to
  the SKA. 25-27 May, 2016 Stellenbosch, South Africa (MeerKAT2016). p.~14
  (\mn@eprint {arXiv} {1708.07371})

\bibitem[\protect\citeauthoryear{{Hani}, {Sparre}, {Ellison}, {Torrey}  \&
  {Vogelsberger}}{{Hani} et~al.}{2018}]{hani2018}
{Hani} M.~H.,  {Sparre} M.,  {Ellison} S.~L.,  {Torrey} P.,   {Vogelsberger}
  M.,  2018, \mn@doi [\mnras] {10.1093/mnras/stx3252}, \href
  {http://adsabs.harvard.edu/abs/2018MNRAS.475.1160H} {475, 1160}

\bibitem[\protect\citeauthoryear{{Healey}, {Romani}, {Taylor}, {Sadler},
  {Ricci}, {Murphy}, {Ulvestad}  \& {Winn}}{{Healey} et~al.}{2007}]{healey2007}
{Healey} S.~E.,  {Romani} R.~W.,  {Taylor} G.~B.,  {Sadler} E.~M.,  {Ricci} R.,
   {Murphy} T.,  {Ulvestad} J.~S.,   {Winn} J.~N.,  2007, \mn@doi [\apjs]
  {10.1086/513742}, \href {http://adsabs.harvard.edu/abs/2007ApJS..171...61H}
  {171, 61}

\bibitem[\protect\citeauthoryear{{Heckman}, {Balick}, {van Breugel}  \&
  {Miley}}{{Heckman} et~al.}{1983}]{heckman1983}
{Heckman} T.~M.,  {Balick} B.,  {van Breugel} W.~J.~W.,   {Miley} G.~K.,  1983,
  \mn@doi [\aj] {10.1086/113347}, \href
  {http://adsabs.harvard.edu/abs/1983AJ.....88..583H} {88, 583}

\bibitem[\protect\citeauthoryear{{Helmboldt} et~al.,}{{Helmboldt}
  et~al.}{2007}]{helmboldt2007}
{Helmboldt} J.~F.,  et~al., 2007, \mn@doi [\apj] {10.1086/511005}, \href
  {http://adsabs.harvard.edu/abs/2007ApJ...658..203H} {658, 203}

\bibitem[\protect\citeauthoryear{{Henkel}, {Gusten}  \& {Baan}}{{Henkel}
  et~al.}{1986}]{henkel1986}
{Henkel} C.,  {Gusten} R.,   {Baan} W.~A.,  1986, in Bulletin of the American
  Astronomical Society. p.~689

\bibitem[\protect\citeauthoryear{{Hopkins}, {Hernquist}, {Cox}, {Di Matteo},
  {Robertson}  \& {Springel}}{{Hopkins} et~al.}{2006}]{hopkins2006}
{Hopkins} P.~F.,  {Hernquist} L.,  {Cox} T.~J.,  {Di Matteo} T.,  {Robertson}
  B.,   {Springel} V.,  2006, \mn@doi [\apjs] {10.1086/499298}, \href
  {http://adsabs.harvard.edu/abs/2006ApJS..163....1H} {163, 1}

\bibitem[\protect\citeauthoryear{{Hopkins}, {Hernquist}, {Cox}  \& {Kere{\v
  s}}}{{Hopkins} et~al.}{2008}]{hopkins2008}
{Hopkins} P.~F.,  {Hernquist} L.,  {Cox} T.~J.,   {Kere{\v s}} D.,  2008,
  \mn@doi [\apjs] {10.1086/524362}, \href
  {http://adsabs.harvard.edu/abs/2008ApJS..175..356H} {175, 356}

\bibitem[\protect\citeauthoryear{{Huchtmeier} \& {Richter}}{{Huchtmeier} \&
  {Richter}}{1989}]{huchtmeier1989}
{Huchtmeier} W.~K.,  {Richter} O.-G.,  1989, {A General Catalog of HI
  Observations of Galaxies. The Reference Catalog.}

\bibitem[\protect\citeauthoryear{{Israel}, {Rosenberg}  \& {van der
  Werf}}{{Israel} et~al.}{2015}]{israel2015}
{Israel} F.~P.,  {Rosenberg} M.~J.~F.,   {van der Werf} P.,  2015, \mn@doi
  [\aap] {10.1051/0004-6361/201425175}, \href
  {http://adsabs.harvard.edu/abs/2015A%26A...578A..95I} {578, A95}

\bibitem[\protect\citeauthoryear{{Kauffmann} et~al.,}{{Kauffmann}
  et~al.}{2003}]{kauffmann2003}
{Kauffmann} G.,  et~al., 2003, \mn@doi [\mnras]
  {10.1111/j.1365-2966.2003.07154.x}, \href
  {http://adsabs.harvard.edu/abs/2003MNRAS.346.1055K} {346, 1055}

\bibitem[\protect\citeauthoryear{{Kazes} \& {Dickey}}{{Kazes} \&
  {Dickey}}{1985}]{kazes1985}
{Kazes} I.,  {Dickey} J.~M.,  1985, \aap, \href
  {http://adsabs.harvard.edu/abs/1985A%26A...152L...9K} {152, L9}

\bibitem[\protect\citeauthoryear{{Kennicutt}}{{Kennicutt}}{1998}]{kennicutt1998}
{Kennicutt} Jr. R.~C.,  1998, \mn@doi [\apj] {10.1086/305588}, \href
  {http://adsabs.harvard.edu/abs/1998ApJ...498..541K} {498, 541}

\bibitem[\protect\citeauthoryear{{Kewley}, {Dopita}, {Sutherland}, {Heisler}
  \& {Trevena}}{{Kewley} et~al.}{2001}]{kewley2001}
{Kewley} L.~J.,  {Dopita} M.~A.,  {Sutherland} R.~S.,  {Heisler} C.~A.,
  {Trevena} J.,  2001, \mn@doi [\apj] {10.1086/321545}, \href
  {http://adsabs.harvard.edu/abs/2001ApJ...556..121K} {556, 121}

\bibitem[\protect\citeauthoryear{{Kewley}, {Geller}  \& {Barton}}{{Kewley}
  et~al.}{2006}]{kewley2006}
{Kewley} L.~J.,  {Geller} M.~J.,   {Barton} E.~J.,  2006, \mn@doi [\aj]
  {10.1086/500295}, \href {http://adsabs.harvard.edu/abs/2006AJ....131.2004K}
  {131, 2004}

\bibitem[\protect\citeauthoryear{{Khabiboulline}, {Steinhardt}, {Silverman},
  {Ellison}, {Mendel}  \& {Patton}}{{Khabiboulline}
  et~al.}{2014}]{khabiboulline2014}
{Khabiboulline} E.~T.,  {Steinhardt} C.~L.,  {Silverman} J.~D.,  {Ellison}
  S.~L.,  {Mendel} J.~T.,   {Patton} D.~R.,  2014, \mn@doi [\apj]
  {10.1088/0004-637X/795/1/62}, \href
  {http://adsabs.harvard.edu/abs/2014ApJ...795...62K} {795, 62}

\bibitem[\protect\citeauthoryear{{Kilerci Eser}, {Goto}  \& {Doi}}{{Kilerci
  Eser} et~al.}{2014}]{kilerci2014}
{Kilerci Eser} E.,  {Goto} T.,   {Doi} Y.,  2014, \mn@doi [\apj]
  {10.1088/0004-637X/797/1/54}, \href
  {http://adsabs.harvard.edu/abs/2014ApJ...797...54K} {797, 54}

\bibitem[\protect\citeauthoryear{{Kim} et~al.,}{{Kim} et~al.}{2013}]{kim2013}
{Kim} D.-C.,  et~al., 2013, \mn@doi [\apj] {10.1088/0004-637X/768/2/102}, \href
  {http://adsabs.harvard.edu/abs/2013ApJ...768..102K} {768, 102}

\bibitem[\protect\citeauthoryear{{Kukula}, {Ghosh}, {Pedlar}  \&
  {Schilizzi}}{{Kukula} et~al.}{1999}]{kukula1999}
{Kukula} M.~J.,  {Ghosh} T.,  {Pedlar} A.,   {Schilizzi} R.~T.,  1999, \mn@doi
  [\apj] {10.1086/307254}, \href
  {http://adsabs.harvard.edu/abs/1999ApJ...518..117K} {518, 117}

\bibitem[\protect\citeauthoryear{{Leech}, {Isaak}, {Papadopoulos}, {Gao}  \&
  {Davis}}{{Leech} et~al.}{2010}]{leech2010}
{Leech} J.,  {Isaak} K.~G.,  {Papadopoulos} P.~P.,  {Gao} Y.,   {Davis} G.~R.,
  2010, \mn@doi [\mnras] {10.1111/j.1365-2966.2010.16775.x}, \href
  {http://adsabs.harvard.edu/abs/2010MNRAS.406.1364L} {406, 1364}

\bibitem[\protect\citeauthoryear{{Liszt} \& {Lucas}}{{Liszt} \&
  {Lucas}}{1996}]{liszt1996}
{Liszt} H.,  {Lucas} R.,  1996, \aap, \href
  {http://adsabs.harvard.edu/abs/1996A%26A...314..917L} {314, 917}

\bibitem[\protect\citeauthoryear{{Liu}, {Shen}, {Strauss}  \& {Greene}}{{Liu}
  et~al.}{2010}]{liu2010}
{Liu} X.,  {Shen} Y.,  {Strauss} M.~A.,   {Greene} J.~E.,  2010, \mn@doi [\apj]
  {10.1088/0004-637X/708/1/427}, \href
  {http://adsabs.harvard.edu/abs/2010ApJ...708..427L} {708, 427}

\bibitem[\protect\citeauthoryear{{Maccagni}, {Morganti}, {Oosterloo},
  {Ger{\'e}b}  \& {Maddox}}{{Maccagni} et~al.}{2017}]{maccagni2017}
{Maccagni} F.~M.,  {Morganti} R.,  {Oosterloo} T.~A.,  {Ger{\'e}b} K.,
  {Maddox} N.,  2017, \mn@doi [\aap] {10.1051/0004-6361/201730563}, \href
  {http://adsabs.harvard.edu/abs/2017A%26A...604A..43M} {604, A43}

\bibitem[\protect\citeauthoryear{{Maccagni}, {Morganti}, {Oosterloo}, {Oonk}
  \& {Emonts}}{{Maccagni} et~al.}{2018}]{maccagni2018}
{Maccagni} F.~M.,  {Morganti} R.,  {Oosterloo} T.~A.,  {Oonk} J.~B.~R.,
  {Emonts} B.~H.~C.,  2018, \mn@doi [\aap] {10.1051/0004-6361/201732269}, \href
  {http://adsabs.harvard.edu/abs/2018A%26A...614A..42M} {614, A42}

\bibitem[\protect\citeauthoryear{{Maiolino} et~al.,}{{Maiolino}
  et~al.}{2017}]{maiolino2017}
{Maiolino} R.,  et~al., 2017, \mn@doi [\nat] {10.1038/nature21677}, \href
  {http://adsabs.harvard.edu/abs/2017Natur.544..202M} {544, 202}

\bibitem[\protect\citeauthoryear{{Mart{\'{\i}}n} et~al.,}{{Mart{\'{\i}}n}
  et~al.}{2016}]{martin2016}
{Mart{\'{\i}}n} S.,  et~al., 2016, \mn@doi [\aap]
  {10.1051/0004-6361/201528064}, \href
  {http://adsabs.harvard.edu/abs/2016A%26A...590A..25M} {590, A25}

\bibitem[\protect\citeauthoryear{{Mazzarella}, {Soifer}, {Graham},
  {Neugebauer}, {Matthews}  \& {Gaume}}{{Mazzarella}
  et~al.}{1991}]{mazzarella1991}
{Mazzarella} J.~M.,  {Soifer} B.~T.,  {Graham} J.~R.,  {Neugebauer} G.,
  {Matthews} K.,   {Gaume} R.~A.,  1991, \mn@doi [\aj] {10.1086/115950}, \href
  {http://adsabs.harvard.edu/abs/1991AJ....102.1241M} {102, 1241}

\bibitem[\protect\citeauthoryear{{Mihos} \& {Hernquist}}{{Mihos} \&
  {Hernquist}}{1996}]{mihos1996}
{Mihos} J.~C.,  {Hernquist} L.,  1996, \mn@doi [\apj] {10.1086/177353}, \href
  {http://adsabs.harvard.edu/abs/1996ApJ...464..641M} {464, 641}

\bibitem[\protect\citeauthoryear{{Moran}, {Halpern}, {Bothun}  \&
  {Becker}}{{Moran} et~al.}{1992}]{moran1992}
{Moran} E.~C.,  {Halpern} J.~P.,  {Bothun} G.~D.,   {Becker} R.~H.,  1992,
  \mn@doi [\aj] {10.1086/116292}, \href
  {http://adsabs.harvard.edu/abs/1992AJ....104..990M} {104, 990}

\bibitem[\protect\citeauthoryear{{Moreno}, {Torrey}, {Ellison}, {Patton},
  {Bluck}, {Bansal}  \& {Hernquist}}{{Moreno} et~al.}{2015}]{moreno2015}
{Moreno} J.,  {Torrey} P.,  {Ellison} S.~L.,  {Patton} D.~R.,  {Bluck}
  A.~F.~L.,  {Bansal} G.,   {Hernquist} L.,  2015, \mn@doi [\mnras]
  {10.1093/mnras/stv094}, \href
  {http://adsabs.harvard.edu/abs/2015MNRAS.448.1107M} {448, 1107}

\bibitem[\protect\citeauthoryear{{Morganti}, {Oosterloo}, {Tadhunter}, {van
  Moorsel}, {Killeen}  \& {Wills}}{{Morganti} et~al.}{2001}]{morganti2001}
{Morganti} R.,  {Oosterloo} T.~A.,  {Tadhunter} C.~N.,  {van Moorsel} G.,
  {Killeen} N.,   {Wills} K.~A.,  2001, \mn@doi [\mnras]
  {10.1046/j.1365-8711.2001.04153.x}, \href
  {http://adsabs.harvard.edu/abs/2001MNRAS.323..331M} {323, 331}

\bibitem[\protect\citeauthoryear{{Morganti}, {Oosterloo}, {Tadhunter},
  {Vermeulen}, {Pihlstr{\"o}m}, {van Moorsel}  \& {Wills}}{{Morganti}
  et~al.}{2004}]{morganti2004}
{Morganti} R.,  {Oosterloo} T.~A.,  {Tadhunter} C.~N.,  {Vermeulen} R.,
  {Pihlstr{\"o}m} Y.~M.,  {van Moorsel} G.,   {Wills} K.~A.,  2004, \mn@doi
  [\aap] {10.1051/0004-6361:20041064}, \href
  {http://adsabs.harvard.edu/abs/2004A%26A...424..119M} {424, 119}

\bibitem[\protect\citeauthoryear{{Morganti}, {Peck}, {Oosterloo}, {van
  Moorsel}, {Capetti}, {Fanti}, {Parma}  \& {de Ruiter}}{{Morganti}
  et~al.}{2009}]{morganti2009}
{Morganti} R.,  {Peck} A.~B.,  {Oosterloo} T.~A.,  {van Moorsel} G.,  {Capetti}
  A.,  {Fanti} R.,  {Parma} P.,   {de Ruiter} H.~R.,  2009, \mn@doi [\aap]
  {10.1051/0004-6361/200912605}, \href
  {http://adsabs.harvard.edu/abs/2009A%26A...505..559M} {505, 559}

\bibitem[\protect\citeauthoryear{{Morganti}, {Sadler}  \& {Curran}}{{Morganti}
  et~al.}{2015}]{morganti2015}
{Morganti} R.,  {Sadler} E.~M.,   {Curran} S.,  2015, Advancing Astrophysics
  with the Square Kilometre Array (AASKA14), \href
  {http://adsabs.harvard.edu/abs/2015aska.confE.134M} {p.~134}

\bibitem[\protect\citeauthoryear{{Morganti}, {Veilleux}, {Oosterloo}, {Teng}
  \& {Rupke}}{{Morganti} et~al.}{2016}]{morganti2016}
{Morganti} R.,  {Veilleux} S.,  {Oosterloo} T.,  {Teng} S.~H.,   {Rupke} D.,
  2016, \mn@doi [\aap] {10.1051/0004-6361/201628978}, \href
  {http://adsabs.harvard.edu/abs/2016A%26A...593A..30M} {593, A30}

\bibitem[\protect\citeauthoryear{{Mortazavi} \& {Lotz}}{{Mortazavi} \&
  {Lotz}}{2018}]{mortazavi2018}
{Mortazavi} S.~A.,  {Lotz} J.~M.,  2018, preprint, \href
  {http://adsabs.harvard.edu/abs/2018arXiv180103981M} {} (\mn@eprint {arXiv}
  {1801.03981})

\bibitem[\protect\citeauthoryear{{M{\"u}ller-S{\'a}nchez}, {Nevin},
  {Comerford}, {Davies}, {Privon}  \& {Treister}}{{M{\"u}ller-S{\'a}nchez}
  et~al.}{2018}]{muller2018}
{M{\"u}ller-S{\'a}nchez} F.,  {Nevin} R.,  {Comerford} J.~M.,  {Davies} R.~I.,
  {Privon} G.~C.,   {Treister} E.,  2018, \mn@doi [\nat]
  {10.1038/s41586-018-0033-2}, \href
  {http://adsabs.harvard.edu/abs/2018Natur.556..345M} {556, 345}

\bibitem[\protect\citeauthoryear{{Mundell}, {Ferruit}  \& {Pedlar}}{{Mundell}
  et~al.}{2001}]{mundell2001}
{Mundell} C.~G.,  {Ferruit} P.,   {Pedlar} A.,  2001, \mn@doi [\apj]
  {10.1086/322508}, \href {http://adsabs.harvard.edu/abs/2001ApJ...560..168M}
  {560, 168}

\bibitem[\protect\citeauthoryear{{Osterbrock} \& {Ferland}}{{Osterbrock} \&
  {Ferland}}{2006}]{osterbrock2006}
{Osterbrock} D.~E.,  {Ferland} G.~J.,  2006, {Astrophysics of gaseous nebulae
  and active galactic nuclei}

\bibitem[\protect\citeauthoryear{{Pearson} et~al.,}{{Pearson}
  et~al.}{2016}]{pearson2016}
{Pearson} C.,  et~al., 2016, \mn@doi [\apjs] {10.3847/0067-0049/227/1/9}, \href
  {http://adsabs.harvard.edu/abs/2016ApJS..227....9P} {227, 9}

\bibitem[\protect\citeauthoryear{{Peck}, {Taylor}, {Fassnacht}, {Readhead}  \&
  {Vermeulen}}{{Peck} et~al.}{2000}]{peck2000}
{Peck} A.~B.,  {Taylor} G.~B.,  {Fassnacht} C.~D.,  {Readhead} A.~C.~S.,
  {Vermeulen} R.~C.,  2000, \mn@doi [\apj] {10.1086/308745}, \href
  {http://adsabs.harvard.edu/abs/2000ApJ...534..104P} {534, 104}

\bibitem[\protect\citeauthoryear{{Pettini} \& {Pagel}}{{Pettini} \&
  {Pagel}}{2004}]{pettini2004}
{Pettini} M.,  {Pagel} B.~E.~J.,  2004, \mn@doi [\mnras]
  {10.1111/j.1365-2966.2004.07591.x}, \href
  {http://adsabs.harvard.edu/abs/2004MNRAS.348L..59P} {348, L59}

\bibitem[\protect\citeauthoryear{{Ramos Almeida} et~al.,}{{Ramos Almeida}
  et~al.}{2012}]{ramos2012}
{Ramos Almeida} C.,  et~al., 2012, \mn@doi [\mnras]
  {10.1111/j.1365-2966.2011.19731.x}, \href
  {http://adsabs.harvard.edu/abs/2012MNRAS.419..687R} {419, 687}

\bibitem[\protect\citeauthoryear{{Richter} \& {Huchtmeier}}{{Richter} \&
  {Huchtmeier}}{1987}]{richter1987}
{Richter} O.-G.,  {Huchtmeier} W.~K.,  1987, \aaps, \href
  {http://adsabs.harvard.edu/abs/1987A%26AS...68..427R} {68, 427}

\bibitem[\protect\citeauthoryear{{Romero-Ca{\~n}izales}
  et~al.,}{{Romero-Ca{\~n}izales} et~al.}{2017}]{romero2017}
{Romero-Ca{\~n}izales} C.,  et~al., 2017, \mn@doi [\mnras]
  {10.1093/mnras/stx224}, \href
  {http://adsabs.harvard.edu/abs/2017MNRAS.467.2504R} {467, 2504}

\bibitem[\protect\citeauthoryear{{Samui}, {Srianand}  \& {Subramanian}}{{Samui}
  et~al.}{2007}]{samui2007}
{Samui} S.,  {Srianand} R.,   {Subramanian} K.,  2007, \mn@doi [\mnras]
  {10.1111/j.1365-2966.2007.11603.x}, \href
  {http://adsabs.harvard.edu/abs/2007MNRAS.377..285S} {377, 285}

\bibitem[\protect\citeauthoryear{{Sanders} \& {Mirabel}}{{Sanders} \&
  {Mirabel}}{1996}]{sanders1996}
{Sanders} D.~B.,  {Mirabel} I.~F.,  1996, \mn@doi [\araa]
  {10.1146/annurev.astro.34.1.749}, \href
  {http://adsabs.harvard.edu/abs/1996ARA%26A..34..749S} {34, 749}

\bibitem[\protect\citeauthoryear{{Sanders}, {Mazzarella}, {Kim}, {Surace}  \&
  {Soifer}}{{Sanders} et~al.}{2003}]{sanders2003}
{Sanders} D.~B.,  {Mazzarella} J.~M.,  {Kim} D.-C.,  {Surace} J.~A.,   {Soifer}
  B.~T.,  2003, \mn@doi [\aj] {10.1086/376841}, \href
  {http://adsabs.harvard.edu/abs/2003AJ....126.1607S} {126, 1607}

\bibitem[\protect\citeauthoryear{{Satyapal}, {Ellison}, {McAlpine}, {Hickox},
  {Patton}  \& {Mendel}}{{Satyapal} et~al.}{2014}]{satyapal2014}
{Satyapal} S.,  {Ellison} S.~L.,  {McAlpine} W.,  {Hickox} R.~C.,  {Patton}
  D.~R.,   {Mendel} J.~T.,  2014, \mn@doi [\mnras] {10.1093/mnras/stu650},
  \href {http://adsabs.harvard.edu/abs/2014MNRAS.441.1297S} {441, 1297}

\bibitem[\protect\citeauthoryear{{Satyapal} et~al.,}{{Satyapal}
  et~al.}{2017}]{satyapal2017}
{Satyapal} S.,  et~al., 2017, \mn@doi [\apj] {10.3847/1538-4357/aa88ca}, \href
  {http://adsabs.harvard.edu/abs/2017ApJ...848..126S} {848, 126}

\bibitem[\protect\citeauthoryear{{Schawinski} et~al.,}{{Schawinski}
  et~al.}{2006}]{schawinski2006}
{Schawinski} K.,  et~al., 2006, \mn@doi [\nat] {10.1038/nature04934}, \href
  {http://adsabs.harvard.edu/abs/2006Natur.442..888S} {442, 888}

\bibitem[\protect\citeauthoryear{{Schmitt}}{{Schmitt}}{2001}]{schmitt2001}
{Schmitt} H.~R.,  2001, \mn@doi [\aj] {10.1086/323547}, \href
  {http://adsabs.harvard.edu/abs/2001AJ....122.2243S} {122, 2243}

\bibitem[\protect\citeauthoryear{{Scott} \& {Kaviraj}}{{Scott} \&
  {Kaviraj}}{2014}]{scott2014}
{Scott} C.,  {Kaviraj} S.,  2014, \mn@doi [\mnras] {10.1093/mnras/stt2014},
  \href {http://adsabs.harvard.edu/abs/2014MNRAS.437.2137S} {437, 2137}

\bibitem[\protect\citeauthoryear{{Scudder}, {Ellison}, {Torrey}, {Patton}  \&
  {Mendel}}{{Scudder} et~al.}{2012}]{scudder2012}
{Scudder} J.~M.,  {Ellison} S.~L.,  {Torrey} P.,  {Patton} D.~R.,   {Mendel}
  J.~T.,  2012, \mn@doi [\mnras] {10.1111/j.1365-2966.2012.21749.x}, \href
  {http://adsabs.harvard.edu/abs/2012MNRAS.426..549S} {426, 549}

\bibitem[\protect\citeauthoryear{{Secrest}, {Schmitt}, {Blecha}, {Rothberg}  \&
  {Fischer}}{{Secrest} et~al.}{2017}]{secrest2017}
{Secrest} N.~J.,  {Schmitt} H.~R.,  {Blecha} L.,  {Rothberg} B.,   {Fischer}
  J.,  2017, \mn@doi [\apj] {10.3847/1538-4357/836/2/183}, \href
  {http://adsabs.harvard.edu/abs/2017ApJ...836..183S} {836, 183}

\bibitem[\protect\citeauthoryear{{Soto} \& {Martin}}{{Soto} \&
  {Martin}}{2012}]{soto2012}
{Soto} K.~T.,  {Martin} C.~L.,  2012, \mn@doi [\apjs]
  {10.1088/0067-0049/203/1/3}, \href
  {http://adsabs.harvard.edu/abs/2012ApJS..203....3S} {203, 3}

\bibitem[\protect\citeauthoryear{{Springel}, {Di Matteo}  \&
  {Hernquist}}{{Springel} et~al.}{2005}]{springel2005}
{Springel} V.,  {Di Matteo} T.,   {Hernquist} L.,  2005, \mn@doi [\mnras]
  {10.1111/j.1365-2966.2005.09238.x}, \href
  {http://adsabs.harvard.edu/abs/2005MNRAS.361..776S} {361, 776}

\bibitem[\protect\citeauthoryear{{Srianand}, {Gupta}, {Momjian}  \&
  {Vivek}}{{Srianand} et~al.}{2015}]{srianand2015}
{Srianand} R.,  {Gupta} N.,  {Momjian} E.,   {Vivek} M.,  2015, \mn@doi
  [\mnras] {10.1093/mnras/stv1004}, \href
  {http://adsabs.harvard.edu/abs/2015MNRAS.451..917S} {451, 917}

\bibitem[\protect\citeauthoryear{{Staveley-Smith}, {Cohen}, {Chapman},
  {Pointon}  \& {Unger}}{{Staveley-Smith} et~al.}{1987}]{staveley1987}
{Staveley-Smith} L.,  {Cohen} R.~J.,  {Chapman} J.~M.,  {Pointon} L.,   {Unger}
  S.~W.,  1987, \mn@doi [\mnras] {10.1093/mnras/226.3.689}, \href
  {http://adsabs.harvard.edu/abs/1987MNRAS.226..689S} {226, 689}

\bibitem[\protect\citeauthoryear{{Strauss}, {Huchra}, {Davis}, {Yahil},
  {Fisher}  \& {Tonry}}{{Strauss} et~al.}{1992}]{strauss1992}
{Strauss} M.~A.,  {Huchra} J.~P.,  {Davis} M.,  {Yahil} A.,  {Fisher} K.~B.,
  {Tonry} J.,  1992, \mn@doi [\apjs] {10.1086/191730}, \href
  {http://adsabs.harvard.edu/abs/1992ApJS...83...29S} {83, 29}

\bibitem[\protect\citeauthoryear{{Sun}, {Greene}, {Zakamska}  \&
  {Nesvadba}}{{Sun} et~al.}{2014}]{sun2014}
{Sun} A.-L.,  {Greene} J.~E.,  {Zakamska} N.~L.,   {Nesvadba} N.~P.~H.,  2014,
  \mn@doi [\apj] {10.1088/0004-637X/790/2/160}, \href
  {http://adsabs.harvard.edu/abs/2014ApJ...790..160S} {790, 160}

\bibitem[\protect\citeauthoryear{{Tadhunter}, {Ramos Almeida}, {Morganti},
  {Holt}, {Rose}, {Dicken}  \& {Inskip}}{{Tadhunter}
  et~al.}{2012}]{tadhunter2012}
{Tadhunter} C.~N.,  {Ramos Almeida} C.,  {Morganti} R.,  {Holt} J.,  {Rose} M.,
   {Dicken} D.,   {Inskip} K.,  2012, \mn@doi [\mnras]
  {10.1111/j.1365-2966.2012.22058.x}, \href
  {http://adsabs.harvard.edu/abs/2012MNRAS.427.1603T} {427, 1603}

\bibitem[\protect\citeauthoryear{{Toomre} \& {Toomre}}{{Toomre} \&
  {Toomre}}{1972}]{toomre1972}
{Toomre} A.,  {Toomre} J.,  1972, \mn@doi [\apj] {10.1086/151823}, \href
  {http://adsabs.harvard.edu/abs/1972ApJ...178..623T} {178, 623}

\bibitem[\protect\citeauthoryear{{Treister}, {Schawinski}, {Urry}  \&
  {Simmons}}{{Treister} et~al.}{2012}]{treister2012}
{Treister} E.,  {Schawinski} K.,  {Urry} C.~M.,   {Simmons} B.~D.,  2012,
  \mn@doi [\apjl] {10.1088/2041-8205/758/2/L39}, \href
  {http://adsabs.harvard.edu/abs/2012ApJ...758L..39T} {758, L39}

\bibitem[\protect\citeauthoryear{{Treister} et~al.,}{{Treister}
  et~al.}{2018}]{treister2018}
{Treister} E.,  et~al., 2018, \mn@doi [\apj] {10.3847/1538-4357/aaa963}, \href
  {http://adsabs.harvard.edu/abs/2018ApJ...854...83T} {854, 83}

\bibitem[\protect\citeauthoryear{{Tremblay} et~al.,}{{Tremblay}
  et~al.}{2016}]{tremblay2016}
{Tremblay} G.~R.,  et~al., 2016, \mn@doi [\nat] {10.1038/nature17969}, \href
  {http://adsabs.harvard.edu/abs/2016Natur.534..218T} {534, 218}

\bibitem[\protect\citeauthoryear{{Vermeulen} et~al.,}{{Vermeulen}
  et~al.}{2003}]{vermeulen2003}
{Vermeulen} R.~C.,  et~al., 2003, \mn@doi [\aap] {10.1051/0004-6361:20030468},
  \href {http://adsabs.harvard.edu/abs/2003A%26A...404..861V} {404, 861}

\bibitem[\protect\citeauthoryear{{Villar-Mart{\'{\i}}n}, {Cabrera Lavers},
  {Bessiere}, {Tadhunter}, {Rose}  \& {de Breuck}}{{Villar-Mart{\'{\i}}n}
  et~al.}{2012}]{villar2012}
{Villar-Mart{\'{\i}}n} M.,  {Cabrera Lavers} A.,  {Bessiere} P.,  {Tadhunter}
  C.,  {Rose} M.,   {de Breuck} C.,  2012, \mn@doi [\mnras]
  {10.1111/j.1365-2966.2012.20652.x}, \href
  {http://adsabs.harvard.edu/abs/2012MNRAS.423...80V} {423, 80}

\bibitem[\protect\citeauthoryear{{Villforth} et~al.,}{{Villforth}
  et~al.}{2014}]{villforth2014}
{Villforth} C.,  et~al., 2014, \mn@doi [\mnras] {10.1093/mnras/stu173}, \href
  {http://adsabs.harvard.edu/abs/2014MNRAS.439.3342V} {439, 3342}

\bibitem[\protect\citeauthoryear{{Weinmann}, {van den Bosch}, {Yang}  \&
  {Mo}}{{Weinmann} et~al.}{2006}]{weinmann2006}
{Weinmann} S.~M.,  {van den Bosch} F.~C.,  {Yang} X.,   {Mo} H.~J.,  2006,
  \mn@doi [\mnras] {10.1111/j.1365-2966.2005.09865.x}, \href
  {http://adsabs.harvard.edu/abs/2006MNRAS.366....2W} {366, 2}

\bibitem[\protect\citeauthoryear{{Weston}, {McIntosh}, {Brodwin}, {Mann},
  {Cooper}, {McConnell}  \& {Nielsen}}{{Weston} et~al.}{2017}]{weston2017}
{Weston} M.~E.,  {McIntosh} D.~H.,  {Brodwin} M.,  {Mann} J.,  {Cooper} A.,
  {McConnell} A.,   {Nielsen} J.~L.,  2017, \mn@doi [\mnras]
  {10.1093/mnras/stw2620}, \href
  {http://adsabs.harvard.edu/abs/2017MNRAS.464.3882W} {464, 3882}

\bibitem[\protect\citeauthoryear{{White} \& {Rees}}{{White} \&
  {Rees}}{1978}]{white1978}
{White} S.~D.~M.,  {Rees} M.~J.,  1978, \mn@doi [\mnras]
  {10.1093/mnras/183.3.341}, \href
  {http://adsabs.harvard.edu/abs/1978MNRAS.183..341W} {183, 341}

\bibitem[\protect\citeauthoryear{{White}, {Becker}, {Helfand}  \&
  {Gregg}}{{White} et~al.}{1997}]{white1997}
{White} R.~L.,  {Becker} R.~H.,  {Helfand} D.~J.,   {Gregg} M.~D.,  1997, \apj,
  \href {http://adsabs.harvard.edu/abs/1997ApJ...475..479W} {475, 479}

\bibitem[\protect\citeauthoryear{{Wild}, {Kauffmann}, {Heckman}, {Charlot},
  {Lemson}, {Brinchmann}, {Reichard}  \& {Pasquali}}{{Wild}
  et~al.}{2007}]{wild2007}
{Wild} V.,  {Kauffmann} G.,  {Heckman} T.,  {Charlot} S.,  {Lemson} G.,
  {Brinchmann} J.,  {Reichard} T.,   {Pasquali} A.,  2007, \mn@doi [\mnras]
  {10.1111/j.1365-2966.2007.12256.x}, \href
  {http://adsabs.harvard.edu/abs/2007MNRAS.381..543W} {381, 543}

\bibitem[\protect\citeauthoryear{{Williams} \& {Brown}}{{Williams} \&
  {Brown}}{1983}]{williams1983}
{Williams} B.~A.,  {Brown} R.~L.,  1983, \mn@doi [\aj] {10.1086/113466}, \href
  {http://adsabs.harvard.edu/abs/1983AJ.....88.1749W} {88, 1749}

\bibitem[\protect\citeauthoryear{{Wright} et~al.,}{{Wright}
  et~al.}{2010}]{wright2010}
{Wright} E.~L.,  et~al., 2010, \mn@doi [\aj] {10.1088/0004-6256/140/6/1868},
  \href {http://adsabs.harvard.edu/abs/2010AJ....140.1868W} {140, 1868}

\bibitem[\protect\citeauthoryear{{York} et~al.,}{{York}
  et~al.}{2000}]{york2000}
{York} D.~G.,  et~al., 2000, \mn@doi [\aj] {10.1086/301513}, \href
  {http://adsabs.harvard.edu/abs/2000AJ....120.1579Y} {120, 1579}

\bibitem[\protect\citeauthoryear{{van Gorkom}, {Knapp}, {Ekers}, {Ekers},
  {Laing}  \& {Polk}}{{van Gorkom} et~al.}{1989}]{vangorkom1989}
{van Gorkom} J.~H.,  {Knapp} G.~R.,  {Ekers} R.~D.,  {Ekers} D.~D.,  {Laing}
  R.~A.,   {Polk} K.~S.,  1989, \mn@doi [\aj] {10.1086/115016}, \href
  {http://adsabs.harvard.edu/abs/1989AJ.....97..708V} {97, 708}

\makeatother
\end{thebibliography}
\bsp
\label{lastpage}
\end{document}